\documentclass[12pt,preprint]{aastex}
\usepackage{epsfig}
\usepackage{natbib}
\usepackage{graphicx}
\usepackage{rotating}
\usepackage{multirow}
\usepackage{color}
\usepackage{lscape}
\usepackage{mathrsfs,amssymb}
\usepackage{amsmath}
\usepackage{morefloats}
\newcommand	\beq	{\begin{equation}}	
\newcommand	\eeq	{\end{equation}}	
\newcommand       \Angstrom     {\,{\rm \AA}}

\newcommand       \cm           {\,{\rm cm}}

\newcommand       \km           {\,{\rm km}}

\newcommand       \g            {\,{\rm g}}
\newcommand       \K            {\,{\rm K}}

\newcommand       \kpc          {\,{\rm kpc}}

\newcommand       \s            {\,{\rm s}}

\newcommand       \simlt        {\lesssim}
\newcommand       \simgt        {\gtrsim}

\newcommand       \mum          {\,{\rm \mu m}}

\newcommand       \simali       {\sim\,}

\newcommand       \Ks            {{K_S}}
\newcommand       \Msil           {M_{\rm sil}}
\newcommand       \Mcarb          {M_{\rm carb}}
\def    \obs		{{\rm obs}}
\def    \mod	{{\rm mod}}
\def    \Nobs	{N_{\rm obs}}
\def    \Nmod	{N_{\rm mod}}
\def    \dof	        {{\rm dof}}

\newcommand{\spitzerirs}{{\em Spitzer}/IRS\ }

\newcommand{\saga}{{\em \rm S$^3$AGA}}
\newcommand{\quasam} {{\em \rm quasar sample}}

\newcommand \Lbol { L_{\rm bol}}
\newcommand \MBH  { M_{\rm BH}}
\newcommand \Ledd { L_{\rm Edd}}
\newcommand{\etal}{\textrm{et al.\ }}

\newcommand{\eg}{\textrm{e.g., }}

\newcommand{\Ha}        {\,{\rm {H\alpha}}}

\newcommand \Tw           {T_{\rm w}}
\newcommand \Tc           {T_{\rm c}}
\newcommand \Twi          {T_{{\rm w},i}}
\newcommand \Tci          {T_{{\rm c},i}}
\newcommand \Mwi          {M_{{\rm w},i}}
\newcommand \Mci          {M_{{\rm c},i}}
\newcommand \knui         {\kappa_{{\rm abs},i}(\nu)}
\newcommand \Twsil        {T_{\rm w}^{\rm S}}
\newcommand \Tcsil        {T_{\rm c}^{\rm S}}
\newcommand \Mwsil        {M_{\rm w}^{\rm S}}
\newcommand \Mcsil        {M_{\rm c}^{\rm S}}
\newcommand \Twcarb        {T_{\rm w}^{\rm G}}
\newcommand \Tccarb        {T_{\rm c}^{\rm G}}
\newcommand \Mwcarb        {M_{\rm w}^{\rm G}}
\newcommand \Mccarb        {M_{\rm c}^{\rm G}}

\renewcommand\i   {\hbox{$i$}}

\countdef\decade=200
\advance\decade by \year
\countdef\hours=201
\advance\hours by \time
\divide\hours by 60
\countdef\mins=202
\advance\mins by \hours
\multiply\mins by 60
\multiply\hours by 100
\countdef\miltime=203
\advance\miltime by \hours
\advance\miltime by \time
\advance\miltime by -\mins
\def\today{\number\decade.\number\month.\number\day.\number\miltime}
\shorttitle{Quasar dust properties}
\title{
Silicate Dust in Active Galactic Nuclei 
\\{\small DRAFT: \today ~~}
}
\author{
Yanxia~Xie\altaffilmark{1,2,3},
Aigen~Li\altaffilmark{2}, 
and Lei~Hao\altaffilmark{1}
}
\altaffiltext{1}{Shanghai Astronomical Observatory,
                       Chinese Academy of Sciences,
                       80 Nandan Road, Shanghai 200030, China;
                       {\sf xieya@missouri.edu, haol@shao.ac.cn}
                       }
\altaffiltext{2}{Department of Physics and Astronomy,
                       University of Missouri,
                       Columbia, MO 65211, USA;
                       {\sf lia@missouri.edu}
                       }
\altaffiltext{3}{Kavli Institute for Astronomy and Astrophysics,
                       Peking University, Beijing 100871, China;
                       {\sf yanxia.xie@pku.edu.cn}
                       }

\begin{document}

\begin{abstract}
The unification theory of active galactic nuclei 
(AGNs) hypothesizes that all AGNs are
surrounded by an anisotropic dust torus
and are essentially the same objects 
but viewed from different angles.
However, little is known about the dust
which plays a central role 
in the unification theory.
There are suggestions that
the AGN dust extinction law
appreciably differs from that of the Galaxy. 
Also, the silicate emission features 
observed in type 1 AGNs appear anomalous
(i.e., their peak wavelengths and widths differ
considerably from that of the Galaxy).
In this work, we explore the dust properties of 
147 AGNs of various types
at redshifts $z\simlt0.5$,
with special attention paid to 93 AGNs
which exhibit the 9.7 and 18$\mum$ 
silicate emission features.
We model their silicate emission spectra 
obtained with the {\it Infrared Spectrograph} 
aboard the {\it Spitzer Space Telescope}.
We find that 60/93 of the observed spectra 
can be well explained with
``astronomical silicate'',
while the remaining sources favor
amorphous olivine or pyroxene.
Most notably, all sources require the dust to be 
$\mu$m-sized (with a typical size of 
$\simali$1.5$\pm$0.1$\mum$),
much larger than sub-$\mu$m-sized 
Galactic interstellar grains,
implying a flat or ``gray'' extinction law for AGNs.
We also find that, while the 9.7$\mum$ emission feature 
arises predominantly from warm silicate dust 
of temperature $T$\,$\simali$270$\K$,
the $\simali$5--8$\mum$ continuum emission
is mostly from carbon dust of $T$\,$\simali$640$\K$.
Finally, the correlations between the dust properties 
(e.g., mass, temperature) and the AGN properties
(e.g., luminosity, black hole mass) have also 
been investigated.
\end{abstract}


\section{Introduction\label{sec:intro}}
The unification theory of active galactic nuclei (AGNs) 
invokes an anisotropic dusty torus to account for 
the observational dichotomy of AGNs 
(Atonucci 1993, Urry \& Padovani 1995). 
This theory assumes that, for type 2 AGNs,
the dust torus blocks the photons from the broad line region 
and accretion disk when they are viewed 
through the edge of the torus.  
For type 1 AGNs, 
the line of sight is perpendicular to 
the dusty torus and allows the detection 
of the broad emission lines. 
The existence of such a dust torus around AGNs
has been confirmed through the detection of 
polarized broad emission lines in type 2 AGNs.
These polarized lines are believed to arise from
the otherwise-blocked broad line regions in type 2 AGNs,
and they are detected just because they are scattered 
by dust into the viewing line-of-sight  
(\eg see Antonucci \& Miller 1985, Lumsden \etal 2004). 

What is the AGN dust torus composed of?
essentially, the torus forms out of 
the original interstellar matter (gas and dust) 
of the host galaxies of AGNs.
These dust grains of interstellar origin 
are processed by the X-ray and ultraviolet (UV) radiation
of the central engine (e.g., see Voit 1991, Li 2007).
They may also undergo coagulational growth 
in the torus (e.g., see Maiolino et al.\ 2001). 
As amorphous silicate and some sorts of
carbonaceous dust are the major dust species
of interstellar grains (e.g., see Mishra \& Li 2015),
one naturally expects amorphous silicate dust 
and carbon dust (e.g., graphite, amorphous carbon) 
to be present in the dust torus around AGNs.
According to the unification theory,
type 1 AGNs are expected to show silicate emission
around 9.7 and 18$\mum$ 
while type 2 AGNs are expected to show silicate absorption.
The detection of silicate emission in a wide variety of type 1 AGNs
ranging from luminous quasars to low-luminosity Seyfert galaxies 
(\eg see Hao \etal 2005a, Siebenmorgen \etal 2005, 
Sturm \etal 2005, Hao \etal 2007, Xie et al.\ 2014)
as well as silicate absorption in type 2 Seyfert galaxies 
(\eg see Rieke \& Low 1975, Jaffe \etal 2004, 
Roche \etal 2007, Shi \etal 2014) and type 2 QSOs 
(\eg see Sturm \etal 2006, Nikutta \etal 2009)
provides further support for the unification theory of AGNs.  

However, the detection of silicate emission 
(and absorption as well) in AGNs 
is found to be rather diverse 
among different AGN types as revealed 
from the rich data set obtained by
the {\it  Infrared Spectrograph} (IRS) 
on board the {\it Spitzer Space Telescope}
(Houck \etal 2004). 
Type 1 Seyfert galaxies are found equally 
displaying silicate emission and weak absorption 
in the mid infrared (IR; Hao \etal 2007),
meanwhile some type 2 AGNs 
exhibit silicate emission rather than absorption 
(e.g., see Sturm \etal 2006, Mason \etal 2009, 
Nikutta \etal 2009). 
For those in which the silicate dust 
is seen in emission, the silicate feature 
often shows an ``anomalous'' spectral profile:
the peak wavelength of the Si--O stretching feature 
which has a canonical wavelength of $\simali$9.7$\mum$
in the Milky Way diffuse interstellar medium (ISM;
e.g., see Kemper et al.\ 2004, 
Chiar \& Tielens 2006, Henning 2010)
often shifts to longer wavelengths 
beyond $\simali$10$\mum$ in AGNs 
(\eg see Sturm \etal 2005, Mason 2015). 
Also, this feature often shows 
a much broader width compared to 
that of the Milky Way diffuse ISM 
(see Li et al.\ 2008, Smith et al.\ 2010 
and references therein).

The spectral profile of the 9.7$\mum$ silicate 
absorption feature seen in AGNs 
also exhibit spatial variations. 
Spatially-resolved observations of
Circinus, a Seyfert 2 galaxy,
made by Tristram \etal (2007) using 
the {\it Mid-Infrared Interferometric} 
(MIDI) instrument 
at the {\it Very Large Telescope} (VLT)
reveals a two-component structure: 
an inner disk-like ($\simali$0.4\,pc)
component showing mild silicate emission 
around $\simali$10$\mum$, 
and an outer, extended, cooler torus 
($\simali$2.0\,pc) exhibiting silicate absorption.
Both the emission and absorption features 
of Circinus resemble the spectral profile of 
the Milky Way diffuse ISM.
On the other hand, the spatially-resolved mid-IR 
spectrum of NGC~1068, a prototypical type 2 AGN,
shows that the silicate absorption profile varies with 
the distance to the nucleus, with the maximum absorption 
occurring around the innermost region
(Rhee \& Larkin 2006, Mason \etal 2006). 
Particularly, the {\it VLT}/MIDI observations 
of the central $\simali$2.0\,pc of NGC~1068 
reveal that the silicate absorption profile
also appears ``anomalous'':
differing from that of the Galactic ISM
and that of common olivine-type silicate dust, 
the 9.7$\mum$ silicate absorption feature of NGC\,1068
shows a relatively flat profile from $\simali$8 to 9$\mum$
and then a sharp drop between $\simali$9 and 10$\mum$, 
while the Galactic silicate absorption profiles 
already begin to drop at $\simali$8$\mum$
(see Figure~1 of K\"ohler \& Li 2010).

Li et al.\ (2008) examined the anomalous redward-shifting 
of the peak wavelength and broadening of the width of
the 9.7$\mum$ emission feature observed in 
the bright quasar 3C\,273 and the low-luminosity AGN
NGC\,3998. They ascribed the anomalous silicate
emission profile of 3C\,273 and NGC\,3998 to porous dust. 
Such an anomalous emission profile is also detected
in the type 1 nucleus of M\,81, a low-ionization nuclear 
emission-line region (LINER) and is explained in terms 
of $\mu$m-sized grains (Smith et al.\ 2010).
In contrast, the anomalous spectral profile of 
the 9.7$\mum$ silicate feature 
observed in the innermost region of NGC\,1068 
was attributed to the presence of 
silicon carbide dust (K\"ohler \& Li 2010)
or to a less extent, gehlenite (Ca$_2$Al$_2$SiO$_7$), 
a high-temperature calcium aluminum silicate species.
However, Nikutta \etal (2009) argued that,
in the framework of a clumpy dust torus, 
the observed anomaly in the silicate emission 
and absorption profiles does not necessarily imply 
anomalies in dust size, structure or composition;
instead, they argued that it could simply be caused 
by radiation transfer effects.
But Xie et al.\ (2015, 2016) noted that the success of
the ``Clumpy'' dust torus model 
(Nenkova \etal 2008a,b, Schartmann \etal 2008,  
H\"{o}nig \& Kishimoto 2010) 
in explaining the much longer peak wavelength of
the silicate Si--O emission feature 
(compared to that of the Galactic diffuse ISM) 
seems to be due to the adoption of a set of 
silicate opacity differing from that commonly adopted: 
Nikutta \etal (2009) adopted 
the silicate opacity of Ossenkopf et al.\ (1992)
which peaks at $\simali$10.0$\mum$ 
while the commonly-adopted opacity profile of 
``astronomical silicate'' of Draine \& Lee (1984)
peaks at $\simali$9.5$\mum$. 
Note that the observed silicate absorption profiles 
of the Galactic diffuse ISM peak at $\simali$9.7$\mum$ 
(e.g., see Kemper et al.\ 2004, 
Chiar \& Tielens 2006, Henning 2010).

The exact properties of the silicate dust component
in AGN torus remain debated and no consensus has 
yet been reached.  
As elaborated above, the current knowledge 
about the silicate dust properties of AGNs 
are mainly derived from several individual sources.
To address the observed silicate diversity among AGNs 
and to gain insight into the origin of the AGN dichotomy,
it is necessary to study the silicate spectral profiles 
for a large and well defined AGN sample,
taking into account a wide range of dust compositions 
and sizes. In this work we will model the {\it Spitzer}/IRS spectra
of a large sample of 147 AGNs, 
including both type 1 and type 2 AGNs 
at both high and low luminosity levels.  
Such a large AGN sample will allow us to obtain 
a better understanding of the size and composition 
of the dust grains in AGN torus.
For simplicity, we will focus on those AGNs showing 
silicate emission (see \S\ref{sec:sampleA}).
 
The structure of this paper is organized 
in the following sequence. 
We briefly describe the sample in \S\ref{sec:sample}
and elaborate the dust model in \S\ref{sec:model}.
We present in \S\ref{sec:results} the results derived from 
modeling the {\it Spitzer}/IRS spectra of 147 AGNs.
Also in \S\ref{sec:results} we discuss the model-derived
dust properties (e.g., composition, size, temperature) 
and their relations with the AGN parameters  
(e.g., accretion rate, luminosity, black hole mass).
We summarize the major results of this paper
in \S\ref{sec:summary}.
Throughout the paper, we take the following cosmology 
parameters: $ \Omega_{m} = 0.3$, $\Omega_{\land} = 0.7$, 
and $H_{0} = 70\,h_{70}$ km\,s$^{-1}$\,Mpc$^{-1}$.


\section{Sample and Data\label{sec:sample}}

\subsection{Samples\label{sec:sampleA}}
Our AGN sample is mainly collected from the literature.  
We consider all 87 PG quasars at $z\simlt0.5$
(Schmidt \& Green 1983, Boroson \& Green 1992) 
and all 52 {\it 2MASS} quasars  at $z\simlt0.3$
(Cutri \etal 2001, Smith \etal 2002).
We also consider all 253 AGNs from 
the {\it Spitzer/IRS-SDSS Spectral Atlas of Galaxies and AGNs}
(\saga; L.~Hao \etal 2016, in preparation) at $z\simlt0.33$.

PG quasars are selected to have an average 
$B$-band absolute magnitude of $\simali$16.16, 
$U-B$ color of $\simlt-0.44$, 
and a dominant point-like sources. 
All these objects show broad emission lines in optical,
and thus are classified as type 1 quasars. 
Due to the large photographic magnitude errors 
and the simple color selection criterion,
the PG sample is incomplete 
(\eg see Goldschmidt \etal 1992, Jester \etal 2005), 
but the incompleteness is independent of the optical 
magnitude and color (Jester \etal 2005).
This indicates that the PG sample is still representative 
of bright optically selected quasars. 
In comparison with PG quasars, the {\it 2MASS} quasars
represent a redder population with $J-\Ks\simgt2$ 
(compared to a typical color of 
$J-\Ks\simgt1.5$ for PG quasars), 
but have a similar $\Ks$-band luminosity
of $\Ks\simlt−23$ (Smith \etal 2002). 
Unlike PG quasars, the {\it 2MASS} sample 
includes objects with narrow, intermediate 
and broad emission lines. 
The {\it 2MASS} sample is 
highly incomplete 
at $\Ks\simgt13$ (Cutri \etal 2001).
Throughout the following text, we will 
refer to this sample as the \quasam.
 
\saga\ is a heterogeneous collection
of galaxies that have {\it Spitzer}/IRS
low-resolution spectra (Houck \etal 2004) 
and {\it Sloan Digital Sky Survey}
(SDSS) spectroscopic observations 
(Data Release 7; Abazajian \etal 2009) 
within a $\simali$3$^{\prime\prime}$ searching radius.
The whole \saga\ sample contains 139 type 1 AGNs, 
114 type 2 AGNs, 241 star-forming (SF) galaxies, 
103 AGN-SF composites, and one quiescent galaxy. 
The galaxy types are classified based on 
the {\it SDSS} optical emission line properties 
(see Hao \etal 2005b).\footnote{%
  Type 1 AGNs are those exhibiting broad $\Ha$ emission lines 
  with an FWHM of $\simgt1200\km\s^{-1}$. 
  Type 2 AGNs are identified with the emission line ratios
   featuring the Baldwin, Phillips, \& Terlevich (1989; BPT) diagram. 
   This sample spans a redshift range of $z$\,$\simali$0.001--0.25, 
   corresponding to a physical size of $\simali$0.06--18$\kpc$ 
   in the {\it SDSS} $3^{\prime\prime}$ aperture.
   }
Throughout the following text, we will refer to this sample as \saga. 



In this work we will disregard those sources 
which show silicate in absorption 
since they do not contain 
a sufficient amount of information 
for constraining the nature of the dust 
(particularly, the temperature of the dust). 
Also, the silicate absorption could have been 
contaminated by the interstellar silicate dust 
of the AGN's host galaxy. 
Therefore, we are left with 147 sources 
(i.e., 85 PG quasars, 18 {\it 2MASS} quasars, 
and 44 \saga\ AGNs).
In Appendix we list the basic parameters
including the redshift ($z$), type, 
black-hole mass ($M_{\rm BH}$),
and luminosity at $\lambda=5100\Angstrom$ 
for each of our 147 sources.
Among these objects, the {\it Spitzer}/IRS
spectra of 62 PG quasars, 13 {\it 2MASS} quasars 
and 18 \saga\ AGNs show silicate in emission.
We will focus on these 93 ``silicate emission'' sources 
which show silicate emission at 9.7 and 18$\mum$.


\subsection{Data\label{sec:data}}
For the selected sample sources, 
we utilize the low resolution mid-IR spectra 
obtained by {\it Spitzer}/IRS. 
The spectral wavelength ranges from 
$\simali$5$\mum$ to $\simali$38$\mum$ 
and the spectral resolution varies between 
$\simali$60 and $\simali$128. 
 
The {\it Spitzer}/IRS spectra 
for the  \quasam\ are compiled 
from Shi \etal (2014). 
The detailed observations and data reduction 
can be found in Shi \etal (2014) and reference therein. 
For the \saga~sample, 
the mid-IR spectra are obtained from 
the {\it Cornell Atlas of Spitzer/IRS Sources} (CASSIS)
which have been processed 
with Pipeline S18.7 (Lebouteiller \etal 2011).
For more details we refer to L.~Hao \etal 
(2016, in preparation).


We have not specifically applied any 
quantitative signal-to-noise (SNR) 
cut to the selected spectra;
instead, the selection is mainly based on
visual inspection and we require
an apparent detection of the 9.7 
and 18$\mum$ silicate emission features
or a flat, featureless emission continuum.\footnote{%
   For some sources, the flat emission continuum
   may be superimposed by several spectral features
   from polycyclic aromatic hydrocarbon (PAH) molecules
   (see \S\ref{sec:other}).
   }
As demonstrated in \S\ref{sec:other},
the ``silicate emission'' sources 
and the ``flat continuum'' sources
appreciably distinguish themselves 
from each other in terms of the flux fraction
of the silicate emission features in the total
mid-IR emission. Finally, we note that
for those sources whose {\it Spitzer}/IRS spectra
are of a rather low SNR, we will exclude them 
when we statistically examine whether (and how)
the dust properties are related to the AGN properties.


\section{The Dust Model\label{sec:model}}
We aim to constrain the dust chemical composition,
size, and temperature through modeling the observed
dust thermal IR emission. 
We will consider two kinds of dust: 
amorphous silicate and carbonaceous dust. 
For the former, we will consider a range of
compositions: (1) the Draine \& Lee (1984)
``astronomical silicate'', 
(2) three pyroxene species
(Mg$_{\rm x}$Fe$_{\rm 1-x}$SiO$_3$
with ${\rm x} = 0.4, 0.7, 1.0$), and 
(3) two olivine species
(Mg$_{\rm 2x}$Fe$_{\rm 2(1-x)}$SiO$_4$
with ${\rm x} = 0.4, 0.5$). 
For the latter, we will consider graphite 
and amorphous carbon. 
Although other dust species
(e.g., SiC, oxides) may also be present 
in AGN torus
(e.g., see Laor \& Draine 1993,
Markwick-Kemper et al.\ 2007,
K\"oher \& Li 2010),
in this work they are not included in our model.
The dust is expected to have a distribution of sizes.
For simplicity, we will only consider 100 discrete sizes:
$a$\,=\,0.1, 0.2, ..., 10.0$\mum$ at a step of 0.1$\mum$
where $a$ is the spherical radius of the dust
(we assume a spherical shape for the dust).
The dust is also expected to
have a distribution of temperatures,
with the dust temperature reaching $\simgt$1500$\K$
--- the sublimation temperature of silicate, graphite and
amorphous carbon and dropping to $\simlt$100$\K$
in the outer boundary of the torus.
Also for simplicity, we will only consider two temperatures
--- a warm component of temperature $\Tw$
and a cold component of temperature $\Tc$--- to
represent the temperature distribution.

Assuming that the torus is optically thin in the IR,
we model the dust IR emission as
\begin{equation}
\label{eq:num_1}
F_{\nu} = \frac{1}{d^2} \times 
          \sum_{i} \left\{
          B_\nu(\Twi) 
          \times \knui
          \times \Mwi 
        +  B_\nu(\Tci) 
          \times \knui
          \times \Mci\right\} ~~,
\end{equation}
where the sum is over the two dust species
(silicate and graphite or amorphous carbon),
$d$ is the luminosity distance of the object,
$\knui$ is the mass absorption coefficient
(in unit of $\cm^{2}\g^{-1}$)
of dust of type $i$,
$B_\nu(T)$ is the Planck function
of temperature $T$ at frequency $\nu$,
$\Twi$ and $\Tci$ are respectively
the temperatures of the warm and cold components
of dust of type $i$, and
$\Mwi$ and $\Mci$ are respectively
the masses of the warm and cold components
of dust of type $i$.
For a given composition and size,
the mass absorption coefficient
$\kappa_{\rm abs}(\nu)$ is obtained
from Mie theory
(Bohren \& Huffman 1983)
using the refractive indices 
(1) of Draine \& Lee (1984) 
for ``astronomical silicate'' and graphite,
(2) of Dorschner  et al.\ (1995) 
for amorphous pyroxene and amorphous olivine, 
and (3) of Rouleau \& Martin (1991) 
for amorphous carbon.
We refer the reader to Figure~2 of Xie et al.\ (2015)
for the computed $\kappa_{\rm abs}(\nu)$ profiles
for different grain materials and sizes.

\section{Results and Discussion\label{sec:results}}
%

%
%

In fitting the observed IR emission,
we have eight parameters:
the temperature ($\Twsil$) and mass ($\Mwsil$)
for the warm silicate component,
the temperature ($\Tcsil$) and mass ($\Mcsil$)
for the cold silicate component,
the temperature ($\Twcarb$) and mass ($\Mwcarb$)
for the warm carbon dust component, and
the temperature ($\Tccarb$) and mass ($\Mccarb$)
for the cold carbon dust component.
We require the dust temperatures not to exceed
the sublimation temperature
($T_{\rm subl}$\,$\simali$1500$\K$)
of silicate and graphite materials.
By applying cosmic  abundance constraints
to $\Mcarb/\Msil$, 
the mass ratio of the silicate component
to the carbon dust component,
we require $0.2<\Mcarb/\Msil<2$
(see Xie et al.\ 2015). 
With these constraints taken into account,
we obtain the best fit for each galaxy
using the MPFIT code,
an IDL $\chi^{2}$-minimization routine
based on the Levenberg-Marquardt
algorithm (Markwardt 2009).
The quality of the fit is measured by
the reduced $\chi^2$ which is defined as follows:
\begin{equation}
\label{eq:num_2}
\chi^2/\dof = \sum_{j=1}^{\Nobs}
\left\{\frac{F_\nu(\mod)-F_\nu(\obs)}
{\sigma_\nu(\obs)}\right\}^2/\left\{\Nobs-\Nmod\right\} ~~,
\end{equation}
where $F_\nu(\mod)$ is the model-calculated
flux density, $F_\nu(\obs)$ is the observed flux density,
$\sigma_\nu(\obs)$ is the observational uncertainty
of the flux density $F_\nu(\obs)$, 
$\Nobs$ is the number of data points,
and $\Nmod=8$ is the number of model parameters.

We note that the {\it Spitzer}/IRS spectra in 
the $\simali$5--14.5$\mum$ wavelength interval
have a lower SNR compared with that in the interval 
of $\simali$14.5--38$\mum$.
This is due to the different observational modules,
i.e., the $\simali$5--14.5$\mum$ {\it Short Low} (SL) 
IRS module has a slit width of $\simali$3.6$^{\prime\prime}$,
while the $\simali$14.5--38$\mum$ {\it Long Low} (LL)
IRS module has a slit width of $\simali$11.2$^{\prime\prime}$.
To fit the $\simali$5--8$\mum$ continuum emission
and the 9.7$\mum$ silicate emission feature,
for those sources of which the SL spectra 
are rather noisy (including 
PG1352+183,
2MASSiJ132917.5+121340, 
2MASSiJ234259.3+134750, 
2MASXJ02335161+0108136,
2MASXJ13495283+0204456, and
SDSSJ115138.24+004946.4), 
we arbitrarily increase the weights
respectively by a factor of ten and two
for the data points at $\simali$5--8$\mum$ 
and $\simali$8--14.5$\mum$
(relative to that at $\simali$14.5--38$\mum$).
These sources will be excluded
when we perform statistical analyses 
of the possible correlations between
the dust properties and the AGN properties.

\begin{figure*}[ht]
\begin{center}
\resizebox{0.8\hsize}{!}{
\includegraphics[angle=90]{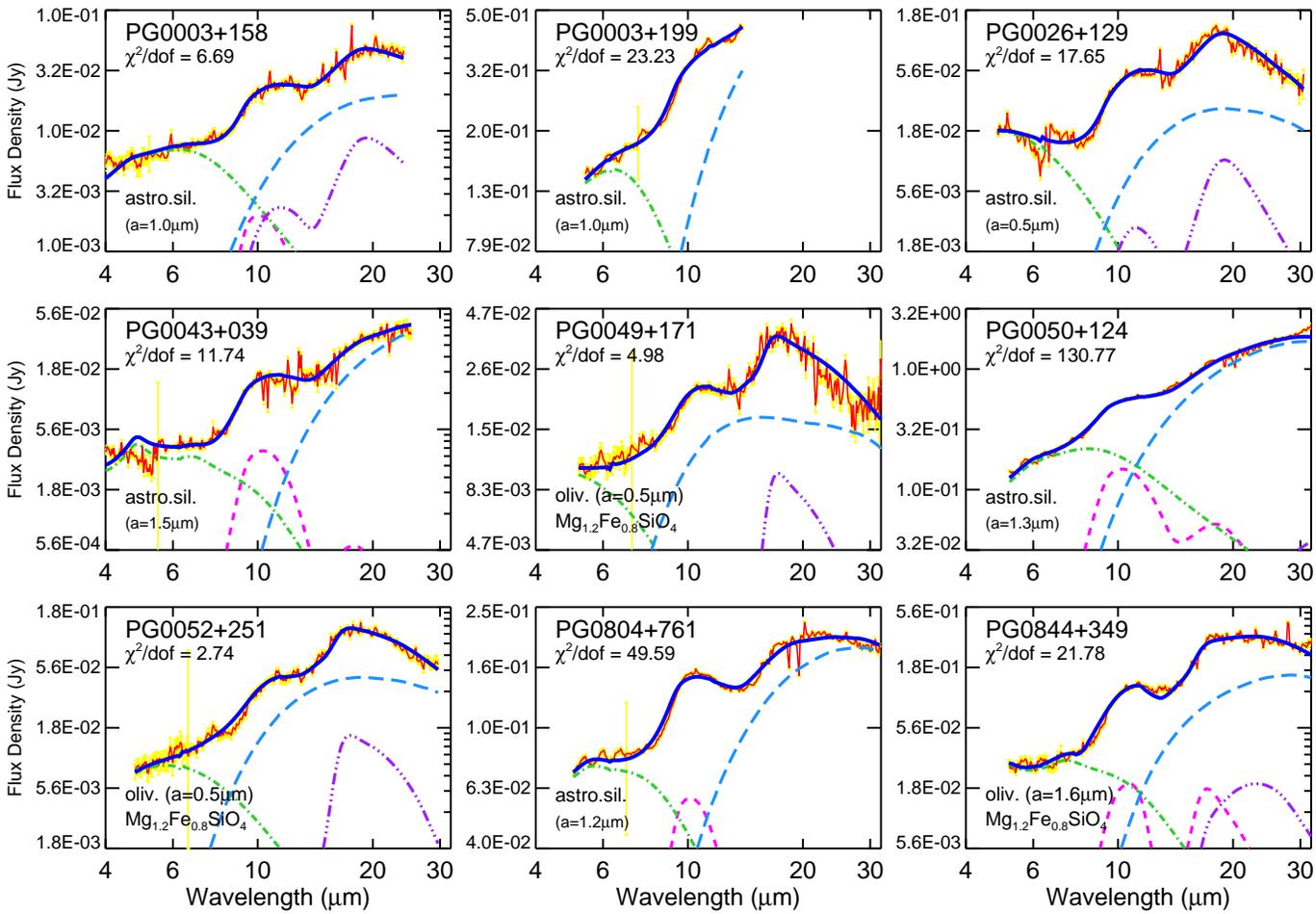}}
\caption{\footnotesize
         \label{fig:sil_em_mod1}
          Comparison of the {\it Spitzer}/IRS spectra (red solid lines) 
          of the PG quasars 
          PG0003+158, PG0003+199, PG0026+129, PG0043+039, PG0049+171, 
          PG0050+124, PG0052+251, PG0804+761, PG0844+349 
          which show silicate emission 
          around 9.7 and 18$\mum$
          with the model spectra (blue solid lines)
          which are the sum of warm silicate
          (magenta short dashed lines),
          cold silicate (purple dash-dot-dotted lines),
          warm graphite (green dash-dotted lines),
          and cold graphite (light blue long dashed lines).
          Also shown are the observed $1\sigma$ errors 
          (yellow vertical lines).
          }
\end{center}
\end{figure*}

In Figure~\ref{fig:sil_em_mod1}, 
we show the best-fit results to
all 93 ``silicate emission'' sources 
which exhibit prominent silicate emission 
at 9.7 and 18$\mum$.
The best-fit model parameters and 
their uncertainties are tabulated in
Table~\ref{tab:sil_em_mod_para}.
The uncertainties for the model parameters
are derived by performing Monte-Carlo simulations.
As illustrated in Figure~\ref{fig:mtcarlo}
with PG 2233+134 as an example,
we assume that the {\it Spitzer}/IRS 
flux density uncertainty 
statistically follows a normal distribution.
The dispersion is taken to be 
the observed 1$\sigma$ error,
composed of the statistical 
and systematic errors,
with the latter arising from
the flux differences between the two nods of
the {\it Spitzer}/IRS spectra,
the sky background contamination,
and the {\it Spitzer}/IRS pointing
and flux calibration errors
(Lebouteiller \etal 2011). 
A new ``observational'' spectrum is generated 
through randomly sampling a point at each wavelength
from the normal distribution.
We then model the new spectrum
and derive a set of model parameters.
We conduct 100 simulations for each source
as the parameters derived from 10,000 simulations
only slightly differ from that derived from 100 simulations
(Xie et al.\ 2015).
The final model spectrum is calculated
from the median values of the model parameters.
The error of each parameter
is derived from the standard deviation
of 100 simulations.

\subsection{Dust Composition\label{sec:composition}}
We find that the combination of the Draine \& Lee (1984)
``astronomical silicate'' and graphite 
can closely reproduce the {\it Spitzer}/IRS spectra
of 60 of our 93 AGNs. 
For 31 AGNs, amorphous olivine 
combined with graphite fits the observed spectra 
better than ``astronomical silicate''.
In contrast, amorphous pyroxene provides 
the best fit to two of our 93 AGNs
(i.e, PG 1535+547 and PG 2214+139). 
%
For illustration, in Figure~\ref{fig:composition} 
we show the best-fit results for three PG quasars
(PG 1004+130, PG 1351+640, and PG 2214+139) 
for which the best-fits are respectively provided by 
``astronomical silicate'',  
amorphous olivine Mg$_{1.2}$Fe$_{0.8}$SiO$_4$, 
and amorphous pyroxene Mg$_{0.3}$Fe$_{0.7}$SiO$_3$, 
again, together with graphite.
Different silicate species have different band widths,
peak wavelengths, and relative strengths for the 9.7 
and 18$\mum$ features. 
For a given grain size, the Draine \& Lee (1984)
``astronomical silicate'' results in an absorption
profile at 9.7$\mum$ much broader than 
amorphous olivine and pyroxene,
while amorphous olivine gives 
the longest peak wavelength 
for the 9.7$\mum$ feature 
and the highest ratio of the 18$\mum$ feature 
to the 9.7$\mum$ feature,
and amorphous pyroxene has the smallest ratio
of the 9.7$\mum$ feature to the ``trough'' 
between the 9.7 and 18$\mum$ features. 
As elaborated in Figure~\ref{fig:composition}, 
the {\it Spitzer}/IRS spectra of PG 1004+130, 
PG 1351+640, and PG 2214+139 show considerable
variations in the spectral profiles of
the 9.7 and 18$\mum$ emission features.
For illustration, we display 
in Figure~\ref{fig:composition_spec}
the {\it Spitzer}/IRS spectra
of several selected AGNs 
for which the best fits favor 
``astronomical silicate'', amorphous olivine,
and amorphous pyroxene, respectively.
Although complicated by 
the dust size and temperature effects, 
a first glance of Figure~\ref{fig:composition_spec}
would already tell that these AGNs differ 
in silicate composition
and their spectral profiles 
appear consistent with 
the feature width and band ratio
expected from ``astronomical silicate'', 
amorphous olivine,
or amorphous pyroxene, respectively.

There are also several AGNs for which their
silicate emission features can not be closely 
fitted in terms of ``astronomical silicate'', 
amorphous olivine, or amorphous pyroxene.
We note that, except a couple of sources
of which the {\it Spitzer}/IRS spectra
are noisy (e.g., PG 1352+183), most of 
these AGNs probably have other dust species 
(e.g., crystalline silicates) present.
In Figure~\ref{fig:crystalline_silicate} 
we display the {\it Spitzer}/IRS spectra 
of those sources which exhibit 
the crystalline silicate emission features
at 11.3, 16.3, 18.5, 23.5, and 27.5$\mum$.

 
Finally, we note that our model fitting is not sensitive
to the choice of graphite or amorphous carbon.
However, the opacity profile of amorphous carbon exhibits
several weak resonant structures in the wavelength range 
of $\simali$5--8$\mum$ (see Figure~3b of Xie et al.\ 2015)
which are not seen in the {\it Spitzer}/IRS spectra of
the AGNs considered here. Therefore, graphite appears
more favorable.

%
%


\subsection{Dust Sizes\label{sec:size}}
From our fitting, we find that the {\it Spitzer}/IRS spectra
of 70 of our 93 AGNs can be well reproduced with 
spherical grains of radii $a=1.5\mum$.
Only three AGNs require grains smaller than $a=1\mum$.
In Figure~\ref{fig:size_distribution} we show the histogram of 
the best-fit grain sizes.
%
%
Roughly speaking, the sizes of the grains in the torus
around these 93 AGNs which show silicate emission 
are constrained to be $\simali$1.5$\pm$0.1$\mum$. 
This is consistent with our previous work that the dust grains 
in AGNs are micrometer-sized (\eg Li \etal 2008, 
Smith \etal 2010, Xie \etal 2015).

We calculate the extinction as a function of 
inverse wavelength ($\lambda^{-1}$) expected 
from mixtures of spherical amorphous silicate 
and graphite of radii $a=1.5\mum$. 
We represent the extinction
by $E(\lambda-V)/E(B-V)$, 
where $E(\lambda-V)\equiv A_\lambda - A_V$,
$E(B-V)\equiv A_B - A_V$,
and $A_\lambda$, $A_B$, and $A_V$ are respectively
the extinction at wavelength $\lambda$, and
at the $B$ and $V$ bands. 
As shown in Figure~\ref{fig:extcurv},
the extinction curve predicted from
a mixture of silicate and graphite grains
of $a=1.5\mum$ is flat or gray 
at $\lambda^{-1}>2.5\mum^{-1}$, 
i.e., the extinction varies little
with $\lambda^{-1}$.
Depending on the mass ratio of
graphite to silicate, the extinction 
displays a gradual rise at $\lambda^{-1}>4.5\mum$
and then flattens off at $\lambda^{-1}>6\mum$.
But overall, the extinction curve is flat.
The resonant structures seen at $\lambda^{-1}<2.5\mum^{-1}$ 
will be smoothed out if we consider a distribution of 
grain sizes instead of single sizes of $a=1.5\mum$. 

The predicted gray extinction curve generally agrees
with that of Gaskell et al.\ (2004) who derived an 
AGN extinction curve based on the composite spectra 
of 72 radio quasars and 1018 radio-quiet AGNs.
Czerny et al.\ (2004) also constructed 
a relatively featureless flat extinction curve for quasars
based on the blue and red composite quasar spectra of
Richards et al.\ (2003) obtained 
from the Sloan Digital Sky Survey (SDSS).
It is interesting to note that, the extinction curve
calculated from spherical silicate dust of $a=1.5\mum$
closely agrees with that of Gaskell et al.\ (2004)
except those resonant structures at $\lambda^{-1}<2.5\mum^{-1}$ 
which are expected to be smoothed out if a distribution 
of grain sizes is considered.
The Galactic extinction curve differs substantially
from our model extinction curve as well as that of
Gaskell et al.\ (2004) in that, the Galactic extinction
curve shows a prominent extinction bump at 2175$\Angstrom$ 
and a steep far-UV rise which is believed to arised 
from small graphite dust grains. 
In contrast, the extinction curve of 
the Small Magellanic Cloud (SMC)
lacks the 2175$\Angstrom$ bump 
and displays an even steeper far-UV rise
than that of the Milky Way.

\subsection{Dust Temperatures\label{sec:temperature}}
Figure~\ref{fig:temperature_se} presents the histograms 
of the dust temperatures derived from our best-fits to
the {\it Spitzer}/IRS spectra of 
these 93 ``silicate emission'' sources 
(which show silicate emission). 
It is seen that the temperatures 
for the warm silicate dust component ($\Twsil$) span 
from $\simali$150$\K$ to $\simali$500$\K$, 
with a median value of $\simali$265$\K$ 
and a dispersion of $\simali$89$\K$. 
The temperatures for the cold silicate component
($\Tcsil$) vary from $\simali$40$\K$ to $\simali$200$\K$.
The median value is  $\simali$66$\K$  for $\Tcsil$
and the dispersion is $\simali$89$\K$. 
For graphite, much higher temperatures are obtained:
$\Twcarb$ is in the range of 
$\simali$200$\K$ to $\simali$1000$\K$,
with a median value of $\simali$638$\K$ 
and a dispersion of $\simali$159$\K$;
$\Tccarb$ is within
$\simali$100$\K$ to $\simali$220$\K$,
with a median value of $\simali$159$\K$ 
and a dispersion of $\simali$22$\K$.
We note that, even if the spatial distributions 
of silicate and graphite are similar in an AGN torus, 
one expects graphite to be much hotter than silicate 
because of the much higher UV/optical absorptivities 
of graphite compared to that of silicate 
(see Draine \& Lee 1984).
Graphite grains could be distributed closer to
the central engine of an AGN than silicate grains
since graphite has a higher sublimation temperature
(see Li 2009).
%
%

The grains in the AGN torus are heated by photons
from the central engine.
Let $R$ be the distance of 
a silicate or graphite grain of size $a$
from the central engine of luminosity $L_\lambda$.
The steady-state temperature of the grain 
can be calculated from the energy balance 
between absorption and emission:
\begin{equation}\label{eq:Teq}
\int_{0}^{\infty} \frac{L_\lambda}{4\pi R^2}
C_{\rm abs}(a,\lambda)\,d\lambda
= \int_{0}^{\infty} 4\pi B_\lambda[T(a,R)]
C_{\rm abs}(a,\lambda)\,d\lambda ~~,
\end{equation} 
where $C_{\rm abs}(a,\lambda)$ is the absorption
cross section of the spherical grain of radii $a$ 
at wavelength $\lambda$, and $B_\lambda(T)$ is 
the Planck function of temperature $T$.
For simplicity, in eq.\,\ref{eq:Teq} we neglect 
the extinction of the illuminating light in the torus. 
If the dust extinction is included, one would expect
a smaller $R$ for the same dust temperature. 
For the AGN luminosity $L_\lambda$,
we take the tabulated $L_\lambda(5100\Angstrom)$
(see Table~\ref{tab:silem93para}) and 
the spectral shape of Rowan-Robinson (1995).
For each AGN and each dust component, 
we derive the distance of the dust
from the central engine where the dust attains
an equilibrium temperature exactly equaling
that derived from the {\it Spitzer}/IRS 
spectral modeling. 
We find that the warm dust components are
at several hundredth to tenth parsecs 
from the central engine
and the cold dust components are at several parsecs
from the central engine 
(see Figure~\ref{fig:radius}).  
The actual distances could be smaller 
since the torus extinction is neglected 
in calculating the dust temperature.
These results are consistent with Elitzur (2006)
who argues for a torus size of 
no more than a few parsecs.

\subsection{Dust Masses\label{sec:mass}}
We show in Figure~\ref{fig:mass_se} 
the mass ratios of 
the warm graphite component to the warm silicate component
($\Mwcarb$/$\Mwsil$),
and the mass ratios of 
the cold graphite component to the cold silicate component
($\Mccarb$/$\Mcsil$).
We derive a mean ratio of $\simali$0.33 for $\Mwcarb$/$\Mwsil$,
and a mean ratio of $\simali$0.93 for $\Mccarb$/$\Mcsil$.
However, these ratios should be treated with caution 
since for a substantial number of sources 
the mass ratio reaches the pre-set limiting values
of $\Mcarb$/$\Msil = 0.2$ and $\Mcarb$/$\Msil = 2.0$
(see Figure~\ref{fig:mass_se}).
We find that reasonably good fits are still 
achievable as long as $\Mcarb$/$\Msil \simgt 0.05$.

For our ``silicate emission'' sources, 
the stellar mass ($M_\star$)
is known for 39 PG quasars 
and two {\it 2MASS} quasars
(see Zhang et al.\ 2016).
For each of these sources, 
we obtain $M_{\rm dust}$, 
the total dust mass summed 
over all four dust components 
(i.e., $M_{\rm dust} = \Mwsil + \Mcsil + \Mwcarb + \Mccarb$).
In Figure~\ref{fig:mass_dust_star} we compare
the dust mass with the stellar mass. 
On average, the dust-to-stellar mass ratio 
of these sources is $\simali$10$^{-7}$,
much smaller than that of the Milky Way
($\simali$10$^{-3}$; see Li 2004).
This ratio appears reasonable 
since the mid-IR emission considered here
only probes the dust in the torus
while the bulk mass is in 
starlight-heated cold dust
in the host galaxy
which emits in the far-IR
and escapes from detection by {\it Spitzer}/IRS.
%

\subsection{Correlations between Dust and AGNs
                    \label{sec:correlation}}
With the dust properties determined, 
we now explore the possible connection 
between the fundamental properties of AGNs
and the properties of the dust  
derived from the silicate emission modeling. 
If the dust in the torus is heated by the photons 
originating from the accretion disk, 
one would expect the dust properties 
to somewhat correlate with the AGN parameters.
Therefore, we examine the correlation between 
the dust temperature and mass 
and the bolometric luminosity ($\Lbol$), 
black hole mass ($\MBH$), 
and Eddington ratio ($\Lbol/\Ledd$) of AGNs. 
%
%
We represent $\Lbol$ by 
$\lambda L_{\lambda}(5100\Angstrom)$, 
since the spectral region at this wavelength 
is barely contaminated by emission lines and 
is assumed to be purely from the AGN accretion disk.
%
%
The Eddington ratio ($\varepsilon$) relates 
the AGN bolometric luminosity 
with the Eddington luminosity: 
$\varepsilon \equiv \Lbol/\Ledd$, 
with $\Ledd = 4\pi cGM m_{\rm H}/\sigma_{T}$,
where $c$ is the speed of light, 
$G$ is the gravitational constant,
and $\sigma_{T}$ is the Thomson cross section. 
We compile these parameters from the literature 
and list them in Table~\ref{tab:silem93para}.

In Figures~\ref{fig:M_Lbol},\,\ref{fig:M_Mbh},\,\ref{fig:M_Lbol2Ledd}
we present the correlations between the dust masses of
the warm silicate, warm graphite, cold silicate
 and cold graphite components
with the bolometric luminosity, the black hole mass, 
and the Eddington ratio of AGNs.  
While the dust masses show no correlation with 
the black hole mass or with the Eddington ratio,
they do show a somewhat correlation 
with the bolometric luminosity 
(see Figure~\ref{fig:M_Lbol}a,b,d). 
This suggests that the covering factor of 
the dust torus (i.e., the fraction of the sky 
covered by the torus as seen from 
the central engine) may increase with
the bolometric luminosity of AGNs.
%
%
Similarly,  
in Figures~\ref{fig:T_Lbol},\,\ref{fig:T_Mbh},\,\ref{fig:T_Lbol2Ledd}
we present the correlations between the temperatures 
($T$) of the warm silicate, warm graphite, cold silicate
and cold graphite components
with the bolometric luminosity, the black hole mass, 
and the Eddington ratio of AGNs.  
No correlations are found.
The lack of correlation between $T$ and $L_{\rm bol}$
is not unexpected since $T$ depends on 
both $L_{\rm bol}$ and $R$, the distance of 
the dust from the central engine.

\subsection{Other Sources\label{sec:other}}

Among our sample of 147 AGNs, 
there are 30 sources 
which do not show prominent 
silicate emission but a featureless 
thermal continuum in the mid-IR
(hereafter we will call these AGNs
``flat continuum'' sources). 
In addition, there are 24 sources 
exhibit weak PAH features 
superimposed on an otherwise 
featureless thermal continuum
(hereafter we will call these AGNs
``PAH\,+\,continuum'' sources). 
In Figure~\ref{fig:se_frac} we show
$F_{\rm silicate}/F_{\rm MIR}$,
the fractional fluxes emitted in the 9.7 
and 18$\mum$ silicate features 
relative to the total mid-IR emission
at $\simali$5--38$\mum$, 
for all three categories of sources.\footnote{%
   For the ``flat continuum'' sources 
   and the ``PAH\,+\,continuum'' sources,
   $F_{\rm silicate}$ is actually an upper limit. 
   }
It is apparent that, with a considerably
larger mean fractional flux 
($\langle F_{\rm silicate}/F_{\rm MIR}\rangle\approx0.13$),
the ``silicate emission'' sources clearly 
distinguish them from those 
``flat continuum'' sources
(with $\langle F_{\rm silicate}/F_{\rm MIR}\rangle\approx0.041$)
and ``PAH\,+\,continuum'' sources
(with $\langle F_{\rm silicate}/F_{\rm MIR}\rangle\approx0.051$).
This confirms that the silicate emission features 
in the 93 ``silicate emission'' sources are indeed 
prominent and therefore the silicate dust properties 
yielded from our modeling have a high level of significance. 
We have also modeled 
the {\it Spitzer}/IRS spectra 
of those 30 ``flat continuum'' sources 
which do not show any silicate emission 
but a featureless thermal continuum.
As shown in Figure~\ref{fig:nosil_em_mod1},
mixtures of $\mu$m-sized silicate and graphite 
also provide close fits to the observed spectra. 
Compared with those ``silicate emission'' sources
(which show prominent silicate emission features; 
see Figure~\ref{fig:sil_em_mod1}),
these sources have lower dust temperatures 
(see Figure~\ref{fig:temperature_fl}).
In Figure~\ref{fig:mass_fl} we show
the graphite-to-silicate mass ratios.

We have also modeled the thermal continuum emission
of these 24 ``PAH\,+\,continuum'' sources 
with mixtures of silicate and graphite grains. 
As shown in Figure~\ref{fig:PAH_em_mod1},
the best-fits to the observed spectra 
are provided by $\mu$m-sized dust grains. 
The graphite-to-silicate mass ratios
are shown in Figure~\ref{fig:mass_pah}.
The derived dust temperatures are
shown in Figure~\ref{fig:temperature_pah}
and they are rather close to that of those 
PAH-lacking ``flat continuum'' sources.
This suggests that the thermal continuum emission
seen in these ``PAH\,+\,continuum'' AGNs 
is not from their host galaxies but from the torus.

A wide variety of Galactic and extragalactic objects
show a distinctive set of emission features 
at 3.3, 6.2, 7.7, 8.6, and 11.3$\mum$.
These features are generally identified as 
the vibrational modes of PAH molecules 
(L\'{e}ger \& Puget 1984, 
Allamandola, Tielens, \& Barker 1985).
However, the PAH features are often absent in AGNs
(e.g., see Roche et al.\ 1991). 
This is generally interpreted as the destruction of PAHs 
by extreme UV and soft X-ray photons in AGNs 
(Roche et al.\ 1991, Voit 1992, 
Siebenmorgen, Kr\"ugel, \& Spoon 2004).
If the PAH emission and the thermal emission continuum
seen in these sources are contaminated by their host galaxies,
one would expect the dust to be smaller and colder than 
that in those ``flat continuum'' sources 
of which the {\it Spitzer}/IRS spectra 
are characterized by a featureless, PAH-lacking 
thermal continuum. This is because the interstellar dust 
grains in the host galaxies of AGNs are believed to 
be around $\simali$0.1$\mum$ in size 
and $\simali$20$\K$ in temperature
(see Li \& Draine 2001).  
But as we see that this is not the case. 
The PAH emission features may as well 
arise from the AGN torus,
i.e., some quantities of PAHs may survive 
in the hostile environments of AGNs
(L.C.~Ho, private communication).

\section{Summary\label{sec:summary}}
We have investigated the dust properties 
of a sample of 147 AGNs compiled from PG quasars,
{\it 2MASS} quasars, and \saga\ AGNs
which do not show silicate absorption features.
Our principal results are as follows:
\begin{enumerate}
\item Through fitting the {\it Spitzer}/IRS spectra
      of 93 AGNs of various types in which
      the 9.7 and 18$\mum$ emission features 
      are seen in emission with mixtures of
      silicate and graphite grains,   
      we find that the majority (60/93) 
      of the observed spectra can be well 
      reproduced by ``astronomical silicate'',
      with the remaining 31 sources favoring
      amorphous olivine and two sources favoring 
      amorphous pyroxene.
\item All sources require the dust to be $\mu$m-sized 
      (with a typical size of $\simali$1.5$\mum$),
      much larger than the sub-$\mu$m-sized Galactic 
      interstellar dust. This implies a flat or ``gray'' 
      extinction curve for AGNs.
\item The 9.7$\mum$ emission feature arises predominantly 
      from warm silicate dust of temperature $T\simali270\K$,
      while the $\simali$5--8$\mum$ continuum emission 
      is mostly from graphite of $T\simali640\K$. 
\item We have examined the possible relations 
      between the dust masses and temperatures
      and the bolometric luminosity, black hole masses,
      and Eddington ratios of AGNs. 
      It is found that except the dust masses are somewhat 
      correlated with the bolometric luminosity,
      we do not see any correlations 
      between any other quantities.
\item We have also modeled the {\it Spitzer}/IRS spectra
      of 30 (of 147) sources which do not show silicate
      emission features but a featureless thermal continuum.
      Compared to those 93 sources which show silicate emission,
      $\mu$m-sized silicate and graphite grains
      with smaller silicate-to-graphite mass ratios
      and lower dust temperatures are preferred.
      We have also modeled the {\it Spitzer}/IRS spectra
      of 24 (of 147) sources which exhibit weak PAH emission 
      features superimposed on an otherwise featureless 
      thermal continuum. It is found that the derived dust 
      sizes and temperatures are not appreciably different
      from that of those 30 sources which emit a featureless 
      thermal continuum.
\end{enumerate}

\acknowledgments
We thank B.T.~Draine, L.C.~Ho and
the anonymous referee
for stimulating discussions and suggestions.
We thank Y.~Shi for kindly providing us
with the {\it Spitzer}/IRS spectra of PG quasars.
A.L. and X.Y.X. are supported in part
by NSF AST-1311804 and NASA NNX14AF68G.
L.H. is supported by NSFC 11473305 and
the CAS Strategic Priority Research Program
XDB09030200.
The Cornell Atlas of \spitzerirs Sources (CASSIS)
is a product of the Infrared Science Center
at Cornell University, supported by NASA and JPL.

\appendix
\section{Basic Parameters for Our Sample of 147 AGNs}
In this appendix we list 
the redshift ($z$), type, 
black-hole mass ($M_{\rm BH}$),
and luminosity at $\lambda=5100\Angstrom$ 
for each of our 147 AGNs,
which are divided into three categories:
the ``silicate emission'' sources
which exhibit the 9.7 and 18$\mum$ 
silicate emission features
(see Table~\ref{tab:silem93para}),
the ``flat continuum'' sources
which exhibit a featureless emission continuum
(see Table~\ref{tab:conti30para}), and
the ``PAH\,+\,continuum'' sources
which show PAH emission features superimposed
on an otherwise featureless continuum
(see Table~\ref{tab:pah24para}).
%

%

\begin{landscape}
\begin{deluxetable}{lllllllll}
\tabletypesize\tiny 
\tablecolumns{9}           
\tablecaption{\footnotesize
              \label{tab:sil_em_mod_para}
              Model Parameters for 93 AGNs 
              Showing Silicate Emission 
              around 9.7 and 18$\mum$
              } 
\tablewidth{0pt}           
\tablehead{                
\multicolumn{1}{l}{Source} &
\multicolumn{1}{c}{$\Twsil$} &
\multicolumn{1}{c}{$\Mwsil$} &
\multicolumn{1}{c}{$\Tcsil$} &
\multicolumn{1}{c}{$\Mcsil$} &
\multicolumn{1}{c}{$\Twcarb$} &
\multicolumn{1}{c}{$\Mwcarb$} &
\multicolumn{1}{c}{$\Tccarb$} &
\multicolumn{1}{c}{$\Mccarb$}  \\ 
 &
\multicolumn{1}{c}{(K)} &
\multicolumn{1}{c}{(${\rm M_{\odot}}$)} &
\multicolumn{1}{c}{(K)} &
\multicolumn{1}{c}{(${\rm M_{\odot}}$)} &
\multicolumn{1}{c}{(K)} &
\multicolumn{1}{c}{(${\rm M_{\odot}}$)} &
\multicolumn{1}{c}{(K)} &
\multicolumn{1}{c}{(${\rm M_{\odot}}$)} \\ 
\multicolumn{1}{l}{(1)} &
\multicolumn{1}{c}{(2)} &
\multicolumn{1}{c}{(3)} &
\multicolumn{1}{c}{(4)} &
\multicolumn{1}{c}{(5)} &
\multicolumn{1}{c}{(6)} &
\multicolumn{1}{c}{(7)} &
\multicolumn{1}{c}{(8)} &
\multicolumn{1}{c}{(9)} 
} 
\startdata 
PG0003+158               &   358.00$\pm$73.60 & 4.32E1  $\pm$  2.00E1 & 174.33$\pm$7.98 & 2.53E3  $\pm$  1.64E2 & 790.57$\pm$59.82 & 8.64$\pm$4.00 & 182.83$\pm$3.77 & 5.06E3  $\pm$  3.28E2 \\   
PG0003+199               &   317.95$\pm$1.06 & 2.77  $\pm$  5.71E-2 &  40.00  $\pm$  3.10 & 1.42E3  $\pm$  2.51E1 & 744.55$\pm$4.43 & 5.54E-1$\pm$1.14E-2 & 187.60  $\pm$  0.66 & 2.84E2  $\pm$  5.05 \\
PG0026+129               &   203.95$\pm$2.70 & 1.32E1  $\pm$  2.59E-1 & 193.68  $\pm$  0.69 & 2.72E2  $\pm$  6.92 & 671.86$\pm$3.34 & 2.65E0$\pm$5.18E-2 & 206.22  $\pm$  0.68 & 4.95E2  $\pm$  2.31E1 \\
PG0043+039               &   482.21 $\pm$  17.11 & 3.17E1  $\pm$  3.67 &  40.00  $\pm$  0.00 & 4.45E4  $\pm$  2.46E3 & 884.88$\pm$3.94 & 7.84E0$\pm$1.27 & 149.29  $\pm$  0.79 & 1.47E4  $\pm$  1.20E3 \\
PG0049+171               &   184.14$\pm$1.05 & 8.60E-1  $\pm$  1.24E-1 & 184.20  $\pm$  1.06 & 2.25E1  $\pm$  5.85E-1 & 683.54  $\pm$  12.93 & 1.72E-1$\pm$4.18E-2 & 219.69  $\pm$  0.82 & 4.44E1  $\pm$  1.92 \\
PG0050+124               &   410.84$\pm$3.06 & 1.64E1  $\pm$  4.40E-1 &  68.29  $\pm$  0.06 & 2.65E4  $\pm$  1.32E3 & 545.32 $\pm$1.11 & 1.42E1$\pm$5.17E-1 & 143.54  $\pm$  0.16 & 1.17E4  $\pm$  8.64E2 \\
PG0052+251               &   171.76$\pm$2.23 & 2.40E1  $\pm$  5.78 & 173.53  $\pm$  2.55 & 5.15E2  $\pm$  3.78E1 & 565.33  $\pm$  10.01 & 5.44E0$\pm$5.27 & 208.04  $\pm$  1.59 & 9.25E2  $\pm$  1.14E2 \\
PG0804+761               &   407.53$\pm$1.43 & 1.86E1  $\pm$  1.98E-1 &  40.00  $\pm$  0.00 & 1.70E4  $\pm$  4.92E1 & 819.90 $\pm$2.87 & 3.72E0$\pm$3.97E-2 & 164.88  $\pm$  0.12 & 3.41E3  $\pm$  9.86 \\
PG0844+349               &   338.14$\pm$2.85 & 5.86  $\pm$  1.66E-1 & 113.33  $\pm$  0.60 & 6.03E2  $\pm$  2.17E1 & 699.13 $\pm$5.87 & 1.17E0$\pm$3.32E-2 & 161.88  $\pm$  0.41 & 5.42E2  $\pm$  3.11E1 \\
PG0921+525               &   182.74$\pm$0.28 & 3.14E-1  $\pm$  7.68E-3 & 182.81  $\pm$  0.23 & 3.74E1  $\pm$  2.66E-1 &1131.80  $\pm$  11.20 & 6.28E-2$\pm$1.58E-3 & 210.67  $\pm$  0.19 & 6.16E1  $\pm$  7.91E-1 \\
PG0923+201               &   380.40$\pm$8.37 & 2.74E1  $\pm$  2.12 &  40.00  $\pm$  0.00 & 2.13E4  $\pm$  3.82E2 & 797.53 $\pm$6.90 & 5.93E0$\pm$6.41E-1 & 154.97  $\pm$  0.45 & 4.37E3  $\pm$  1.20E2 \\
PG0947+396               &   266.70$\pm$3.48 & 5.09E1  $\pm$  2.32 &  63.40  $\pm$  1.72 & 2.57E4  $\pm$  4.82E2 & 657.53 $\pm$4.39 & 1.02E1$\pm$5.84E-1 & 155.23  $\pm$  0.45 & 5.14E3  $\pm$  1.47E2 \\
PG0953+414               &   340.04$\pm$5.05 & 3.66E1  $\pm$  1.42 &  40.00  $\pm$  0.00 & 1.93E4  $\pm$  8.95E2 & 842.36  $\pm$  10.77 & 7.33E0$\pm$2.83E-1 & 168.92  $\pm$  0.79 & 4.10E3  $\pm$  2.31E2 \\
PG1001+054               &   290.92$\pm$9.86 & 1.82E1  $\pm$  2.80 &  71.50  $\pm$  3.89 & 9.85E3  $\pm$  6.15E2 & 735.20  $\pm$  24.25 & 3.63E0$\pm$5.61E-1 & 160.26  $\pm$  1.94 & 1.97E3  $\pm$  1.23E2 \\
PG1004+130               &   468.54$\pm$3.29 & 4.54E1  $\pm$  9.22E-1 &  40.00  $\pm$  0.00 & 3.86E4  $\pm$  4.71E2 & 698.83 $\pm$3.42 & 9.09E0$\pm$1.84E-1 & 146.45  $\pm$  0.21 & 1.51E4  $\pm$  2.87E2 \\
PG1011-040               &   308.07$\pm$1.50 & 5.34  $\pm$  1.00E-1 &  71.55  $\pm$  0.64 & 3.08E3  $\pm$  3.22E1 & 611.19 $\pm$3.07 & 1.07E0$\pm$2.01E-2 & 161.83  $\pm$  0.32 & 6.17E2  $\pm$  6.45 \\
PG1012+008               &   373.64 $\pm$  13.91 & 1.22E1  $\pm$  1.74 &  70.35  $\pm$  0.88 & 2.29E4  $\pm$  1.69E3 & 608.69 $\pm$4.53 & 1.06E1$\pm$2.04 & 157.00  $\pm$  0.51 & 4.58E3  $\pm$  6.27E2 \\
PG1048-090               &   361.13 $\pm$  62.98 & 1.80E1  $\pm$  6.59 & 176.05  $\pm$  6.21 & 1.34E3  $\pm$  1.32E2 & 893.65  $\pm$  63.66 & 3.59E0$\pm$1.32 & 184.11  $\pm$  2.88 & 2.68E3  $\pm$  3.31E2 \\
PG1049-005               &   288.49$\pm$5.18 & 2.57E2  $\pm$  2.02E1 &  72.93  $\pm$  0.99 & 2.33E5  $\pm$  5.33E3 & 633.50 $\pm$1.73 & 5.93E1$\pm$6.37 & 148.37  $\pm$  0.49 & 4.66E4  $\pm$  1.07E3 \\
PG1048+342               &   150.00 $\pm$  29.39 & 1.15E1  $\pm$  4.29 & 143.79  $\pm$ 50.92 & 3.80E2  $\pm$  2.04E2 & 716.78  $\pm$  11.33 & 2.31E0$\pm$8.24 & 185.77  $\pm$  4.41 & 7.60E2  $\pm$  4.25E2 \\
PG1100+772               &   349.45$\pm$4.07 & 6.98E1  $\pm$  2.22 &  40.00  $\pm$ 12.94 & 6.06E3  $\pm$  3.06E2 & 754.47 $\pm$7.05 & 1.40E1$\pm$4.43E-1 & 163.32  $\pm$  1.24 & 9.94E3  $\pm$  8.30E2 \\
PG1103-006               &   387.23 $\pm$  13.03 & 8.87E1  $\pm$  7.85 & 117.73  $\pm$  8.59 & 7.08E3  $\pm$  7.01E2 & 763.72  $\pm$  17.04 & 1.77E1$\pm$1.57 & 163.36  $\pm$  1.82 & 1.42E4  $\pm$  1.40E3 \\
PG1114+445               &   262.72$\pm$2.61 & 2.93E1  $\pm$  5.88E-1 & 143.85  $\pm$  1.88 & 1.57E3  $\pm$  8.30E1 & 667.64 $\pm$3.48 & 5.87E0$\pm$1.18E-1 & 182.72  $\pm$  0.30 & 2.90E3  $\pm$  2.25E2 \\
PG1116+215               &   366.03$\pm$6.11 & 3.29E1  $\pm$  1.52 &  40.00  $\pm$  0.00 & 2.80E4  $\pm$  9.90E3 & 902.47  $\pm$  15.08 & 6.59E0$\pm$3.04E-1 & 164.80  $\pm$  0.54 & 5.80E3  $\pm$  7.06E3 \\
PG1121+422               &   280.10$\pm$5.52 & 1.86E1  $\pm$  7.36E-1 & 172.63  $\pm$  3.21 & 2.83E2  $\pm$  9.47 & 801.03 $\pm$9.87 & 3.72E0$\pm$1.47E-1 & 181.64  $\pm$  1.68 & 5.65E2  $\pm$  1.89E1 \\
PG1151+117               &   332.48 $\pm$  95.07 & 1.70E1  $\pm$  1.09E1 &  40.00  $\pm$  9.65 & 9.87E3  $\pm$  2.35E3 & 725.01  $\pm$  72.48 & 3.40E0$\pm$7.46 & 158.87  $\pm$  3.43 & 2.02E3  $\pm$  2.07E3 \\
PG1202+281               &   320.80$\pm$1.82 & 2.91E1  $\pm$  5.72E-1 &  47.34  $\pm$  2.63 & 3.09E4  $\pm$  1.51E2 & 644.61 $\pm$3.43 & 5.83E0$\pm$1.14E-1 & 151.39  $\pm$  0.17 & 6.18E3  $\pm$  3.02E1 \\
PG1211+143               &   267.00$\pm$4.38 & 3.99E1  $\pm$  2.28 & 118.08  $\pm$  0.68 & 1.28E3  $\pm$  4.38E1 & 632.66  $\pm$  10.45 & 7.99E0$\pm$4.56E-1 & 167.74  $\pm$  0.36 & 2.45E3  $\pm$  1.17E2 \\
PG1216+069               &   315.83 $\pm$  67.65 & 2.68E1  $\pm$  2.13 & 148.30  $\pm$ 27.37 & 1.38E3  $\pm$  6.14E2 & 814.50  $\pm$  15.37 & 5.38E0$\pm$4.51E-1 & 204.81  $\pm$  5.60 & 2.19E3  $\pm$  1.07E3 \\
PG1229+204               &   276.58$\pm$5.27 & 3.94  $\pm$  2.56E-1 & 115.50  $\pm$  0.41 & 5.63E2  $\pm$  2.02 & 696.31  $\pm$  10.99 & 7.89E-1$\pm$5.13E-2 & 161.76  $\pm$  0.11 & 1.13E3  $\pm$  4.03 \\
PG1259+593               &   431.78$\pm$4.93 & 3.94E1  $\pm$  1.63 & 117.00  $\pm$  4.31 & 9.87E3  $\pm$  1.87E3 &1014.46 $\pm$7.85 & 7.88E0$\pm$4.16E-1 & 226.28  $\pm$  1.54 & 2.36E3  $\pm$  6.97E2 \\
PG1302-102               &   324.63$\pm$2.81 & 1.23E2  $\pm$  3.97 &  55.72  $\pm$  1.44 & 1.11E5  $\pm$  6.85E2 & 653.52 $\pm$2.95 & 2.46E1$\pm$1.08 & 155.28  $\pm$  0.20 & 2.22E4  $\pm$  1.37E2 \\
PG1307+085               &   161.35$\pm$3.15 & 1.87E1  $\pm$  5.40 & 161.47  $\pm$  3.09 & 8.20E2  $\pm$  9.29E1 & 708.55  $\pm$  27.93 & 3.73E0$\pm$8.96 & 185.39  $\pm$  1.91 & 1.43E3  $\pm$  2.59E2 \\
PG1309+355               &   441.88 $\pm$ 118.63 & 1.50E1  $\pm$  5.49E1 & 105.79  $\pm$  1.67 & 9.14E3  $\pm$  3.75E3 & 658.17  $\pm$  68.58 & 1.18E1$\pm$4.37E1 & 167.12  $\pm$  8.80 & 5.37E3  $\pm$  5.52E3 \\
PG1310-108               &   160.55$\pm$0.49 & 1.00  $\pm$  2.31E-2 & 160.64  $\pm$  0.32 & 6.96E1  $\pm$  8.27E-1 & 679.46 $\pm$4.40 & 2.00E-1$\pm$4.62E-3 & 199.48  $\pm$  0.19 & 1.03E2  $\pm$  2.03 \\
PG1322+659               &   266.61$\pm$1.67 & 2.94E1  $\pm$  4.52E-1 &  51.66  $\pm$  1.37 & 1.85E4  $\pm$  1.11E2 & 676.31 $\pm$3.00 & 5.89E0$\pm$9.04E-2 & 151.36  $\pm$  0.20 & 3.70E3  $\pm$  2.21E1 \\
PG1341+258               &   284.24$\pm$2.73 & 5.82  $\pm$  1.70E-1 &  58.28  $\pm$  0.85 & 4.45E3  $\pm$  3.63E1 & 658.15 $\pm$5.67 & 1.16E0$\pm$3.40E-2 & 152.81  $\pm$  0.27 & 8.90E2  $\pm$  7.25 \\
PG1351+640               &   351.66$\pm$0.60 & 4.59E1  $\pm$  2.74E-1 &  84.16  $\pm$  0.38 & 7.07E3  $\pm$  2.18E1 & 540.18 $\pm$1.75 & 9.88E0$\pm$1.30E-1 & 133.05  $\pm$  0.07 & 1.41E4  $\pm$  4.37E1 \\
PG1352+183               &   248.21 $\pm$  25.87 & 5.38E1  $\pm$  3.09E1 &  40.00  $\pm$  1.97 & 6.43E3  $\pm$  1.36E3 & 335.40  $\pm$  86.50 & 1.08E1$\pm$6.19 & 161.73  $\pm$  5.22 & 1.29E3  $\pm$  5.77E2 \\
PG1402+261               &   403.94$\pm$8.90 & 1.98E1  $\pm$  1.58 &  69.50  $\pm$  0.24 & 5.03E4  $\pm$  5.33E2 & 661.52 $\pm$5.08 & 1.38E1$\pm$1.57 & 142.40  $\pm$  0.36 & 1.01E4  $\pm$  1.07E2 \\
PG1404+226               &   315.15$\pm$5.32 & 3.37  $\pm$  2.28E-1 &  64.89  $\pm$  0.50 & 2.53E3  $\pm$  2.80E1 & 600.81 $\pm$4.64 & 1.69E0$\pm$1.59E-1 & 161.12  $\pm$  0.43 & 5.07E2  $\pm$  5.60 \\
PG1411+442               &   265.36 $\pm$  17.48 & 2.35E1  $\pm$  8.18 &  59.66  $\pm$  2.18 & 7.16E3  $\pm$  1.37E2 & 721.91  $\pm$  49.45 & 4.70E0$\pm$1.65 & 171.08  $\pm$  1.53 & 1.43E3  $\pm$  2.75E1 \\
PG1416-129               &   183.99 $\pm$  19.61 & 3.69E-1  $\pm$  1.47 & 172.33  $\pm$  1.38 & 1.54E2  $\pm$  2.75 & 924.34  $\pm$  29.30 & 7.38E-1$\pm$2.95 & 202.23  $\pm$  0.96 & 3.08E2  $\pm$  5.49 \\
PG1426+015               &   330.84$\pm$2.55 & 9.90  $\pm$  2.96E-1 &  61.10  $\pm$  0.13 & 1.33E4  $\pm$  3.35E1 & 640.93 $\pm$2.12 & 5.92E0$\pm$2.41E-1 & 163.35  $\pm$  0.11 & 2.66E3  $\pm$  6.71 \\
PG1435-067               &   319.81$\pm$3.71 & 9.28  $\pm$  5.78E-1 &  40.00  $\pm$ 30.10 & 7.73E2  $\pm$  1.09E2 & 729.28  $\pm$  14.00 & 1.86E0$\pm$1.16E-1 & 182.86  $\pm$  4.48 & 7.92E2  $\pm$  1.23E2 \\
PG1444+407               &   434.56$\pm$6.63 & 2.75E1  $\pm$  1.57 &  65.56  $\pm$  0.92 & 4.88E4  $\pm$  5.13E2 & 662.35 $\pm$2.16 & 1.68E1$\pm$1.30 & 161.15  $\pm$  0.36 & 9.77E3  $\pm$  1.03E2 \\
PG1512+370               &   367.38$\pm$4.50 & 4.52E1  $\pm$  1.80 &  90.10  $\pm$  0.84 & 2.75E4  $\pm$  6.92E2 & 818.86 $\pm$5.81 & 9.26E0$\pm$5.43E-1 & 184.86  $\pm$  0.85 & 5.50E3  $\pm$  1.38E2 \\
PG1534+580               &   240.61 $\pm$  31.24 & 1.59  $\pm$  8.29 & 128.85  $\pm$ 43.61 & 1.07E2  $\pm$  1.60E2 & 694.01  $\pm$  85.59 & 3.18E-1$\pm$1.66 & 170.48  $\pm$  6.01 & 2.14E2  $\pm$  3.22E2 \\
PG1535+547               &   239.98$\pm$0.79 & 2.23  $\pm$  1.89E-2 &  49.95  $\pm$  0.54 & 6.48E2  $\pm$  7.14E1 & 776.61 $\pm$2.20 & 4.46E-1$\pm$3.78E-3 & 183.34  $\pm$  0.15 & 1.30E2  $\pm$  4.26E1 \\
PG1545+210               &   173.68 $\pm$  38.91 & 2.39E1  $\pm$  1.04E1 & 172.48  $\pm$  3.50 & 7.86E2  $\pm$  2.87E1 & 857.38  $\pm$  19.41 & 4.84E0$\pm$1.92E1 & 199.93  $\pm$  1.23 & 1.57E3  $\pm$  7.94E1 \\
PG1552+085               &   401.77 $\pm$  61.66 & 1.67  $\pm$  6.81 &  72.47  $\pm$  1.39 & 2.93E3  $\pm$  9.96E1 & 602.90  $\pm$  16.89 & 3.34E0$\pm$1.37E1 & 173.79  $\pm$  0.89 & 5.87E2  $\pm$  1.99E1 \\
PG1617+175               &   339.70 $\pm$  26.44 & 1.25E1  $\pm$  6.77 &  53.90  $\pm$  2.02 & 5.92E3  $\pm$  2.25E2 & 757.72  $\pm$  55.33 & 2.50E0$\pm$1.36 & 163.43  $\pm$  2.53 & 1.18E3  $\pm$  4.50E1 \\
PG1626+554               &   314.90 $\pm$  20.66 & 3.68  $\pm$  1.27 & 225.05  $\pm$  5.27 & 6.18E1  $\pm$  5.03 & 834.28  $\pm$  77.72 & 7.35E-1$\pm$1.01 & 214.70  $\pm$  2.58 & 1.24E2  $\pm$  1.37E1 \\
PG1700+518               &   365.49$\pm$8.00 & 1.43E2  $\pm$  9.85 &  76.48  $\pm$  0.34 & 2.07E5  $\pm$  2.29E3 & 640.92 $\pm$7.26 & 7.55E1$\pm$7.23 & 153.06  $\pm$  0.34 & 4.14E4  $\pm$  4.58E2 \\
PG1704+608               &   395.20$\pm$2.70 & 1.16E2  $\pm$  2.15 &  40.00  $\pm$  0.00 & 1.93E5  $\pm$  6.08E3 & 837.88 $\pm$4.76 & 2.31E1$\pm$4.31E-1 & 138.84  $\pm$  0.19 & 8.70E4  $\pm$  3.76E3 \\
PG2112+059               &   352.84$\pm$2.46 & 2.15E2  $\pm$  8.04 & 100.75  $\pm$  2.60 & 8.25E4  $\pm$  8.20E3 & 819.08 $\pm$3.03 & 4.71E1$\pm$2.20 & 190.72  $\pm$  2.63 & 1.65E4  $\pm$  1.64E3 \\
PG2209+184               &   288.22$\pm$2.85 & 3.16  $\pm$  9.52E-2 &  60.48  $\pm$  1.03 & 8.10E2  $\pm$  1.07E1 & 716.31 $\pm$6.56 & 6.32E-1$\pm$1.90E-2 & 176.17  $\pm$  0.62 & 1.62E2  $\pm$  2.14 \\
PG2214+139               &   314.14$\pm$1.03 & 7.78  $\pm$  8.73E-2 &  51.91  $\pm$  1.17 & 6.18E2  $\pm$  9.14E1 & 827.31 $\pm$3.17 & 1.56E0$\pm$1.75E-2 & 182.26  $\pm$  0.18 & 5.02E2  $\pm$  1.05E2 \\
PG2233+134               &   253.00 $\pm$  26.06 & 1.55E2  $\pm$  2.18E1 & 143.65  $\pm$  4.18 & 5.98E3  $\pm$  2.95E2 & 592.41  $\pm$  15.16 & 3.11E1$\pm$4.93 & 155.43  $\pm$  1.25 & 1.20E4  $\pm$  5.89E2 \\
PG2251+113               &   334.92 $\pm$  34.85 & 5.47E1  $\pm$  8.93E1 &  40.00  $\pm$ 26.51 & 3.00E3  $\pm$  2.09E2 & 735.66  $\pm$ 100.48 & 1.09E1$\pm$1.79E1 & 183.65  $\pm$  7.21 & 6.01E3  $\pm$  4.17E2 \\
PG2304+042               &   150.00 $\pm$  17.07 & 5.04E-2  $\pm$  3.08E-1 & 194.77  $\pm$  1.46 & 1.22E1  $\pm$  6.53E-1 &1500.00  $\pm$ 356.30 & 1.20E-2$\pm$8.08E-2 & 217.36  $\pm$  4.98 & 2.44E1  $\pm$  1.31 \\
PG2308+098               &   197.34 $\pm$  12.19 & 4.42  $\pm$  1.14E1 & 197.39  $\pm$  0.72 & 1.67E3  $\pm$  3.68E1 & 871.36  $\pm$  12.40 & 8.84E0$\pm$2.29E1 & 192.43  $\pm$  1.63 & 3.34E3  $\pm$  7.36E1 \\
2MASSiJ081652.2+425829   &   319.66 $\pm$  19.84 & 1.48E1  $\pm$  5.86 &  40.00  $\pm$ 35.45 & 3.71E3  $\pm$  1.42E3 & 719.47  $\pm$  65.88 & 2.96E0$\pm$1.17 & 176.42  $\pm$ 18.26 & 7.64E2  $\pm$  7.09E2 \\
2MASSiJ095504.5+170556   &   178.45 $\pm$  40.51 & 7.34  $\pm$  3.20 & 172.85  $\pm$  5.58 & 1.64E2  $\pm$  8.58 & 731.20  $\pm$  19.56 & 1.47E0$\pm$6.04 & 181.01  $\pm$  1.92 & 3.27E2  $\pm$  2.05E1 \\
2MASSiJ130005.3+163214   &   150.00$\pm$0.00 & 1.74E1  $\pm$  3.99 & 137.78  $\pm$  2.74 & 5.40E2  $\pm$  4.89E1 & 715.68  $\pm$  14.96 & 3.48E0$\pm$7.52 & 189.57  $\pm$  0.51 & 9.41E2  $\pm$  1.26E2 \\
2MASSiJ132917.5+121340   &   255.49 $\pm$  82.96 & 2.08E1  $\pm$  9.70 &  81.99  $\pm$  9.98 & 3.25E3  $\pm$  8.51E2 & 562.71  $\pm$  37.76 & 5.22E0$\pm$9.53 & 179.31  $\pm$  8.73 & 6.49E2  $\pm$  1.70E2 \\
2MASSiJ1402511+263117    &   191.28 $\pm$  27.29 & 2.33  $\pm$  8.62 & 156.37  $\pm$  9.90 & 3.51E2  $\pm$  3.52E1 & 729.98  $\pm$  14.12 & 4.65E0$\pm$1.73E1 & 197.07  $\pm$  2.02 & 7.01E2  $\pm$  7.56E1 \\
2MASSiJ145608.6+275008   &   330.00 $\pm$  63.97 & 2.58E1  $\pm$  5.95E1 &  84.98  $\pm$ 15.50 & 1.72E4  $\pm$  1.34E4 & 698.81  $\pm$  95.40 & 8.58E0$\pm$2.04E1 & 164.89  $\pm$ 13.87 & 3.44E3  $\pm$  2.67E3 \\
2MASSiJ151653.2+190048   &   323.31 $\pm$  12.31 & 1.03E2  $\pm$  1.58E1 &  40.00  $\pm$  0.00 & 4.75E4  $\pm$  6.65E2 & 683.70 $\pm$6.98 & 2.86E1$\pm$6.09 & 156.70  $\pm$  0.76 & 1.02E4  $\pm$  1.49E2 \\
2MASSiJ151901.5+183804   &   350.92 $\pm$  23.98 & 4.37  $\pm$  9.66E-1 &  83.62  $\pm$  3.96 & 3.59E3  $\pm$  4.69E2 & 821.46  $\pm$  59.21 & 8.74E-1$\pm$2.51E-1 & 158.54  $\pm$  3.72 & 7.18E2  $\pm$  9.39E1 \\
2MASSiJ154307.7+193751   &   246.49 $\pm$  36.20 & 1.63E2  $\pm$  3.89E1 &  40.00  $\pm$  0.00 & 5.70E4  $\pm$  7.63E3 & 597.44  $\pm$  20.57 & 3.25E1$\pm$3.81E1 & 152.52  $\pm$  1.94 & 1.14E4  $\pm$  3.35E3 \\
2MASSiJ222221.1+195947   &   312.89$\pm$6.62 & 3.64E1  $\pm$  2.55 &  40.00  $\pm$  0.00 & 2.03E4  $\pm$  2.51E2 & 719.87  $\pm$  13.54 & 7.29E0$\pm$5.11E-1 & 166.80  $\pm$  0.58 & 4.07E3  $\pm$  5.03E1 \\
2MASSiJ223742.6+145614   &   262.58 $\pm$  11.27 & 6.18E1  $\pm$  7.75 &  71.60  $\pm$ 24.54 & 2.63E4  $\pm$  8.85E3 & 586.67  $\pm$  18.74 & 1.24E1$\pm$1.55 & 144.68  $\pm$  9.83 & 5.25E3  $\pm$  1.77E3 \\
2MASSiJ234259.3+134750   &   338.21 $\pm$  50.33 & 3.61E1  $\pm$  1.94E1 &  96.94  $\pm$  5.53 & 1.95E4  $\pm$  3.72E3 & 606.38  $\pm$  67.85 & 7.22E0$\pm$3.88 & 146.57  $\pm$  6.14 & 3.90E3  $\pm$  1.17E3 \\
2MASSiJ234449.5+122143   &   296.00 $\pm$  40.78 & 3.96E1  $\pm$  1.77E1 &  71.45  $\pm$  2.01 & 4.13E4  $\pm$  2.45E3 & 625.95  $\pm$  31.90 & 9.22E0$\pm$4.26 & 144.47  $\pm$  1.91 & 8.25E3  $\pm$  4.89E2 \\
2MASXJ09210862+4538575   &   219.07 $\pm$  49.13 & 6.97  $\pm$  8.66 & 163.39  $\pm$  9.07 & 1.55E2  $\pm$  1.96E1 & 666.39  $\pm$ 131.72 & 1.39E0$\pm$4.00 & 200.06  $\pm$  7.72 & 3.10E2  $\pm$  4.21E1 \\
2MASXJ00370409-0109081   &   295.44 $\pm$  32.02 & 1.20  $\pm$  3.08E-1 & 127.84  $\pm$ 37.46 & 1.89E1  $\pm$  7.88 & 637.13  $\pm$  52.61 & 2.39E-1$\pm$8.98E-2 & 178.78  $\pm$  3.99 & 3.78E1  $\pm$  1.61E1 \\
2MASXJ02335161+0108136   &   634.16 $\pm$ 198.50 & 1.18E-2  $\pm$  4.37E-3 &  64.09  $\pm$  0.66 & 1.10E2  $\pm$  7.82 &1288.37  $\pm$ 140.98 & 2.35E-3$\pm$1.08E-3 & 132.70  $\pm$  2.15 & 2.21E1  $\pm$  1.56 \\
2MASXJ07582810+3747121   &   345.77 $\pm$ 106.23 & 8.08E-2  $\pm$  1.84E-1 &  58.13  $\pm$ 23.66 & 5.79E1  $\pm$  4.24E1 & 931.25  $\pm$ 185.08 & 2.20E-2$\pm$5.18E-2 & 152.34  $\pm$ 11.18 & 2.93E1  $\pm$  4.80E1 \\
2MASXiJ0208238-002000    &   236.18$\pm$8.17 & 5.12  $\pm$  6.54E-1 & 126.35  $\pm$  8.32 & 9.27E1  $\pm$  2.88E1 & 492.74  $\pm$  17.14 & 1.02E0$\pm$1.31E-1 & 168.73  $\pm$  2.05 & 1.85E2  $\pm$  6.52E1 \\
2MASXJ02061600-0017292   &   296.29$\pm$3.04 & 3.39  $\pm$  1.13E-1 &  58.80  $\pm$  4.61 & 5.57E2  $\pm$  1.95E2 & 746.93 $\pm$7.76 & 6.79E-1$\pm$2.25E-2 & 175.24  $\pm$  0.81 & 1.87E2  $\pm$  8.45E1 \\
2MASXJ10493088+2257523   &   150.00 $\pm$  10.34 & 1.22E-1  $\pm$  3.72E-1 & 139.69  $\pm$  9.98 & 1.04E2  $\pm$  4.15 & 712.89 $\pm$8.52 & 2.45E-1$\pm$7.50E-1 & 173.89  $\pm$  0.57 & 2.08E2  $\pm$  8.30 \\
2MASXJ12485992-0109353   &   301.01$\pm$5.02 & 2.33E1  $\pm$  1.59 &  42.06  $\pm$  1.43 & 3.54E4  $\pm$  2.62E2 & 490.22 $\pm$6.89 & 4.66E0$\pm$3.18E-1 & 136.12  $\pm$  0.29 & 7.07E3  $\pm$  5.26E1 \\
2MASXJ14070036+2827141   &   334.20$\pm$2.02 & 4.63E1  $\pm$  9.03E-1 &  62.77  $\pm$  0.08 & 5.47E4  $\pm$  2.11E2 & 581.09 $\pm$2.49 & 9.27E0$\pm$1.81E-1 & 134.13  $\pm$  0.14 & 1.09E4  $\pm$  4.23E1 \\
2MASXJ02143357-0046002   &   267.91$\pm$3.67 & 2.45  $\pm$  1.31E-1 &  56.11  $\pm$  0.49 & 2.10E3  $\pm$  1.18E1 & 539.30 $\pm$6.50 & 4.89E-1$\pm$2.62E-2 & 153.40  $\pm$  0.25 & 4.20E2  $\pm$  2.36 \\
2MASXJ09234300+2254324   &   237.82$\pm$4.77 & 6.93  $\pm$  8.41E-1 & 112.27  $\pm$  3.56 & 1.38E2  $\pm$  3.10E1 & 575.01  $\pm$  13.89 & 1.39E0$\pm$1.68E-1 & 171.89  $\pm$  0.95 & 2.77E2  $\pm$  6.95E1 \\
2MASXJ12170991+0711299   &   302.86 $\pm$  26.60 & 5.82E-2  $\pm$  2.24E-1 &  57.72  $\pm$  3.07 & 6.02E1  $\pm$  1.30E1 & 566.17  $\pm$  47.07 & 3.81E-2$\pm$1.47E-1 & 143.98  $\pm$ 13.98 & 1.20E1  $\pm$  1.37E1 \\
2MASXJ12232410+0240449   &   270.26 $\pm$  29.61 & 4.20E-1  $\pm$  7.78E-2 & 170.29  $\pm$  9.67 & 5.35  $\pm$  9.59E-1 & 710.58  $\pm$  31.95 & 8.39E-2$\pm$1.56E-2 & 190.11  $\pm$  1.87 & 1.07E1  $\pm$  2.27 \\
2MASXJ13381586+0432330   &   370.52$\pm$4.78 & 5.01E-1  $\pm$  2.07E-2 &  63.94  $\pm$  0.22 & 3.38E2  $\pm$  9.51 & 985.22  $\pm$  15.52 & 1.00E-1$\pm$4.15E-3 & 159.66  $\pm$  0.33 & 1.22E2  $\pm$  4.98 \\
2MASXJ13495283+0204456   &   207.83$\pm$4.27 & 2.52  $\pm$  1.09E-1 & 129.11  $\pm$  1.38 & 4.83E1  $\pm$  7.08E-1 & 534.18 $\pm$5.39 & 5.04E-1$\pm$2.18E-2 & 168.67  $\pm$  0.53 & 9.65E1  $\pm$  1.42 \\
2MASXJ23044349-0841084   &   295.60$\pm$1.63 & 5.41  $\pm$  9.96E-2 &  62.65  $\pm$  0.12 & 6.90E3  $\pm$  2.83E1 & 721.66 $\pm$4.11 & 1.08E0$\pm$1.99E-2 & 150.29  $\pm$  0.16 & 1.38E3  $\pm$  5.65 \\
SDSSJ115138.24+004946.4  &   334.95 $\pm$ 148.97 & 3.62  $\pm$  4.29 & 155.57  $\pm$  7.42 & 1.77E2  $\pm$  1.85E1 & 696.58  $\pm$ 154.45 & 7.23E-1$\pm$1.58 & 193.56  $\pm$  6.91 & 3.54E2  $\pm$  3.71E1 \\
SDSSJ170246.09+602818.8  &   325.32 $\pm$ 165.85 & 3.21E-1  $\pm$  4.02E-1 &  51.97  $\pm$ 22.73 & 1.62E2  $\pm$  6.62E1 & 512.92  $\pm$  77.13 & 8.84E-2$\pm$2.60E-1 & 137.93  $\pm$  7.20 & 3.85E1  $\pm$  7.19E1 \\
\enddata
\end{deluxetable}
\end{landscape}

\begin{landscape}
\begin{deluxetable}{lllllllll}
\tabletypesize\tiny 
\tablecolumns{9}           
\tablecaption{\footnotesize
              \label{tab:flat_continuum_mod_para}
              Model Parameters for 30 AGNs 
              Which Show No Silicate Emission
              but a Featureless Thermal Continuum
              } 
\tablewidth{0pt}           
\tablehead{                
\multicolumn{1}{l}{Source} &
\multicolumn{1}{c}{$\Twsil$} &
\multicolumn{1}{c}{$\Mwsil$} &
\multicolumn{1}{c}{$\Tcsil$} &
\multicolumn{1}{c}{$\Mcsil$} &
\multicolumn{1}{c}{$\Twcarb$} &
\multicolumn{1}{c}{$\Mwcarb$} &
\multicolumn{1}{c}{$\Tccarb$} &
\multicolumn{1}{c}{$\Mccarb$}  \\ 
 &
\multicolumn{1}{c}{(K)} &
\multicolumn{1}{c}{(${\rm M_{\odot}}$)} &
\multicolumn{1}{c}{(K)} &
\multicolumn{1}{c}{(${\rm M_{\odot}}$)} &
\multicolumn{1}{c}{(K)} &
\multicolumn{1}{c}{(${\rm M_{\odot}}$)} &
\multicolumn{1}{c}{(K)} &
\multicolumn{1}{c}{(${\rm M_{\odot}}$)} \\ 
\multicolumn{1}{l}{(1)} &
\multicolumn{1}{c}{(2)} &
\multicolumn{1}{c}{(3)} &
\multicolumn{1}{c}{(4)} &
\multicolumn{1}{c}{(5)} &
\multicolumn{1}{c}{(6)} &
\multicolumn{1}{c}{(7)} &
\multicolumn{1}{c}{(8)} &
\multicolumn{1}{c}{(9)} 
} 
\startdata 
PG0838+770               &  383.27  $\pm$  23.92  &  4.53 $\pm$  8.96E-1  &   65.69 $\pm$0.16  &  6.98E3 $\pm$  3.89E2  &  552.66 $\pm$9.14  &  5.89 $\pm$  1.19  &  150.94 $\pm$  0.18  &  3.83E3 $\pm$  3.02E2 \\  
PG1226+023               &  245.43 $\pm$0.53  &  3.09E2 $\pm$  1.01  &   40.00 $\pm$0.00  &  1.34E5 $\pm$  2.51E3  &  686.41 $\pm$0.65  &  6.18E1 $\pm$  2.01E-1  &  152.17 $\pm$  0.06  &  3.90E4 $\pm$  1.13E3 \\
PG1354+213               &  521.75  $\pm$  22.11  &  3.69 $\pm$  5.67E-1  &   52.09 $\pm$4.04  &  4.14E4 $\pm$  9.19E2  &  713.85 $\pm$5.54  &  7.39 $\pm$  1.32  &  155.59 $\pm$  0.39  &  8.28E3 $\pm$  2.66E2 \\
PG1427+480               &  150.00  $\pm$  75.01  &  1.44E1 $\pm$  1.48E1  &   72.71 $\pm$3.53  &  3.66E4 $\pm$  3.27E3  &  624.51  $\pm$  26.38  &  7.88 $\pm$  8.23  &  149.61 $\pm$  2.12  &  7.33E3 $\pm$  6.55E2 \\
PG1448+273               &  239.30 $\pm$1.20  &  5.19 $\pm$  6.88E-2  &   40.00 $\pm$0.00  &  3.74E3 $\pm$  9.85  &  607.59 $\pm$2.16  &  1.04 $\pm$  1.38E-2  &  171.10 $\pm$  0.10  &  7.50E2 $\pm$  2.12 \\
PG1501+106               &  178.91 $\pm$1.10  &  4.25 $\pm$  5.59E-2  &   48.99 $\pm$0.27  &  5.04E3 $\pm$  6.90  &  606.14 $\pm$2.14  &  8.51E-1 $\pm$  1.12E-2  &  166.72 $\pm$  0.06  &  1.01E3 $\pm$  1.38 \\
PG1543+489               &  291.97 $\pm$2.11  &  1.18E2 $\pm$  2.60  &   85.60 $\pm$0.15  &  1.98E5 $\pm$  1.30E3  &  570.60 $\pm$1.07  &  9.47E1 $\pm$  2.86  &  163.69 $\pm$  0.21  &  3.96E4 $\pm$  2.60E2 \\
2MASSiJ010835.1+214818   &  184.20  $\pm$  10.74  &  1.68E2 $\pm$  7.13  &   73.28 $\pm$1.07  &  7.98E4 $\pm$  1.53E3  &  599.78 $\pm$6.37  &  3.35E1 $\pm$  1.43  &  158.67 $\pm$  0.71  &  1.60E4 $\pm$  6.32E2 \\
2MASSiJ024807.3+145957   &  150.00  $\pm$  32.16  &  1.71E1 $\pm$  3.71  &   70.70 $\pm$1.11  &  6.23E3 $\pm$  4.01E2  &  485.88  $\pm$  23.01  &  4.75 $\pm$  1.89  &  146.55 $\pm$  2.27  &  1.25E3 $\pm$  8.02E1 \\
2MASSiJ082311.3+435318   &  150.00  $\pm$  13.68  &  6.49E1 $\pm$  1.09E1  &   74.37 $\pm$2.94  &  2.76E4 $\pm$  1.70E3  &  566.30  $\pm$  24.85  &  1.30E1 $\pm$  2.20  &  158.99 $\pm$  2.54  &  5.75E3 $\pm$  5.99E2 \\
2MASSiJ145410.1+195648   &  150.00 $\pm$0.52  &  8.04E1 $\pm$  2.71E1  &   98.49  $\pm$  25.43  &  2.98E3 $\pm$  4.03E3  &  620.35  $\pm$  21.97  &  1.61E1 $\pm$  4.76E1  &  162.54 $\pm$  3.15  &  3.54E3 $\pm$  4.90E3 \\
2MASXJ17223993+3052521   &  150.00 $\pm$0.05  &  1.62 $\pm$  2.77  &   64.56 $\pm$3.48  &  2.82E3 $\pm$  1.94E2  &  476.09  $\pm$  15.79  &  1.53 $\pm$  2.65  &  143.04 $\pm$  2.05  &  5.63E2 $\pm$  3.88E1 \\
2MASXJ13130577+0127561   &  150.00  $\pm$  37.35  &  2.55E-2 $\pm$  5.32E-2  &  154.25  $\pm$  30.08  &  3.13 $\pm$  7.66E-1  &  813.23  $\pm$ 193.63  &  5.19E-3 $\pm$  2.33E-2  &  170.28 $\pm$  4.49  &  6.26 $\pm$  1.58 \\
SDSSJ090738.71+564358.2  &  150.00 $\pm$0.00  &  1.41 $\pm$  7.77E-1  &   68.76 $\pm$1.55  &  1.80E3 $\pm$  2.63E2  &  433.34  $\pm$  65.60  &  1.24 $\pm$  9.42E-1  &  138.17 $\pm$  5.04  &  3.60E2 $\pm$  5.27E1 \\
2MASXJ13130565-0210390   &  239.94  $\pm$  21.18  &  1.79 $\pm$  9.33E-1  &   59.06 $\pm$1.65  &  1.00E3 $\pm$  5.75E2  &  605.56  $\pm$  13.36  &  8.50E-1 $\pm$  5.75E-1  &  158.14 $\pm$  0.96  &  4.29E2 $\pm$  3.53E2 \\
SDSSJ124035.81-002919.4  &  234.95  $\pm$  15.11  &  6.42 $\pm$  3.82  &   47.57 $\pm$4.20  &  4.77E3 $\pm$  1.24E2  &  523.27  $\pm$  27.32  &  1.28 $\pm$  7.67E-1  &  144.28 $\pm$  1.16  &  9.55E2 $\pm$  2.50E1 \\
2MASXJ15055659+0342267   &  260.53 $\pm$2.21  &  3.37 $\pm$  9.97E-2  &   60.44 $\pm$0.18  &  3.39E3 $\pm$  1.32E1  &  588.28 $\pm$4.49  &  6.75E-1 $\pm$  1.99E-2  &  151.76 $\pm$  0.17  &  6.77E2 $\pm$  2.63 \\
2MASXJ09191322+5527552   &  237.54  $\pm$  41.58  &  3.25 $\pm$  7.86E-1  &   58.53 $\pm$0.72  &  2.48E3 $\pm$  1.98E1  &  565.31  $\pm$  11.69  &  7.43E-1 $\pm$  1.09  &  152.32 $\pm$  0.48  &  4.96E2 $\pm$  3.96 \\
SDSSJ101536.21+005459.3  &  150.00 $\pm$0.00  &  1.92E1 $\pm$  7.46E-1  &  137.51 $\pm$1.52  &  4.15E2 $\pm$  2.02E1  &  489.77 $\pm$5.63  &  3.84 $\pm$  1.49E-1  &  163.49 $\pm$  0.46  &  8.29E2 $\pm$  5.15E1 \\
SDSSJ164840.15+425547.6  &  150.00  $\pm$  63.25  &  1.78 $\pm$  5.35  &   62.28 $\pm$5.46  &  3.68E3 $\pm$  5.79E2  &  473.91  $\pm$  36.12  &  1.58 $\pm$  4.83  &  139.21 $\pm$  4.10  &  7.36E2 $\pm$  1.16E2 \\
SDSSJ091414.34+023801.7  &  186.93  $\pm$  39.59  &  1.37 $\pm$  2.68  &   63.66 $\pm$7.25  &  1.43E3 $\pm$  2.01E2  &  461.95  $\pm$  51.19  &  4.55E-1 $\pm$  9.29E-1  &  138.76 $\pm$  4.10  &  2.87E2 $\pm$  4.02E1 \\
2MASXJ12384342+0927362   &  150.00 $\pm$0.00  &  2.27E1 $\pm$  2.26  &  106.73 $\pm$1.92  &  1.25E3 $\pm$  6.21E1  &  472.61 $\pm$3.80  &  4.53 $\pm$  4.10  &  151.59 $\pm$  0.32  &  2.50E3 $\pm$  1.58E2 \\
2MASXJ16164729+3716209   &  172.57 $\pm$7.27  &  7.62E1 $\pm$  5.10  &   40.00  $\pm$  25.60  &  5.73E3 $\pm$  3.38E2  &  458.47 $\pm$4.28  &  1.54E1 $\pm$  1.39  &  140.99 $\pm$  1.24  &  1.15E4 $\pm$  6.76E2 \\
2MASXJ11230133+4703088   &  265.83  $\pm$  24.90  &  7.07E-2 $\pm$  1.59E-2  &   59.62 $\pm$1.05  &  1.23E2 $\pm$  3.12  &  781.16  $\pm$  65.83  &  1.41E-2 $\pm$  3.18E-3  &  161.03 $\pm$  1.10  &  2.46E1 $\pm$  6.23E-1 \\
2MASXJ11110693+0228477   &  150.00  $\pm$  25.49  &  2.65 $\pm$  4.10E-1  &   67.55 $\pm$0.52  &  5.76E2 $\pm$  1.15E1  &  467.19 $\pm$7.51  &  7.11E-1 $\pm$  2.17E-1  &  165.19 $\pm$  1.31  &  1.15E2 $\pm$  2.30 \\
2MASSiJ1448250+355946    &  280.42 $\pm$3.96  &  1.70E1 $\pm$  1.25  &   68.10 $\pm$0.23  &  9.34E3 $\pm$  2.77E3  &  499.67 $\pm$5.73  &  6.40 $\pm$  7.16E-1  &  133.97 $\pm$  0.62  &  6.25E3 $\pm$  2.33E3 \\
SDSSJ164019.66+403744.4  &  150.00 $\pm$2.41  &  6.47 $\pm$  2.37E1  &  137.20 $\pm$5.40  &  4.33E2 $\pm$  1.12E2  &  474.61  $\pm$  78.43  &  1.29 $\pm$  5.42  &  151.75 $\pm$  8.34  &  4.30E2 $\pm$  1.70E2 \\
SDSSJ104058.79+581703.3  &  229.68  $\pm$  96.48  &  1.45 $\pm$  9.23E-1  &   40.00 $\pm$3.30  &  8.74E2 $\pm$  8.44E1  &  514.64  $\pm$  44.30  &  2.91E-1 $\pm$  9.53E-1  &  147.45 $\pm$  3.01  &  1.75E2 $\pm$  1.98E1 \\
UGC05984                 &  235.37  $\pm$  22.94  &  3.49E-1 $\pm$  1.55  &   59.41 $\pm$1.00  &  9.82E2 $\pm$  5.76E1  &  535.89  $\pm$  79.68  &  6.98E-2 $\pm$  3.11E-1  &  138.12 $\pm$  2.54  &  1.96E2 $\pm$  1.15E1 \\
UGC06527                 &  218.61 $\pm$1.50  &  3.05 $\pm$  4.19E-2  &   60.02 $\pm$0.23  &  1.51E3 $\pm$  5.12  &  567.10 $\pm$2.08  &  6.10E-1 $\pm$  8.38E-3  &  159.92 $\pm$  0.14  &  3.01E2 $\pm$  1.02 \\
\enddata
\end{deluxetable}
\end{landscape}

\begin{landscape}
\begin{deluxetable}{lllllllll}
\tabletypesize\tiny 
\tablecolumns{9}           
\tablecaption{\footnotesize
              \label{tab:pah_em_mod_para}
              Model Parameters for 24 AGNs 
              Which Show a Thermal Continuum
              Superimposed with PAH Features
              } 
\tablewidth{0pt}           
\tablehead{                
\multicolumn{1}{l}{Source} &
\multicolumn{1}{c}{$\Twsil$} &
\multicolumn{1}{c}{$\Mwsil$} &
\multicolumn{1}{c}{$\Tcsil$} &
\multicolumn{1}{c}{$\Mcsil$} &
\multicolumn{1}{c}{$\Twcarb$} &
\multicolumn{1}{c}{$\Mwcarb$} &
\multicolumn{1}{c}{$\Tccarb$} &
\multicolumn{1}{c}{$\Mccarb$}  \\ 
 &
\multicolumn{1}{c}{(K)} &
\multicolumn{1}{c}{(${\rm M_{\odot}}$)} &
\multicolumn{1}{c}{(K)} &
\multicolumn{1}{c}{(${\rm M_{\odot}}$)} &
\multicolumn{1}{c}{(K)} &
\multicolumn{1}{c}{(${\rm M_{\odot}}$)} &
\multicolumn{1}{c}{(K)} &
\multicolumn{1}{c}{(${\rm M_{\odot}}$)} \\ 
\multicolumn{1}{l}{(1)} &
\multicolumn{1}{c}{(2)} &
\multicolumn{1}{c}{(3)} &
\multicolumn{1}{c}{(4)} &
\multicolumn{1}{c}{(5)} &
\multicolumn{1}{c}{(6)} &
\multicolumn{1}{c}{(7)} &
\multicolumn{1}{c}{(8)} &
\multicolumn{1}{c}{(9)} 
} 
\startdata 
PG0007+106               &  264.22 $\pm$  3.52  &  1.34E1 $\pm$  7.21E-1  &  59.13 $\pm$  0.23  &  1.71E4 $\pm$  5.29E1  &  624.16$\pm$3.28  &  4.60 $\pm$  3.29E-1  &  146.49  $\pm$  0.12  &  3.42E3 $\pm$  1.06E1 \\  
PG0157+001               &  342.59 $\pm$  0.69  &  1.07E2 $\pm$  7.54E-1  &  63.67 $\pm$  0.14  &  9.94E5 $\pm$  4.06E3  &  558.03$\pm$0.90  &  2.14E1 $\pm$  1.51E-1  &  116.62  $\pm$  0.07  &  1.99E5 $\pm$  8.12E2 \\
PG0923+129               &  184.57 $\pm$  0.82  &  4.91 $\pm$  3.59E-2  &  57.39 $\pm$  0.14  &  2.33E3 $\pm$  4.81  &  549.11$\pm$1.06  &  9.81E-1 $\pm$  7.19E-3  &  156.94  $\pm$  0.09  &  4.66E2 $\pm$  9.62E-1 \\
PG0934+013               &  207.87 $\pm$  2.91  &  6.91 $\pm$  1.95E-1  &  65.89 $\pm$  0.09  &  3.55E3 $\pm$  1.54E1  &  505.09$\pm$3.45  &  1.38 $\pm$  3.89E-2  &  142.56  $\pm$  0.15  &  7.10E2 $\pm$  3.08 \\
PG1022+519               &  199.99 $\pm$  2.04  &  2.46 $\pm$  2.52E-1  &  64.38 $\pm$  0.14  &  1.75E3 $\pm$  1.52E1  &  509.99$\pm$1.94  &  1.05 $\pm$  1.46E-1  &  138.17  $\pm$  0.34  &  3.49E2 $\pm$  3.03 \\
PG1115+407               &  150.00 $\pm$  7.09  &  1.36E1 $\pm$  8.85  &  73.56 $\pm$  6.62  &  1.22E4 $\pm$  2.04E3  &  617.87 $\pm$  17.67  &  6.25 $\pm$  4.18  &  158.95  $\pm$  3.95  &  2.44E3 $\pm$  4.08E2 \\
PG1119+120               &  239.27 $\pm$  3.08  &  7.45 $\pm$  5.97E-1  &  65.26 $\pm$  0.18  &  7.74E3 $\pm$  3.41E1  &  478.03$\pm$1.95  &  4.98 $\pm$  5.27E-1  &  147.76  $\pm$  0.18  &  1.55E3 $\pm$  6.82 \\
PG1126-041               &  150.00 $\pm$  0.00  &  2.88 $\pm$  2.01E-1  &  67.04 $\pm$  0.06  &  1.16E4 $\pm$  2.23E1  &  770.16$\pm$3.85  &  1.94 $\pm$  1.99E-1  &  155.06  $\pm$  0.07  &  2.32E3 $\pm$  4.45 \\
PG1149-110               &  210.57 $\pm$  2.44  &  3.45 $\pm$  6.47E-2  &  49.91 $\pm$  0.39  &  3.89E3 $\pm$  1.08E1  &  605.99$\pm$3.26  &  6.91E-1 $\pm$  1.29E-2  &  151.38  $\pm$  0.10  &  7.78E2 $\pm$  2.15 \\
PG1244+026               &  267.05 $\pm$  2.18  &  2.61 $\pm$  9.20E-2  &  68.36 $\pm$  0.15  &  1.79E3 $\pm$  1.12E2  &  482.88$\pm$2.54  &  1.75 $\pm$  8.76E-2  &  143.43  $\pm$  0.26  &  7.83E2 $\pm$  7.03E1 \\
PG1415+451               &  215.73 $\pm$  2.14  &  2.16E1 $\pm$  1.49  &  66.63 $\pm$  0.37  &  9.46E3 $\pm$  4.82E1  &  573.56$\pm$1.88  &  5.25 $\pm$  5.13E-1  &  154.77  $\pm$  0.19  &  1.89E3 $\pm$  9.64 \\
PG1425+267               &  288.67 $\pm$  9.53  &  1.31E2 $\pm$  1.62E1  &  73.65 $\pm$  1.09  &  8.95E4 $\pm$  2.30E3  &  655.00$\pm$3.73  &  2.74E1 $\pm$  5.37  &  156.77  $\pm$  0.56  &  1.79E4 $\pm$  4.90E2 \\
PG1519+226               &  247.49 $\pm$  2.56  &  1.57E1 $\pm$  7.42E-1  &  70.14 $\pm$  0.47  &  1.11E4 $\pm$  8.56E1  &  637.60$\pm$1.84  &  6.68 $\pm$  4.77E-1  &  163.31  $\pm$  0.28  &  2.21E3 $\pm$  1.71E1 \\
PG1612+261               &  223.24 $\pm$  1.19  &  3.15E1 $\pm$  2.78E-1  &  40.00 $\pm$  0.00  &  2.51E4 $\pm$  7.41E1  &  610.56$\pm$1.51  &  6.30 $\pm$  5.56E-2  &  150.61  $\pm$  0.10  &  5.01E3 $\pm$  1.50E1 \\
PG1613+658               &  150.00 $\pm$  0.00  &  1.26E2 $\pm$  8.70  &  68.19 $\pm$  0.07  &  5.41E4 $\pm$  1.36E2  &  571.74$\pm$1.05  &  2.94E1 $\pm$  3.54  &  155.68  $\pm$  0.08  &  1.08E4 $\pm$  2.72E1 \\
PG2130+099               &  150.00 $\pm$  0.00  &  1.06E1 $\pm$  1.70E-1  &  66.86 $\pm$  0.19  &  1.21E4 $\pm$  3.47E1  &  609.57$\pm$1.15  &  5.88 $\pm$  1.34E-1  &  159.42  $\pm$  0.10  &  2.42E3 $\pm$  6.95 \\
2MASSiJ165939.7+183436   &  150.00 $\pm$  2.60  &  5.45E1 $\pm$  9.40E1  &  76.87 $\pm$  8.09  &  5.25E4 $\pm$  1.25E4  &  492.49 $\pm$  21.82  &  3.57E1 $\pm$  6.19E1  &  153.14  $\pm$  5.03  &  1.05E4 $\pm$  2.49E3 \\
2MASXJ08381094+2453427   &  250.83 $\pm$  3.62  &  1.04 $\pm$  4.96E-2  &  62.80 $\pm$  0.20  &  1.15E3 $\pm$  7.04  &  595.64$\pm$7.61  &  2.07E-1 $\pm$  9.93E-3  &  152.54  $\pm$  0.25  &  2.30E2 $\pm$  1.41 \\
2MASXJ22533142+0048252   &  191.06 $\pm$  6.43  &  8.76 $\pm$  8.54E-1  &  89.88 $\pm$ 31.24  &  2.15E2 $\pm$  2.44E1  &  457.59$\pm$9.73  &  1.75 $\pm$  2.23E-1  &  154.25  $\pm$  2.55  &  4.30E2 $\pm$  4.89E1 \\
2MASXJ15085397-0011486   &  244.12 $\pm$  4.23  &  2.24 $\pm$  7.37E-2  &  45.04 $\pm$  0.75  &  3.80E3 $\pm$  1.75E1  &  707.77$\pm$7.34  &  4.47E-1 $\pm$  1.47E-2  &  154.11  $\pm$  0.17  &  7.59E2 $\pm$  3.50 \\
2MASXJ14175951+2508124   &  216.15 $\pm$  1.76  &  4.55 $\pm$  1.21E-1  &  63.11 $\pm$  0.06  &  2.34E3 $\pm$  6.80  &  477.36$\pm$3.04  &  9.10E-1 $\pm$  2.41E-2  &  140.46  $\pm$  0.12  &  4.67E2 $\pm$  1.36 \\
2MASXJ12042964+2018581   &  281.38 $\pm$ 15.51  &  1.12 $\pm$  1.34E-1  &  65.10 $\pm$  0.14  &  1.55E3 $\pm$  1.16E1  &  538.85$\pm$7.81  &  5.90E-1 $\pm$  1.15E-1  &  145.77  $\pm$  0.35  &  3.10E2 $\pm$  2.32 \\
2MASXJ10032788+5541535   &  218.90 $\pm$ 15.85  &  4.83 $\pm$  1.15 &  58.48 $\pm$  3.03  &  2.01E3 $\pm$  8.48E1  &  502.65 $\pm$  26.73  &  9.65E-1 $\pm$  2.85E-1  &  158.26  $\pm$  1.69  &  4.02E2 $\pm$  1.76E1 \\
2MASSJ16593976+1834367   &  237.46 $\pm$ 29.89  &  6.83E1 $\pm$  1.40E2  &  73.17 $\pm$  7.24  &  7.24E4 $\pm$  1.06E4  &  507.00 $\pm$  24.51  &  4.55E1 $\pm$  9.60E1  &  150.69  $\pm$  3.81  &  1.45E4 $\pm$  2.12E3 \\
\enddata
\end{deluxetable}
\end{landscape}

\clearpage

\begin{figure*}[ht]
\figurenum{\ref{fig:sil_em_mod1}}
\leavevmode
\begin{center}
\resizebox{0.8\hsize}{!}{
\includegraphics[angle=90]{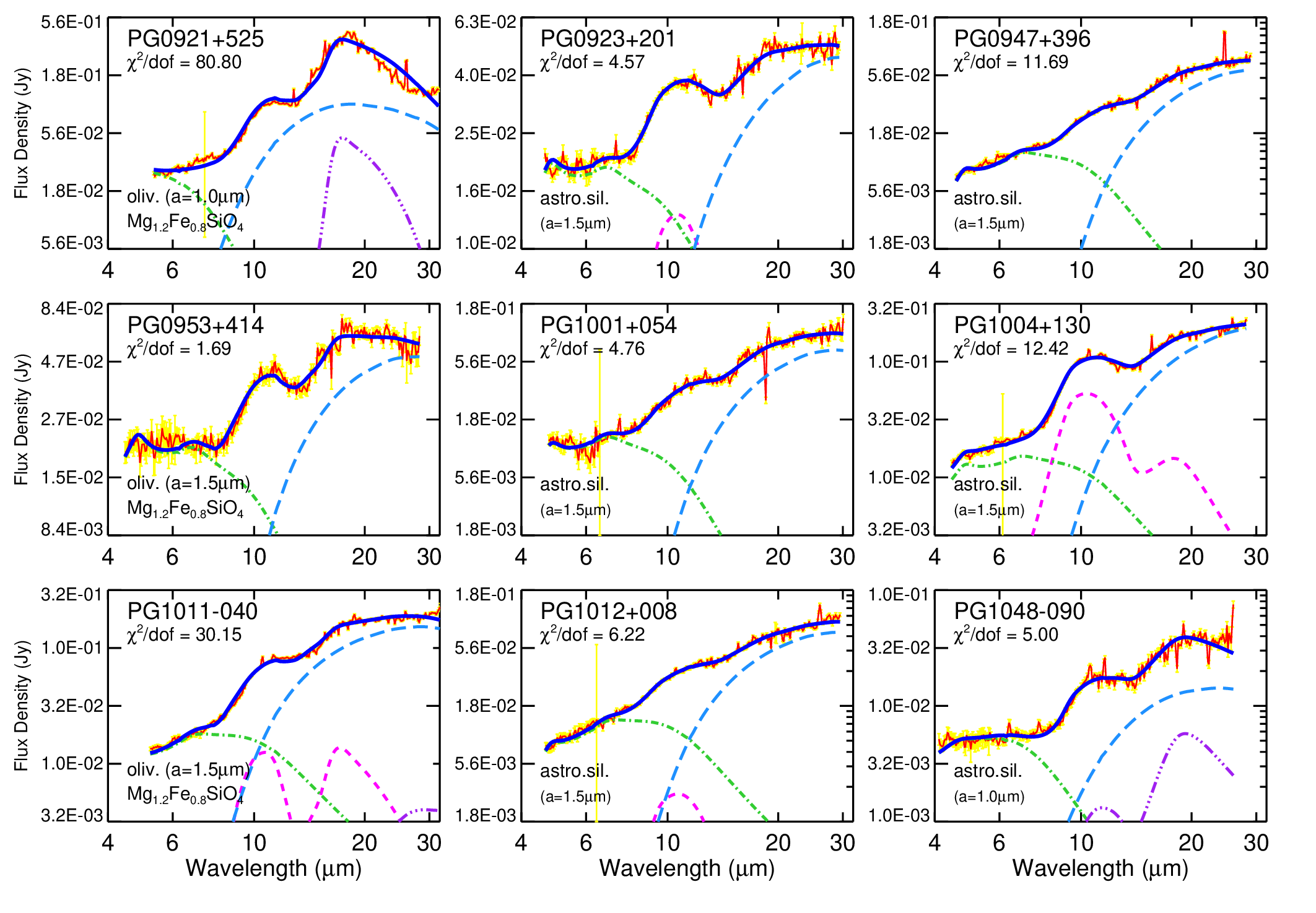}}
\caption{\footnotesize
           Continued, but for
           the PG quasars 
           PG0921+525, PG0923+201, PG0947+396, PG0953+414, PG1001+054, 
           PG1004+130, PG1011-040, PG1012+008, and PG1048-090.           
}
\end{center}
\end{figure*}

\begin{figure*}[ht]
\figurenum{\ref{fig:sil_em_mod1}}
\leavevmode
\begin{center}
\resizebox{0.8\hsize}{!}{
\includegraphics[angle=90]{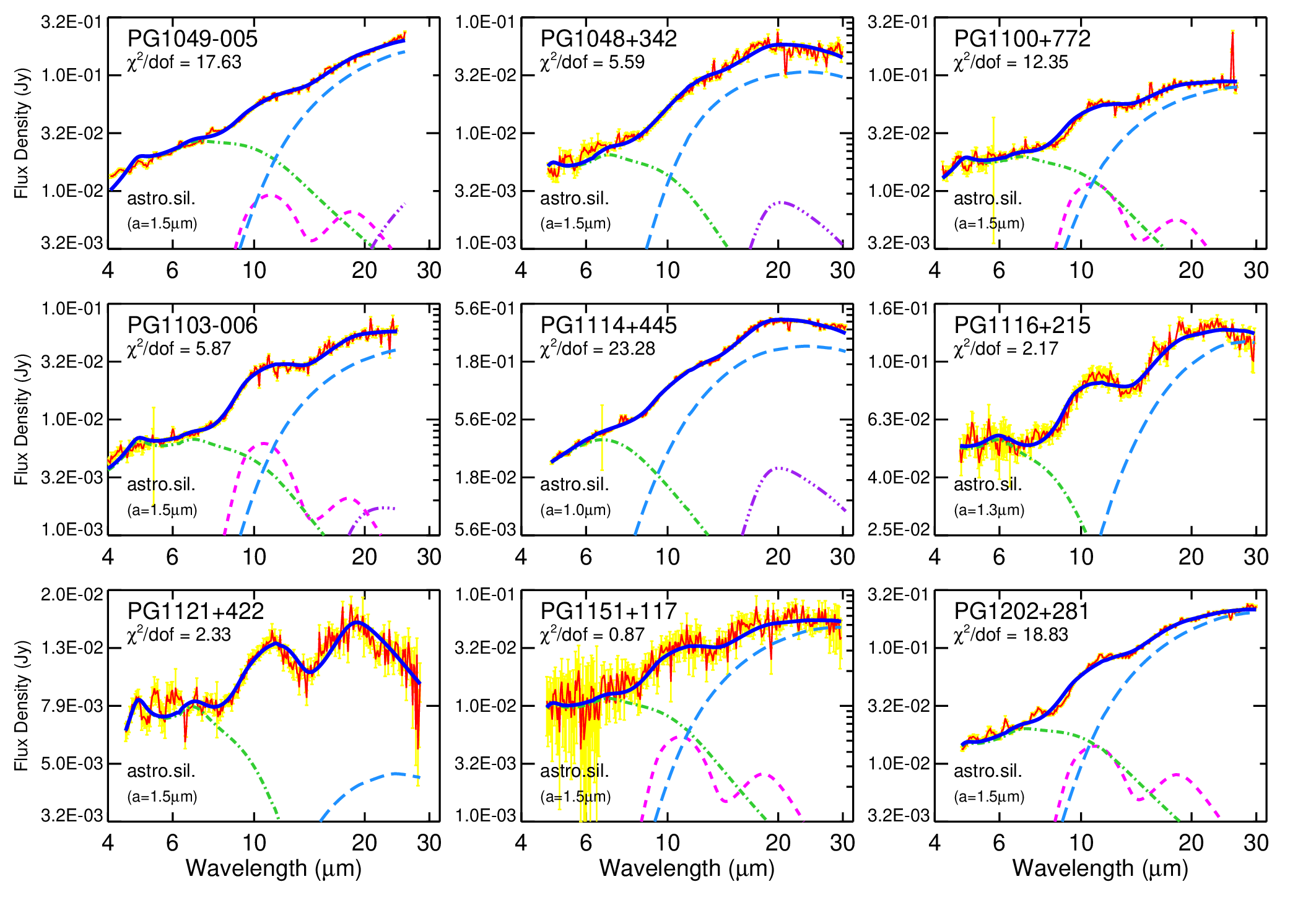}}
\caption{\footnotesize
           Continued, but for
           the PG quasars 
           PG1049-005, PG1048+342, PG1100+772, PG1103-006, 
           PG1114+445, PG1116+215, PG1121+422, PG1151+117, 
           and PG1202+281. 
           }
\end{center}
\end{figure*}

\begin{figure*}[ht]
\figurenum{\ref{fig:sil_em_mod1}}
\leavevmode
\begin{center}
\resizebox{0.8\hsize}{!}{
\includegraphics[angle=90]{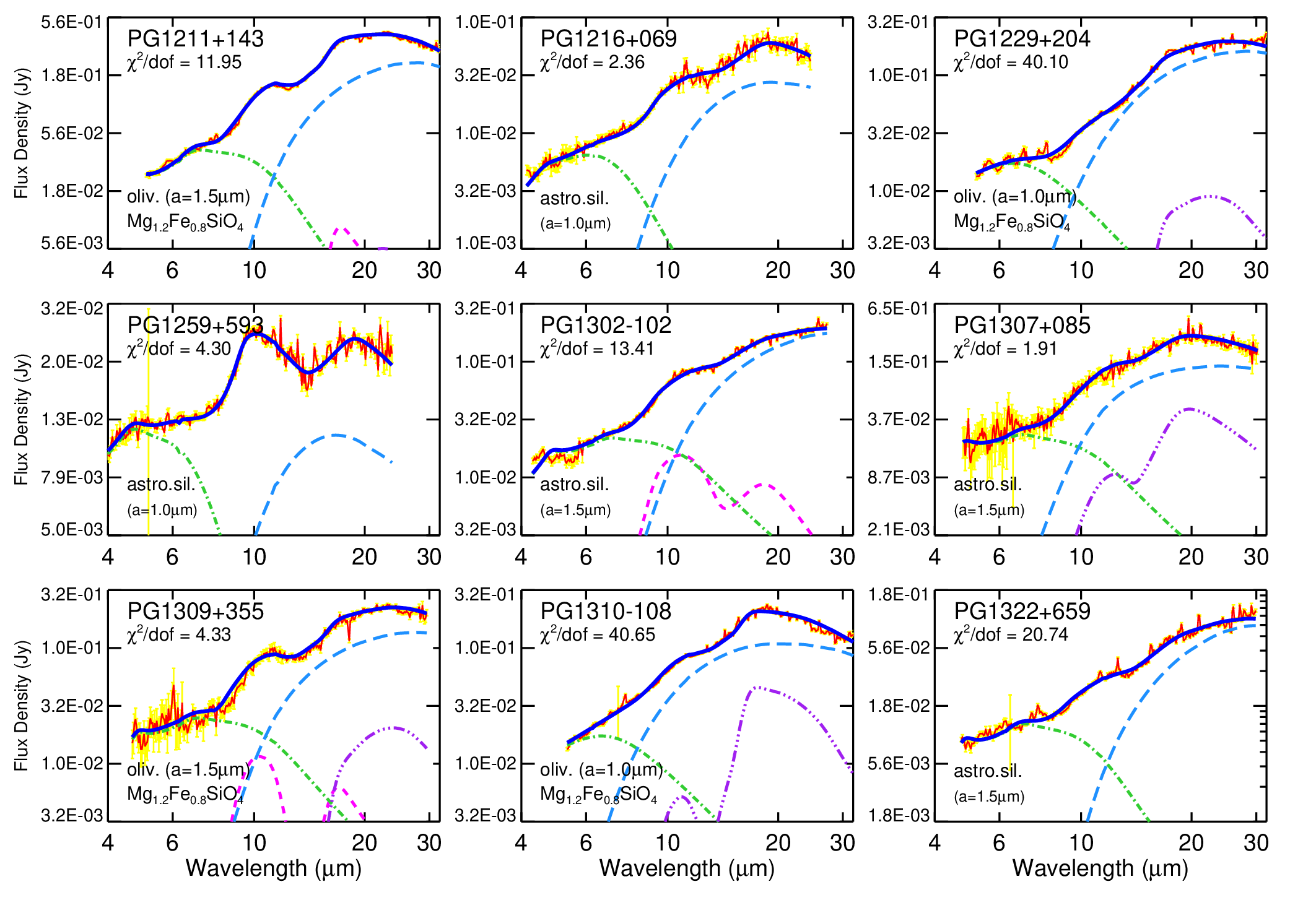}}
\caption{\footnotesize
           Continued, but for
           the PG quasars PG1211+143, PG1216+069, PG1229+204, 
           PG1259+593, PG1302-102, PG1307+085, PG1309+355, 
           PG1310-108, and PG1322+659.   
}
\end{center}
\end{figure*}

\begin{figure*}[ht]
\figurenum{\ref{fig:sil_em_mod1}}
\leavevmode
\begin{center}
\resizebox{0.8\hsize}{!}{
\includegraphics[angle=90]{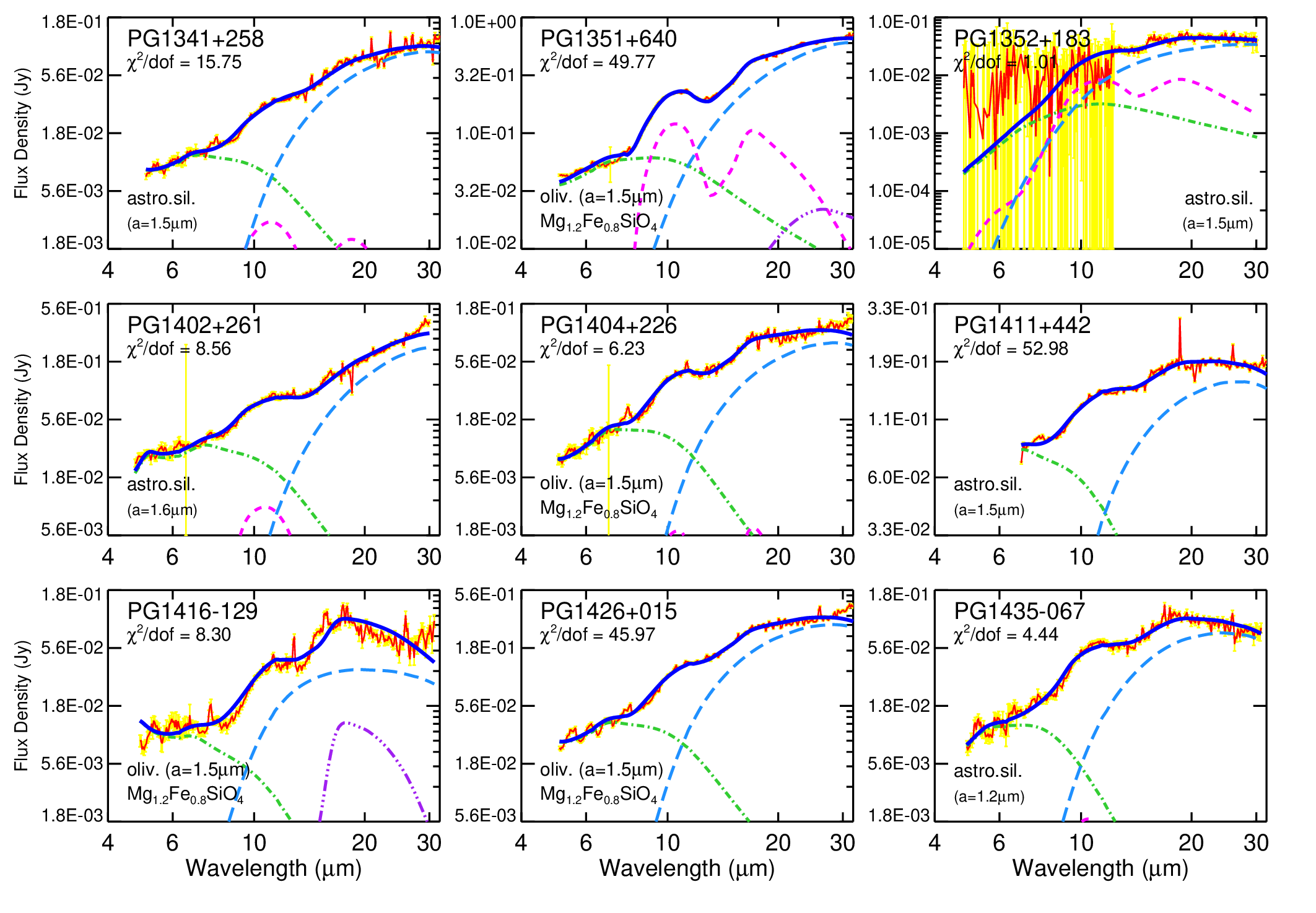}}
\caption{\footnotesize
           Continued, but for
           the PG quasars PG1341+258, PG1351+640, PG1352+183, 
           PG1402+261, PG1404+226, PG1411+442, PG1416-129, 
           PG1426+015, and PG1435-067. 
}
\end{center}
\end{figure*}

\begin{figure*}[ht]
\figurenum{\ref{fig:sil_em_mod1}}
\leavevmode
\begin{center}
\resizebox{0.8\hsize}{!}{
\includegraphics[angle=90]{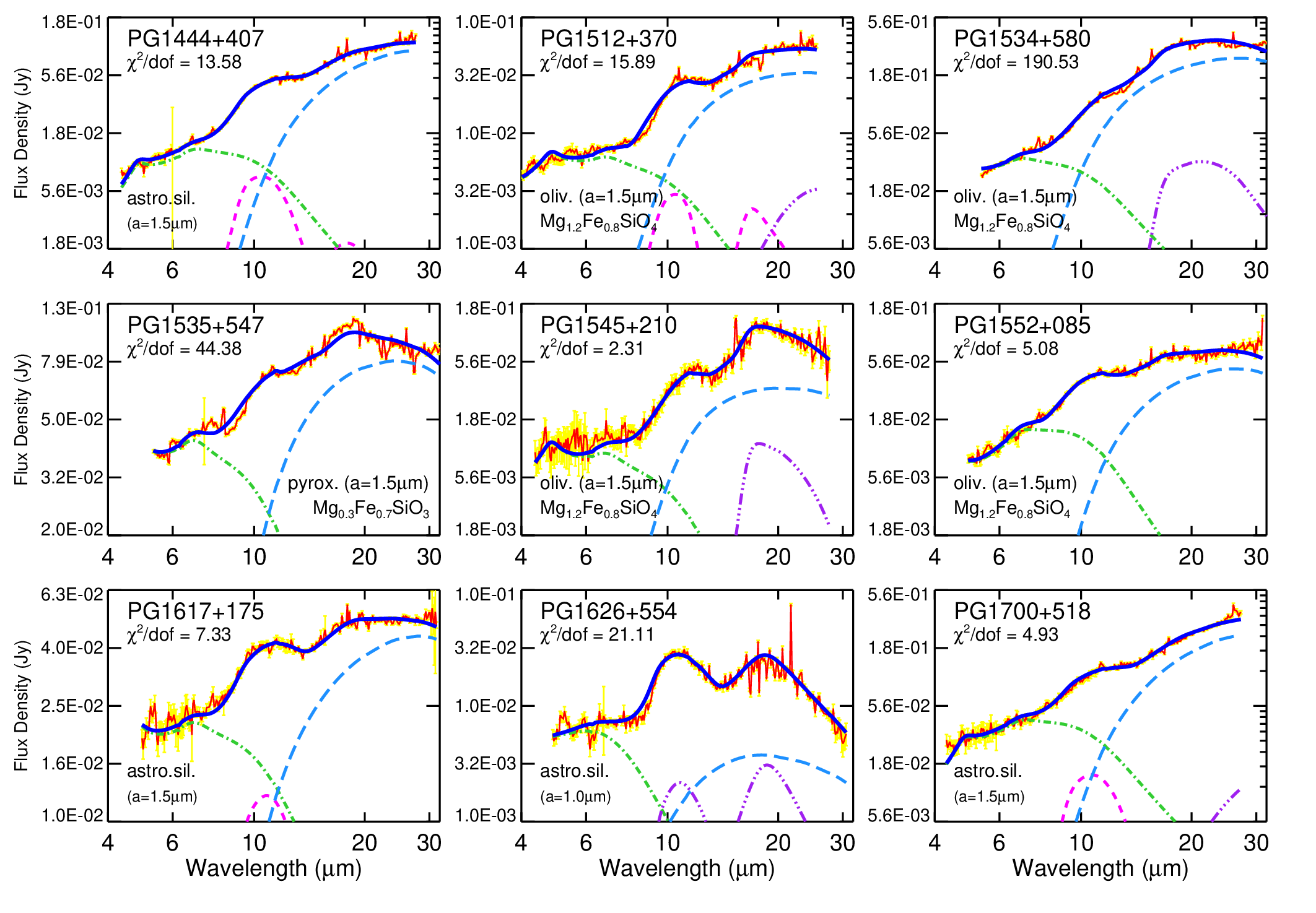}}
\caption{\footnotesize
           Continued, but for
           the PG quasars PG1444+407, PG1512+370, PG1534+580, 
           PG1535+547, PG1545+210, PG1552+085, PG1617+175, 
           PG1626+554, and PG1700+518.
}
\end{center}
\end{figure*}

\begin{figure*}[ht]
\figurenum{\ref{fig:sil_em_mod1}}
\leavevmode
\begin{center}
\resizebox{0.8\hsize}{!}{
\includegraphics[angle=90]{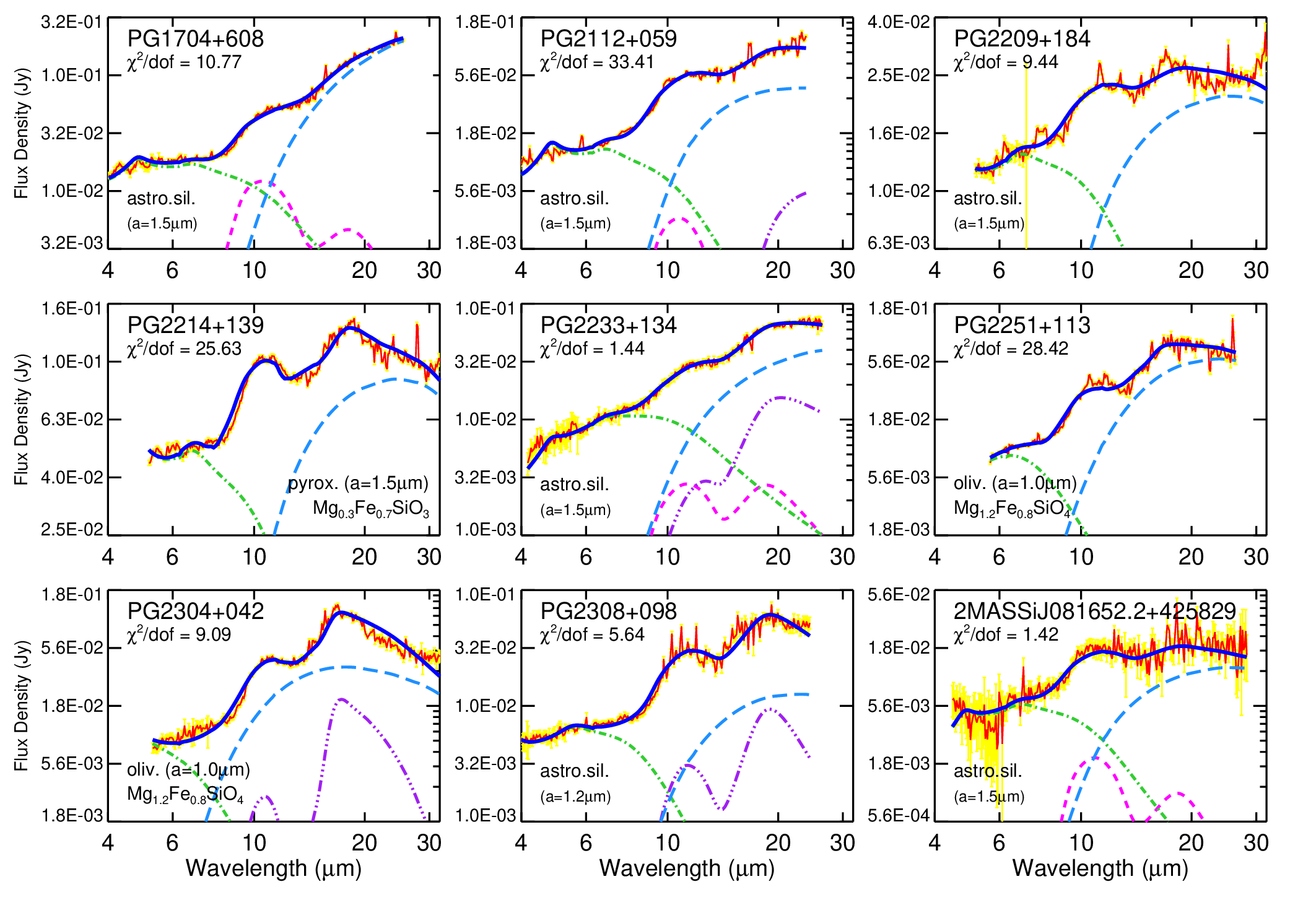}}
\caption{\footnotesize
           Continued, but for
           the PG quasars PG1704+608, PG2112+059, PG2209+184, 
           PG2214+139, PG2233+134, PG2251+113, PG2304+042, PG2308+098  
           and the {\it 2MASS} quasar 2MASSiJ081652.2+425829.  
}
\end{center}
\end{figure*}

\begin{figure*}[ht]
\figurenum{\ref{fig:sil_em_mod1}}
\leavevmode
\begin{center}
\resizebox{0.8\hsize}{!}{
\includegraphics[angle=90]{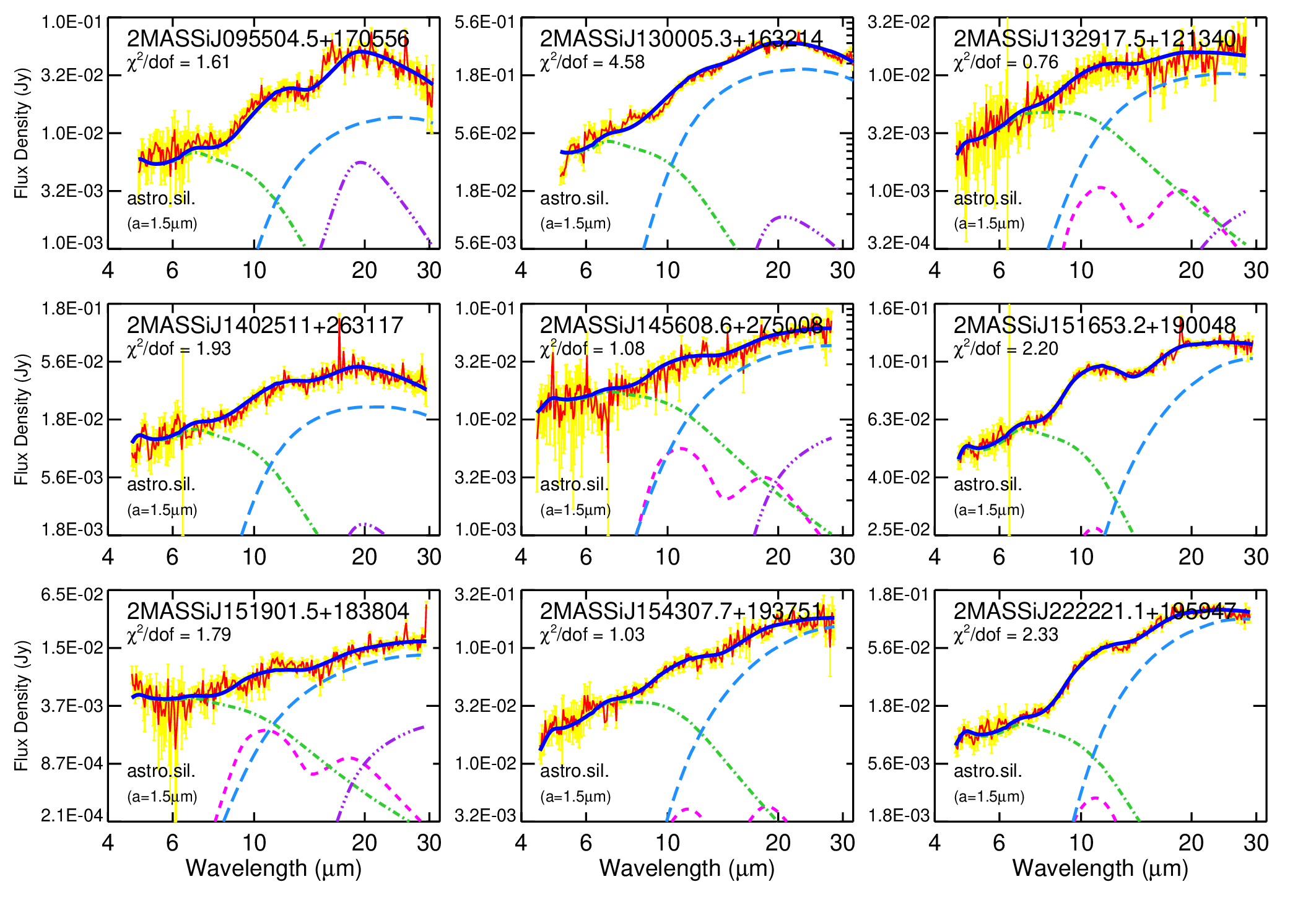}}
\caption{\footnotesize
           Continued, but for
           the {\it 2MASS} quasars 2MASSiJ095504.5+170556, 
           2MASSiJ130005.3+163214, 2MASSiJ132917.5+121340, 
           2MASSiJ1402511+263117, 2MASSiJ145608.6+275008, 
           2MASSiJ151653.2+190048, 2MASSiJ151901.5+183804, 
           2MASSiJ154307.7+193751, and 2MASSiJ222221.1+195947.        
}
\end{center}
\end{figure*}

\begin{figure*}[ht]
\figurenum{\ref{fig:sil_em_mod1}}
\leavevmode
\begin{center}
\resizebox{0.8\hsize}{!}{
\includegraphics[angle=90]{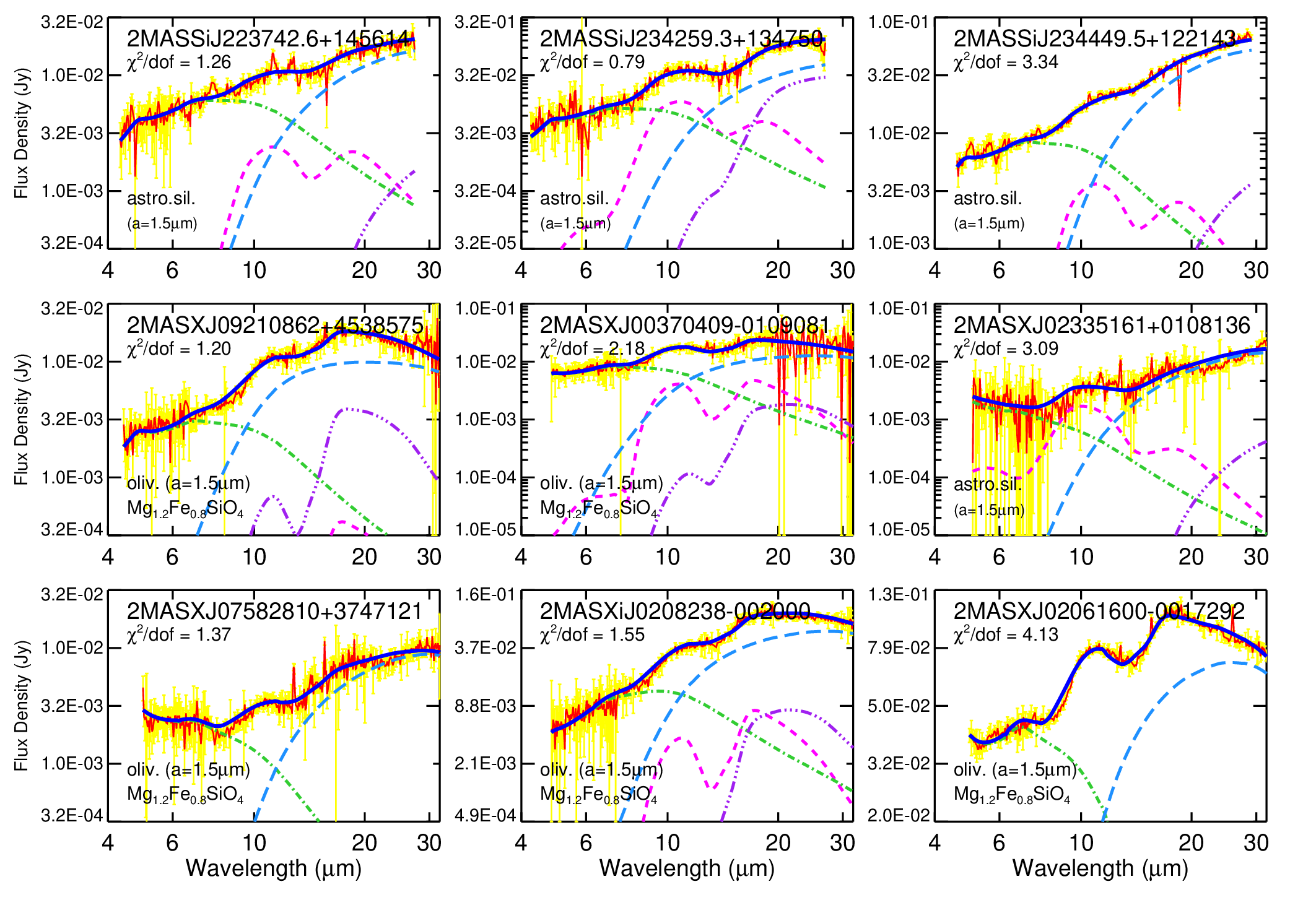}}
\caption{\footnotesize
           Continued, but for
           the {\it 2MASS} quasars 2MASSiJ223742.6+145614, 
           2MASSiJ234259.3+134750, 2MASSiJ234449.5+122143 
           and the \saga\ AGNs 2MASXJ09210862+4538575, 
           2MASXJ00370409-0109081, 2MASXJ02335161+0108136, 
           2MASXJ07582810+3747121, 2MASXiJ0208238-002000, 
           and 2MASXJ02061600-0017292.      
}
\end{center}
\end{figure*}

\begin{figure*}[ht]
\figurenum{\ref{fig:sil_em_mod1}}
\leavevmode
\begin{center}
\resizebox{0.8\hsize}{!}{
\includegraphics[angle=90]{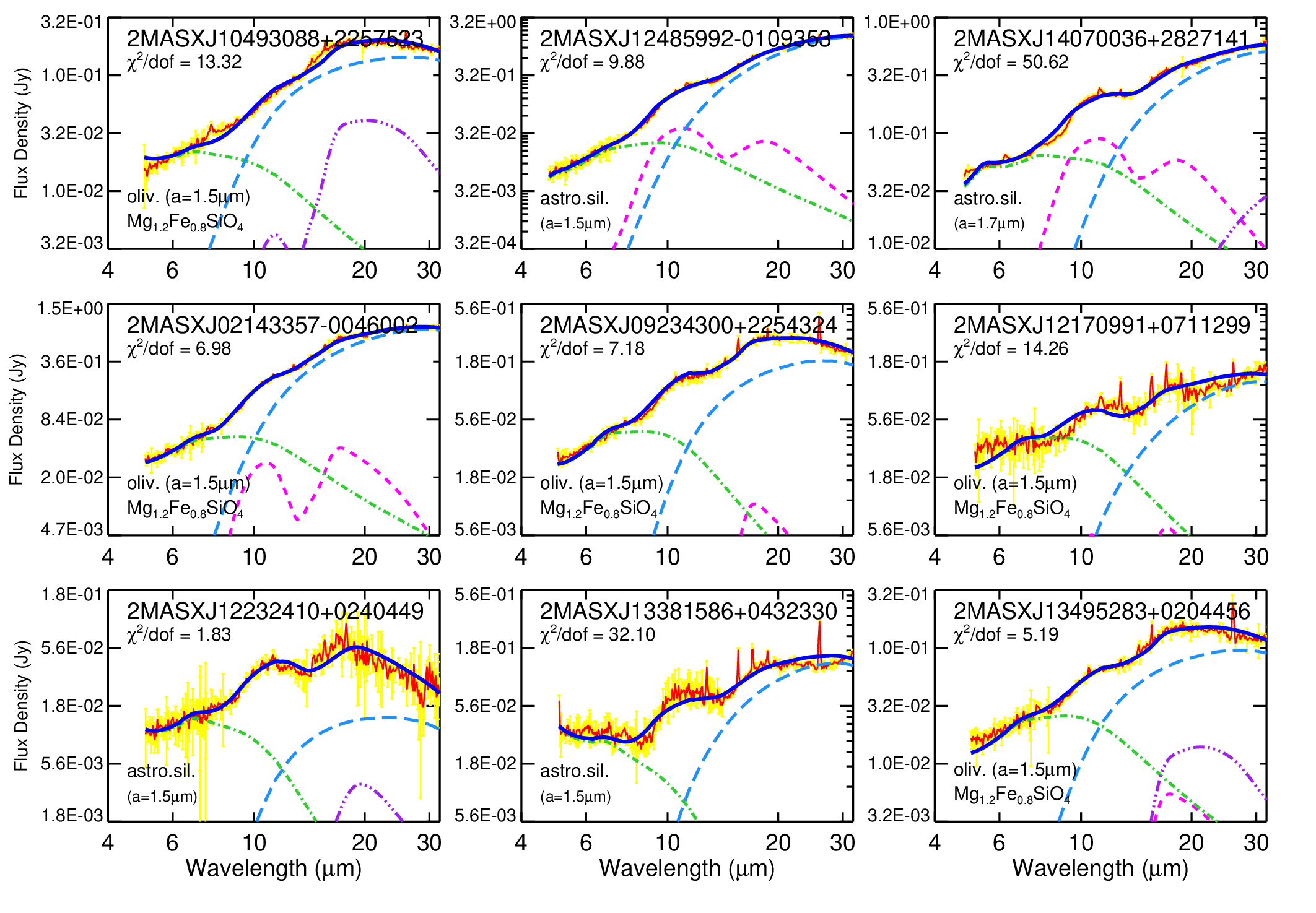}}
\caption{\footnotesize
           Continued, but for
           the \saga\ AGNs 
           2MASXJ10493088+2257523, 2MASXJ12485992-0109353, 
           2MASXJ14070036+2827141, 
           2MASXJ02143357-0046002, 2MASXJ09234300+2254324, 
           2MASXJ12170991+0711299, 2MASXJ12232410+0240449,   
           2MASXJ13381586+0432330, and 2MASXJ13495283+0204456.  
}
\end{center}
\end{figure*}

\begin{figure*}[ht]
\figurenum{\ref{fig:sil_em_mod1}}
\leavevmode
\begin{center}
\resizebox{0.8\hsize}{!}{
\includegraphics[angle=90]{f1k.eps}}
\caption{\footnotesize
           \label{fig:sil_em_mod11}
           Continued, but for
           the \saga\ AGNs 
           2MASXJ23044349-0841084,  
           SDSSJ115138.24+004946.4, and SDSSJ170246.09+602818.8. 
}
\end{center}
\end{figure*}

\begin{figure*}
\begin{center}
\resizebox{0.75 \vsize}{!}{
\includegraphics{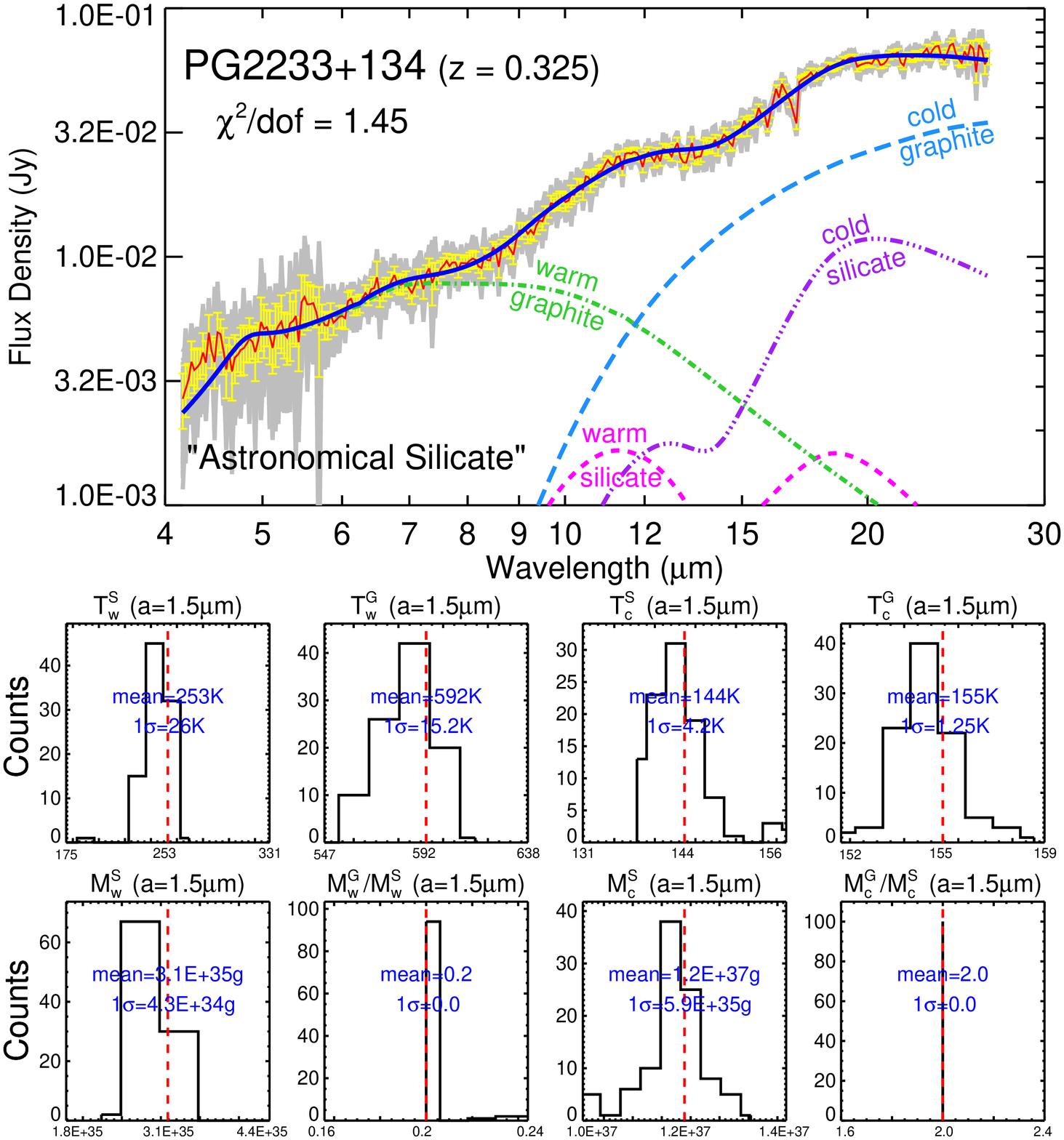}}
\caption{\footnotesize
         \label{fig:mtcarlo}
         Estimating the uncertainties of the model parameters from
         Monte-Carlo simulation, with PG 2233+134 as an example. 
         The top panel compares the {\it Spitzer}/IRS spectrum 
         (red solid line) of PG 2233+134,
         as well as the observed $1\sigma$ error (yellow vertical lines) 
         and the random spectrum generated from 
         Monte-Carlo simulation (gray lines), 
         with the model spectrum (blue solid line)
          which is the sum of warm silicate
          (magenta short dashed lines),
          cold silicate (purple dash-dot-dotted lines),
          warm graphite (green dash-dotted lines),
          and cold graphite (light blue long dashed lines).
         The middle panels show the distributions of
         the dust temperatures 
         derived from 100 Monte-Carlo simulations for 
         warm silicate, warm graphite, 
         cold silicate and cold graphite. 
         The bottom panels show the mass and mass-ratio distributions  
         derived from 100 Monte-Carlo simulations for 
         warm silicate or graphite, and cold silicate or graphite.
         }
\end{center}
\end{figure*}

\begin{figure*}
\begin{center}
$
\begin{array}{c}
\resizebox{0.5\vsize}{!}{\includegraphics{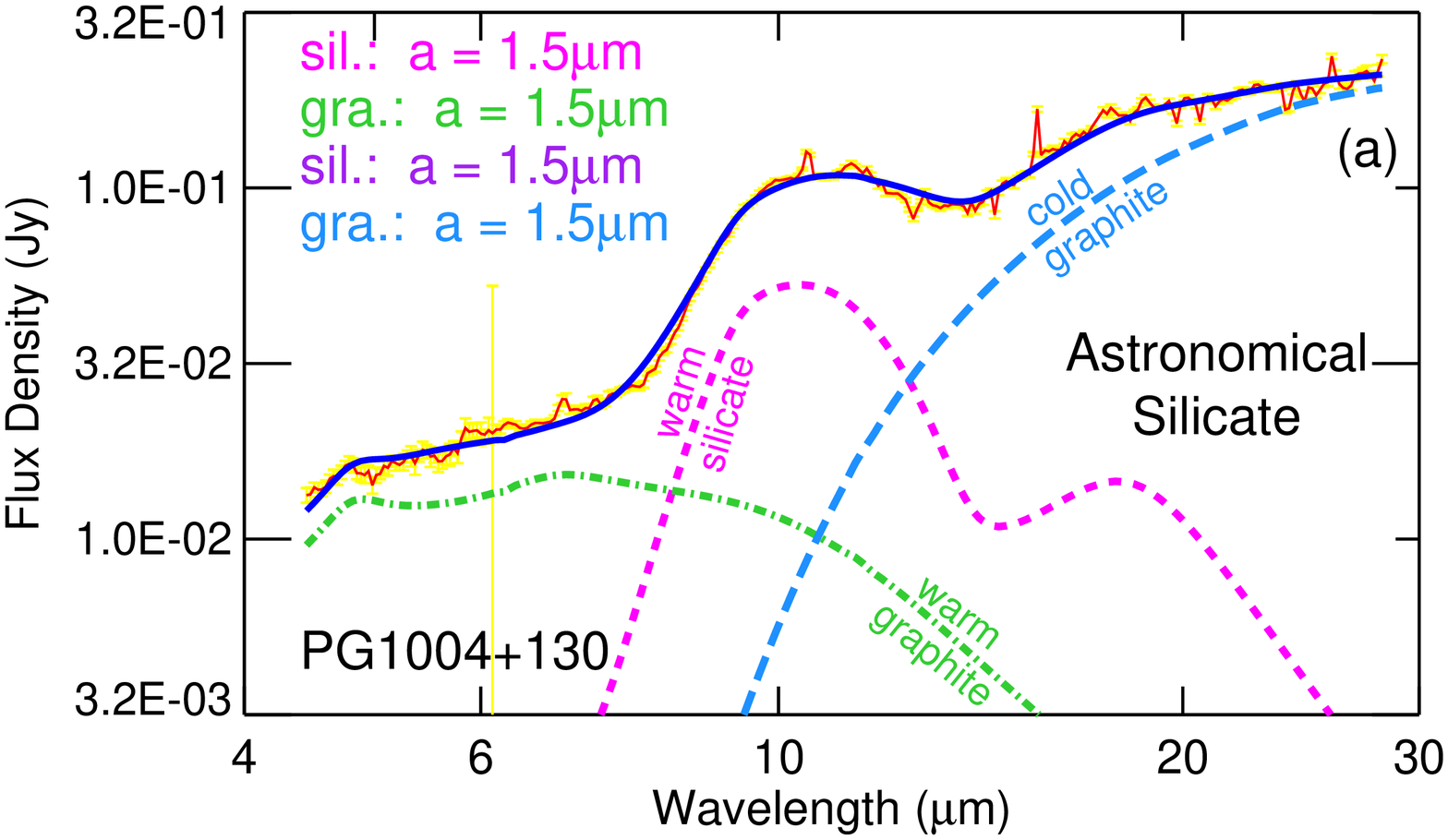}} \\
\resizebox{0.5\vsize}{!}{\includegraphics{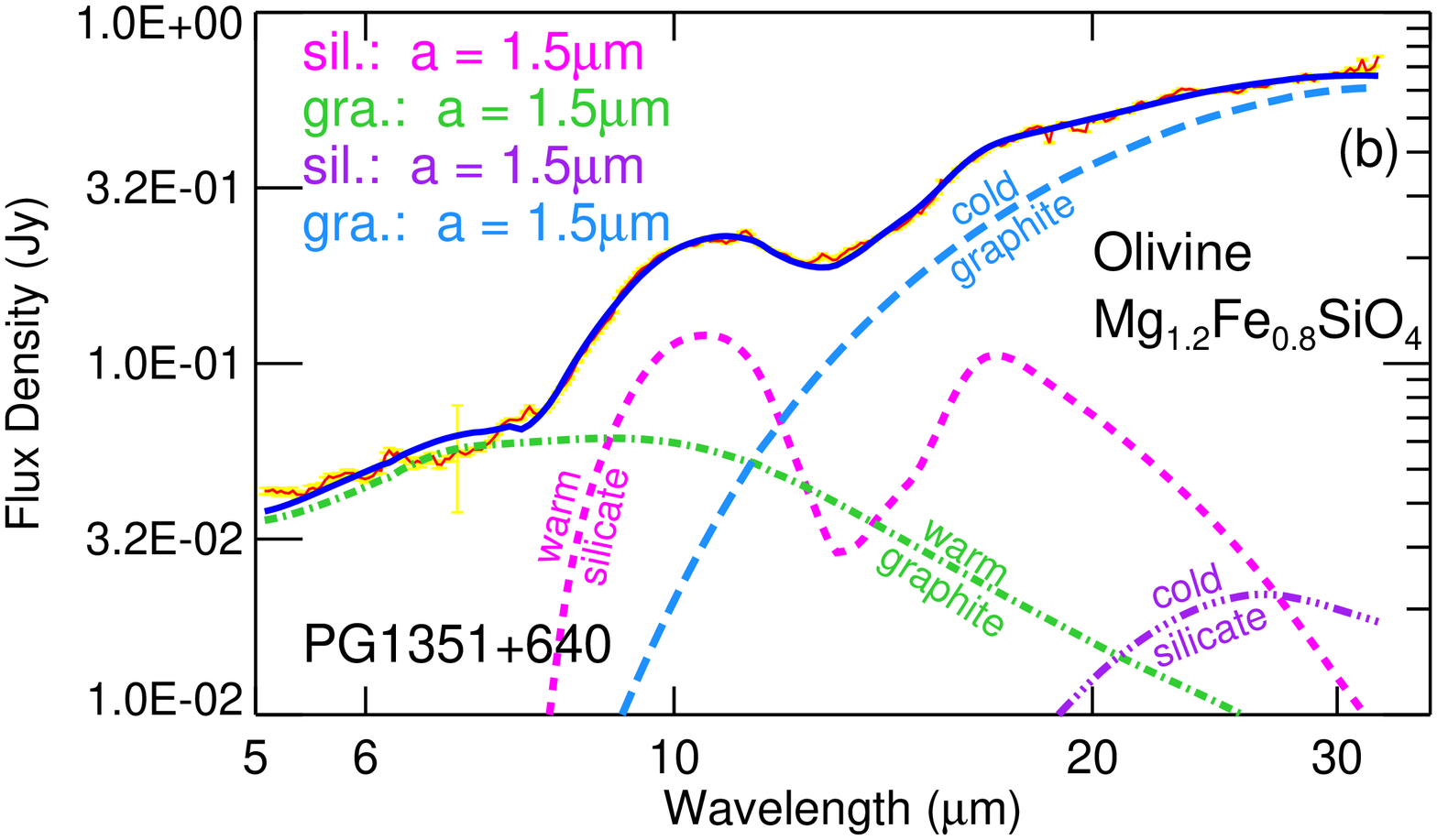}} \\
\resizebox{0.5\vsize}{!}{\includegraphics{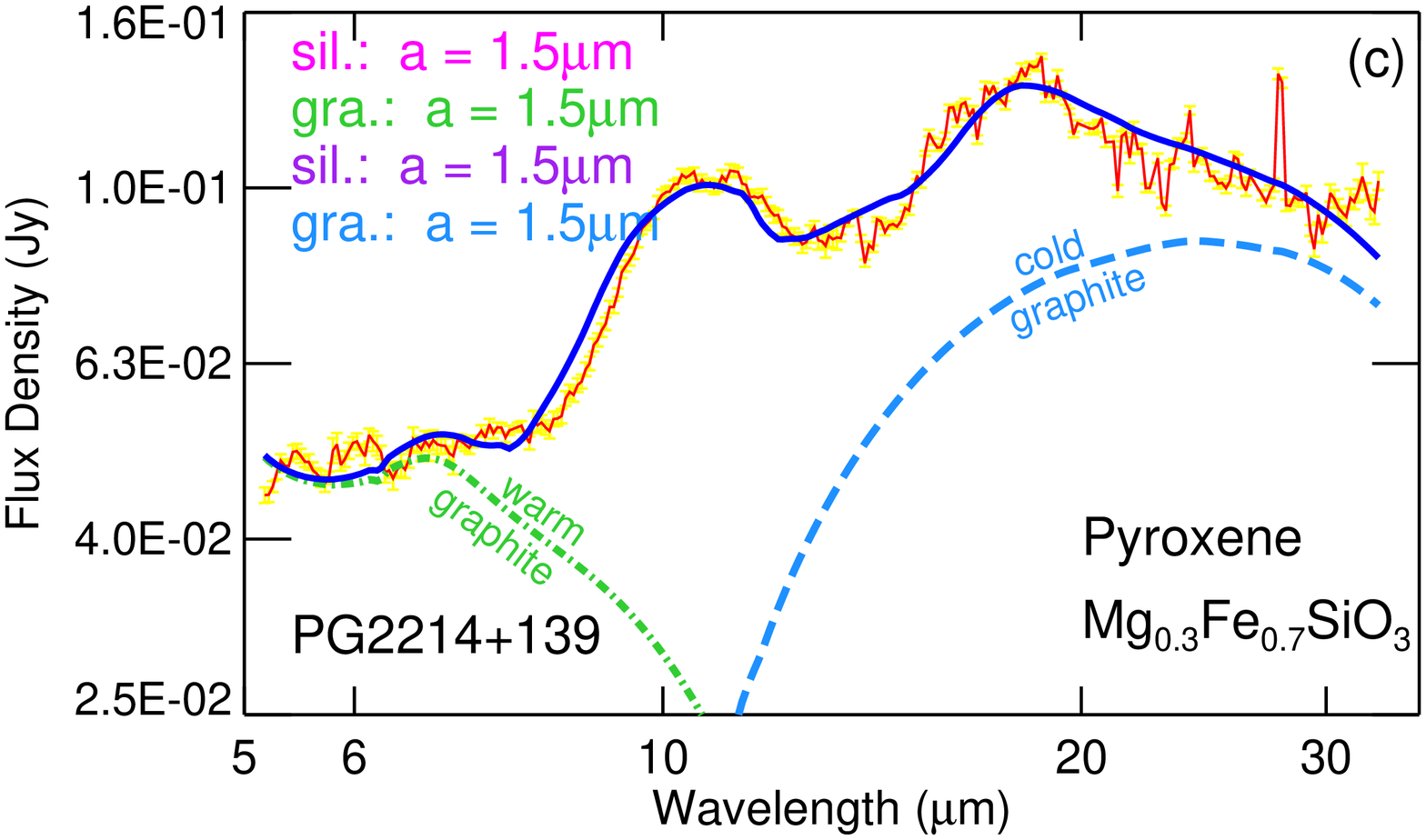}}
\end{array}
$
\caption{\footnotesize
         \label{fig:composition}
         Illustrating the model fits to the {\it Spitzer}/IRS
         spectra of three PG quasars which require different
         silicate compositions:
         ``astronomical silicate'' for PG\,1004+130 (a),
         olivine Mg$_{1.2}$Fe$_{0.8}$SiO$_4$ for PG\,1351+640 (b),
         and pyroxene Mg$_{0.3}$Fe$_{0.7}$SiO$_3$
         for PG\,2214+139 (c).  
         In each sub-figure, we plot 
         the {\it Spitzer}/IRS spectrum (solid red line), 
         the observed $1\sigma$ error (yellow vertical lines),
         the model spectrum (blue solid line) and 
         the four fitting components:
         warm silicate (magenta short dashed line),
         cold silicate (purple dash-dot-dotted line),
         warm graphite (green dash-dotted line),
        and cold graphite (light blue long dashed line). 
        }
\end{center}
\end{figure*}

\begin{figure*}
\begin{center}
\resizebox{0.85\vsize}{!}{\includegraphics{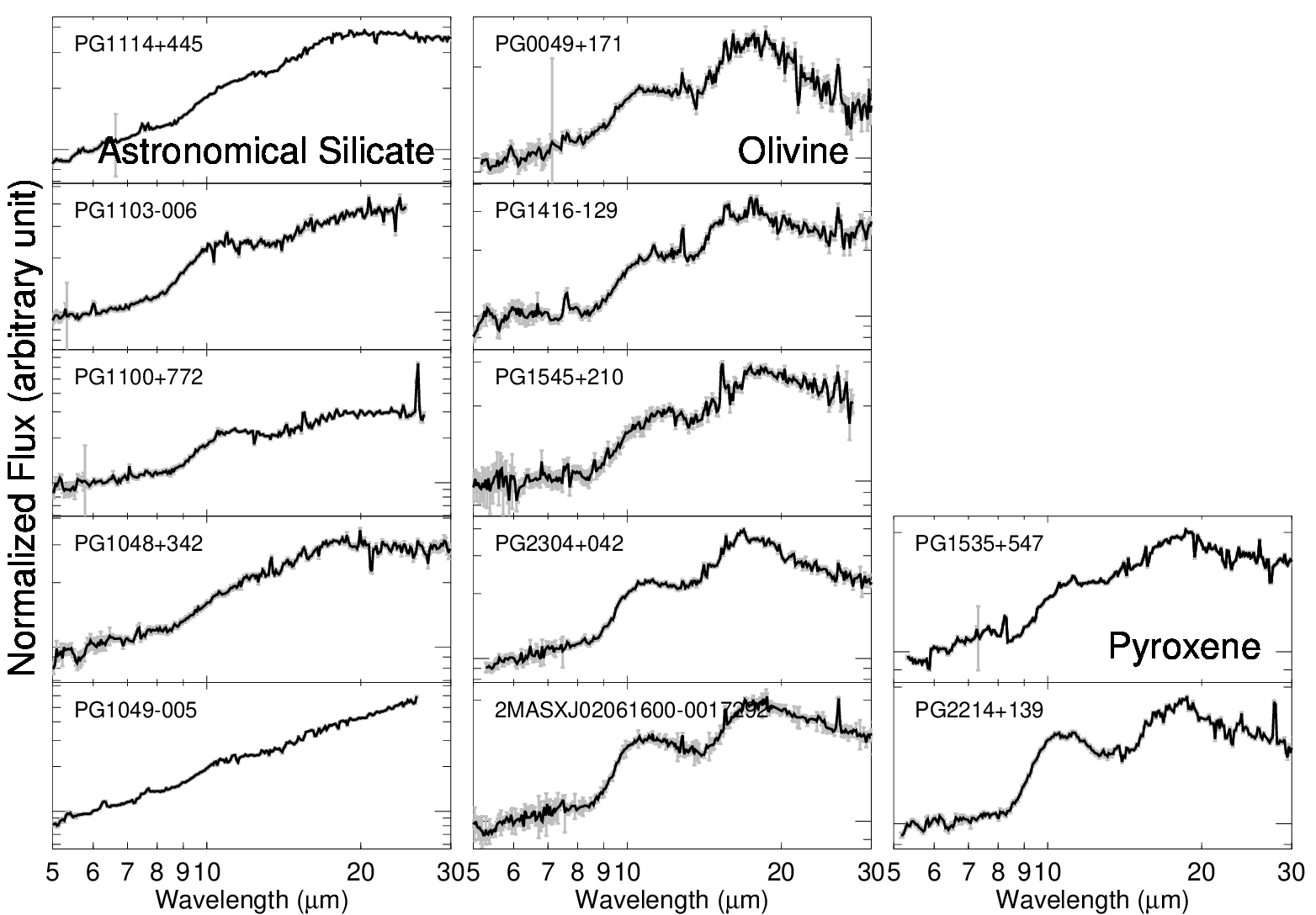} }
\caption{\footnotesize
         \label{fig:composition_spec}
         Illustrations of the {\it Spitzer}/IRS spectra
         of those AGNs for which the best fits favor 
         ``astronomical silicate'' (left),
          amorphous olivine (middle),
          and amorphous pyroxene (right).
          }
\end{center}
\end{figure*}

\begin{figure*}
\begin{center}
\resizebox{0.8\hsize}{!}{
\includegraphics[angle=90]{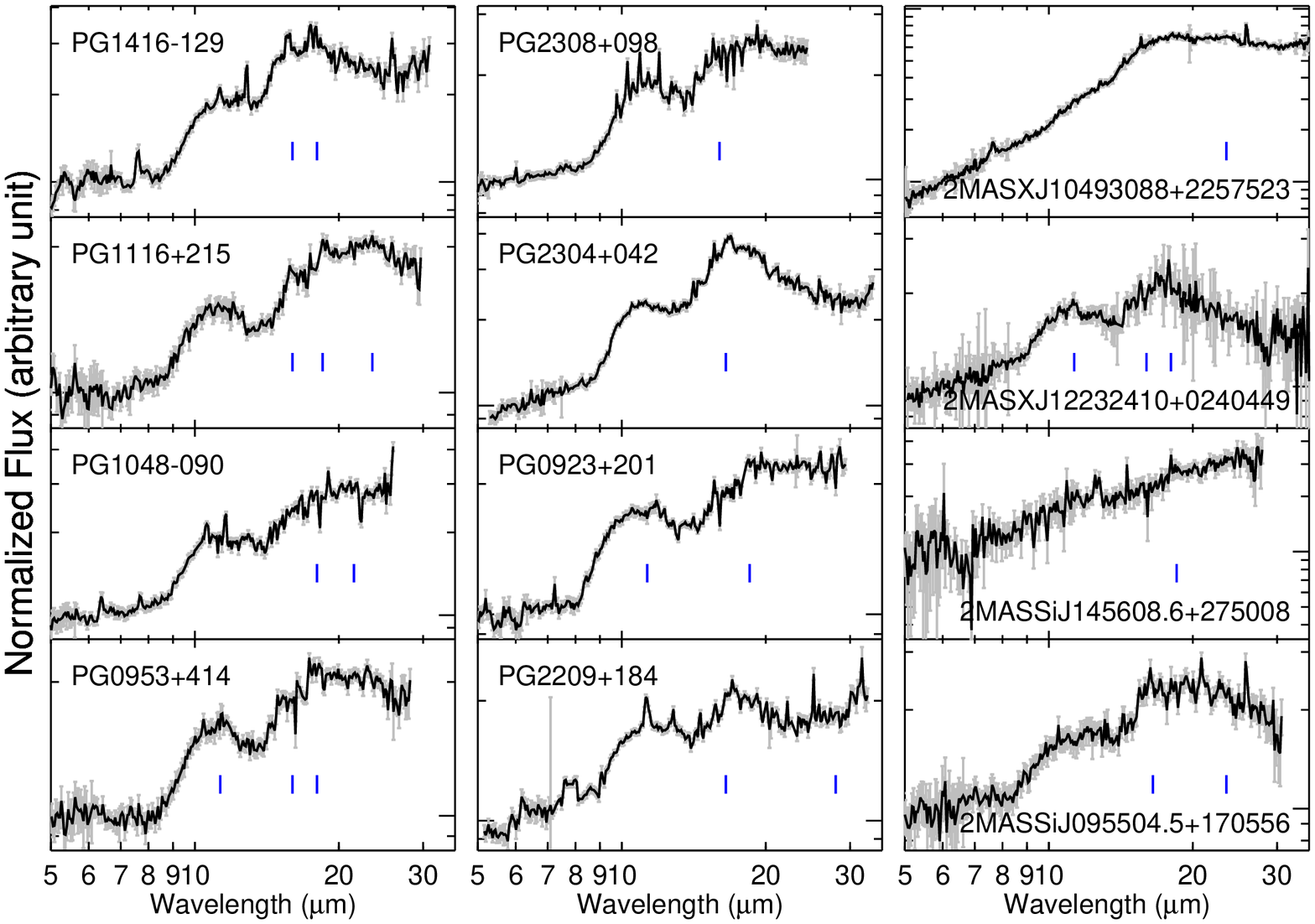}}
\caption{\footnotesize
         \label{fig:crystalline_silicate} 
         {\it Spitzer}/IRS spectra of those sources
         which exhibit crystalline silicate emission features.
         }
\end{center}
\end{figure*}

\begin{figure*}
\begin{center}
\resizebox{0.8\hsize}{!}{
\includegraphics{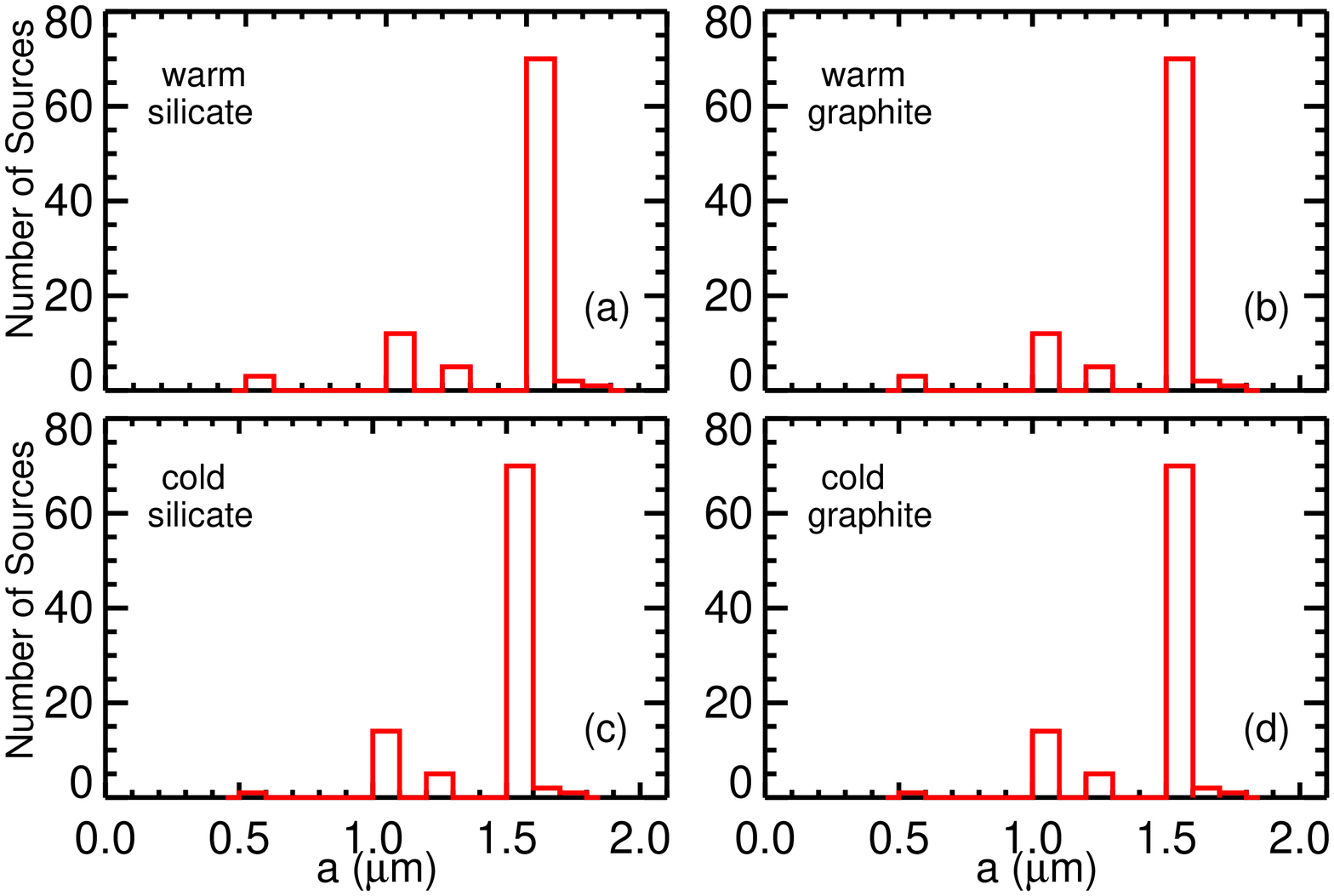}}
\caption{\footnotesize
         \label{fig:size_distribution}
         Histogram of the sizes of
         (a) the warm silicate component, 
         (b) the warm graphite component, 
         (c) the cold silicate component, and
         (d) the cold graphite component
         derived for 93 sources which exhibit
         the 9.7 and 18$\mum$ silicate emission features.
         }
\end{center}
\end{figure*}

\begin{figure*}
\begin{center}
\resizebox{0.8\hsize}{!}{
\includegraphics{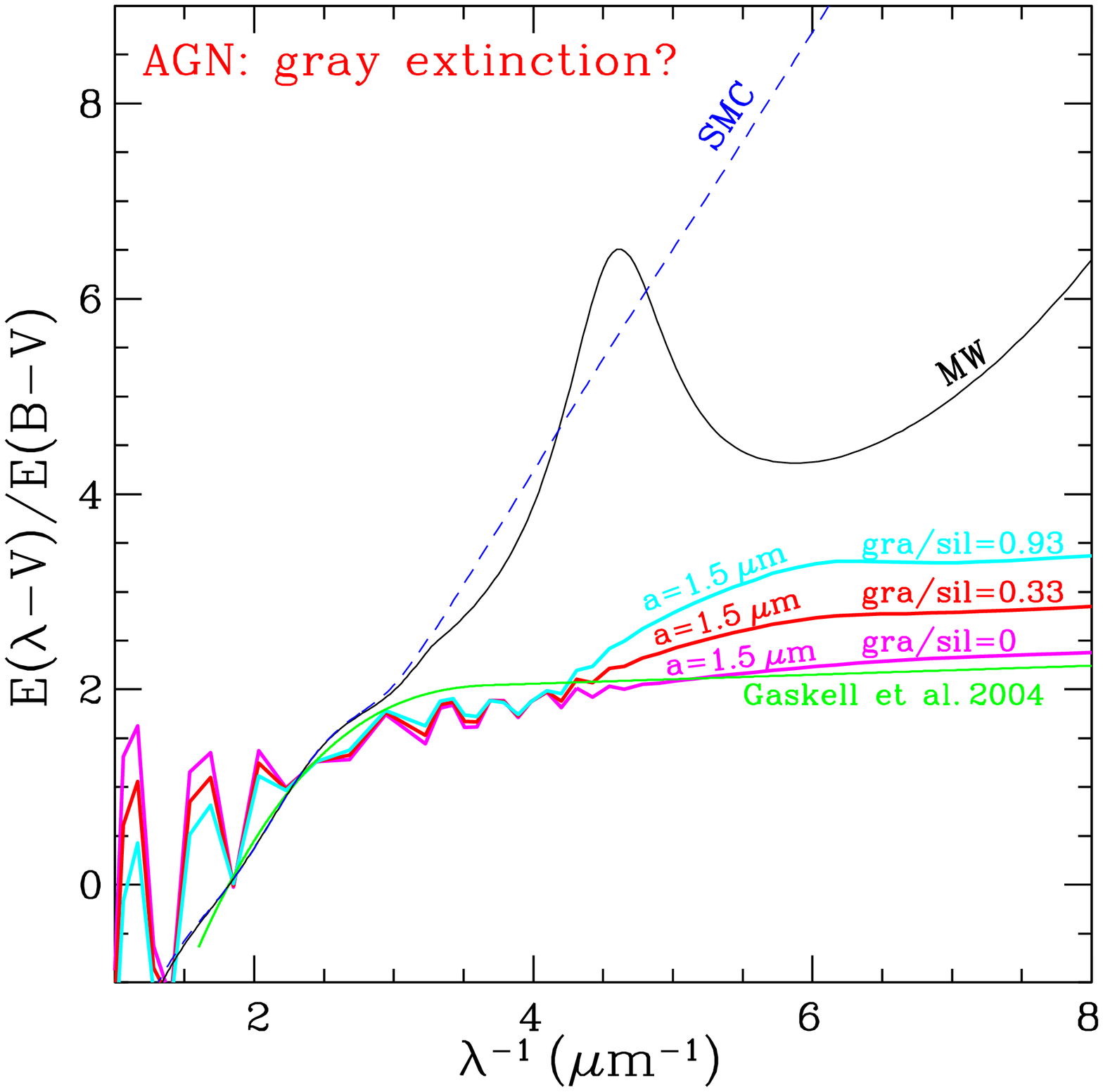}}
\caption{\footnotesize
         \label{fig:extcurv}
         Comparison of the extinction curves
         of the Milky Way (black solid line)
         and the SMC (blue dashed line)
         with that of Gaskell et al.\ (2004)
         derived from composite quasar spectra
         (green solid line) and that calculated 
         from spherical silicate and graphite grains 
         of radii $a=1.5\mum$ with a mass mixing ratio
         of $\Mcarb/\Msil=0$ (magenta solid line),
         $\Mcarb/\Msil=0.33$ (red solid line), and
         $\Mcarb/\Msil=0.93$ (cyan solid line).
         The sawtooth-like structures seen in
         the calculated extinction curves 
         at $\lambda^{-1}<2.5\mum^{-1}$ will be smoothed out
         if a distribution of grain sizes is considered.
         Note that at $\lambda^{-1}>2.5\mum^{-1}$
         the extinction curve of spherical silicate dust
         of radii $a=1.5\mum$ (magenta solid line)
         closely agrees with the AGN extinction curve 
         of Gaskell et al.\ (2004).
         }
\end{center}
\end{figure*}

\begin{figure*}
\begin{center}
\resizebox{0.6\hsize}{!}{\includegraphics{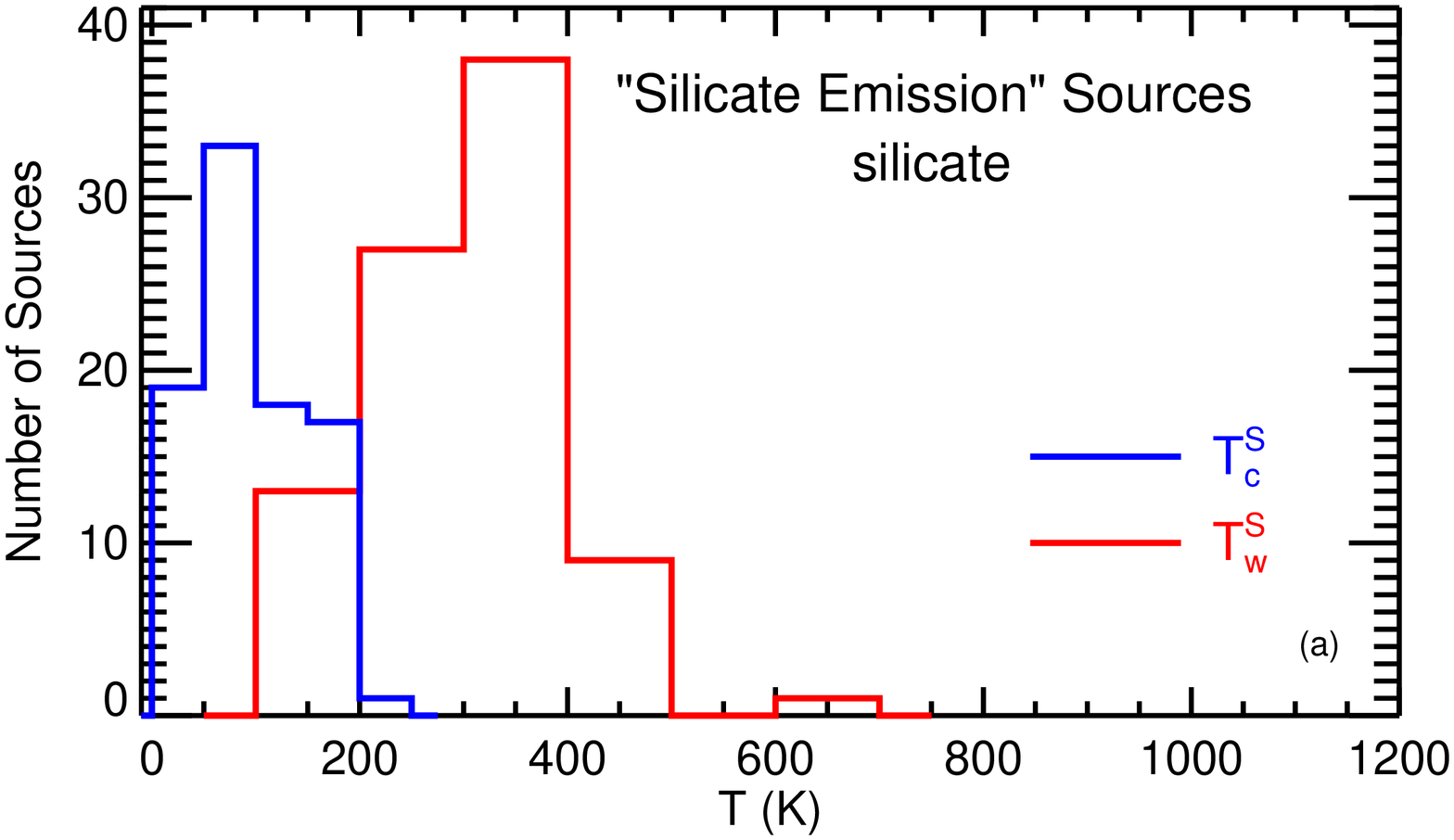}}
\resizebox{0.6\hsize}{!}{\includegraphics{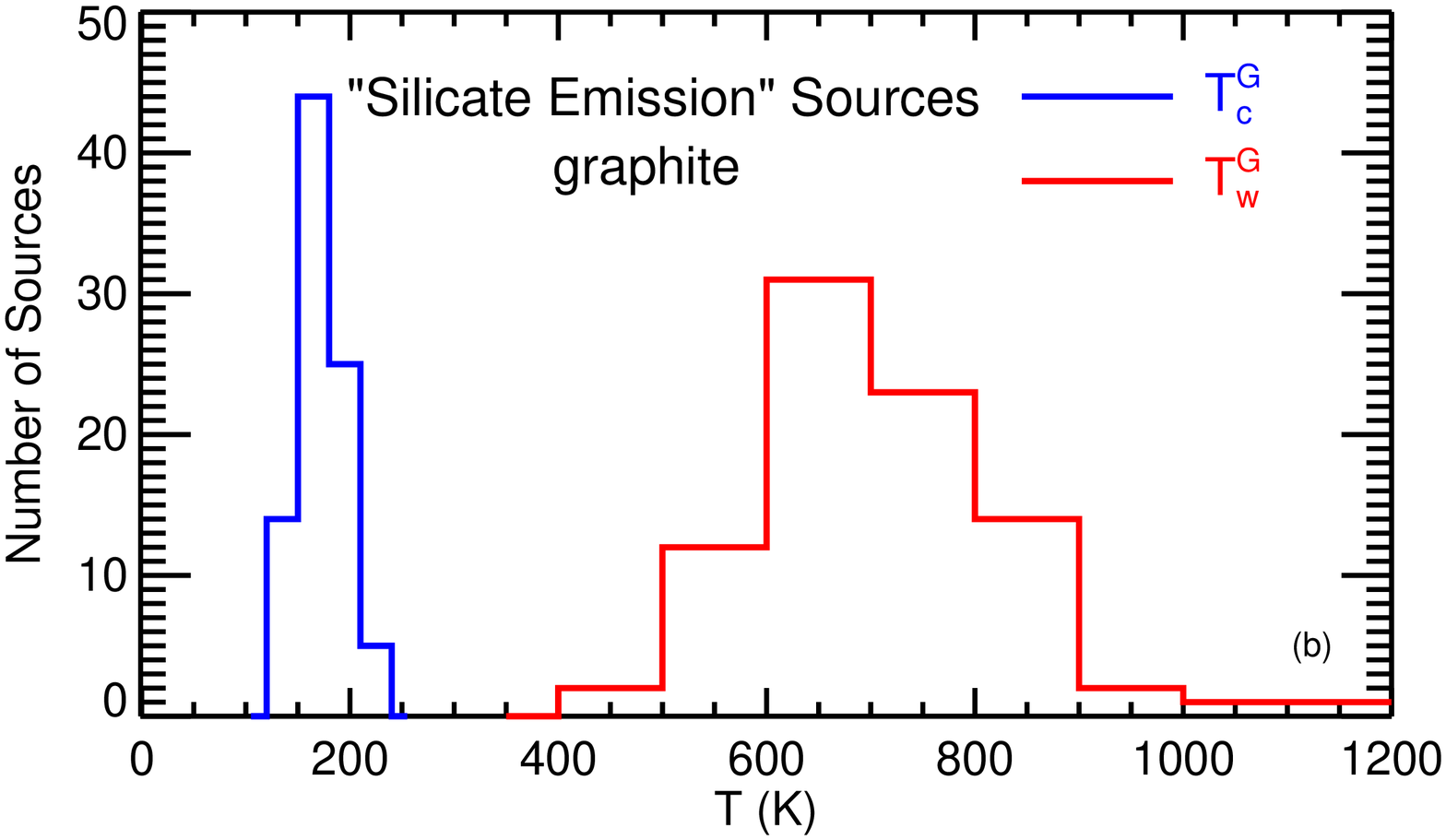}}
\caption{\footnotesize
         \label{fig:temperature_se}
         Histograms of dust temperatures of
         the warm silicate component (red line in upper panel),
         the cold silicate component (blue line in upper panel),
         the warm graphite component (red line in bottom panel),
         and the cold graphite component (blue line in bottom panel)
         derived for 93 sources which exhibit
         the 9.7 and 18$\mum$ silicate emission features.
         }
\end{center}
\end{figure*}

\begin{figure*}
\begin{center}
\resizebox{\hsize}{!}{\includegraphics{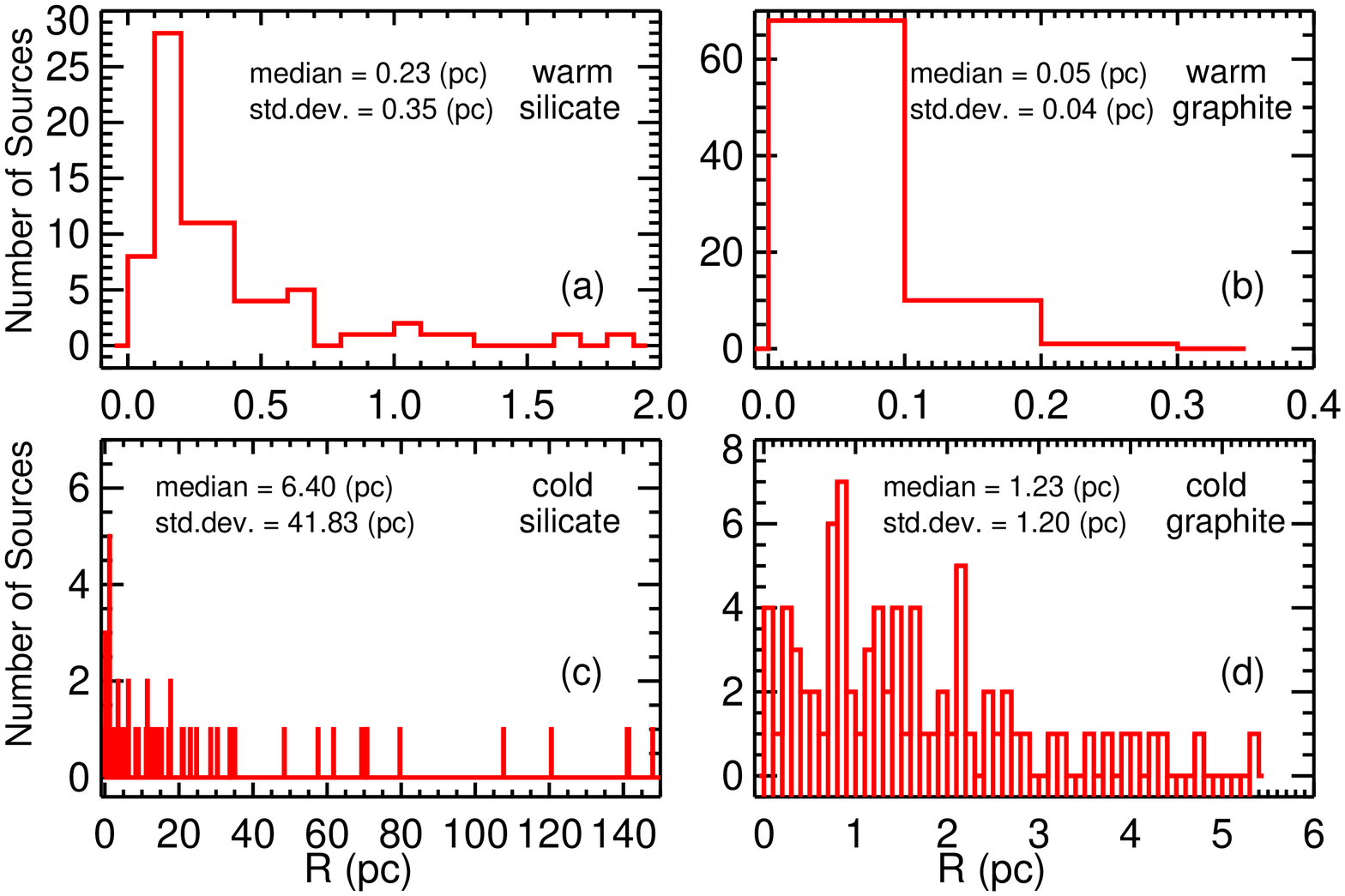}}
\caption{\footnotesize
         \label{fig:radius}
         Histograms of the distances of the dust
         from the central engine for those 
         93 ``silicate emission'' sources:
         warm silicate (a), warm graphite (b),
         cold silicate (c), and cold graphite (d).
         }
\end{center}
\end{figure*}
\begin{figure*}
\begin{center}
\resizebox{0.6\hsize}{!}{\includegraphics{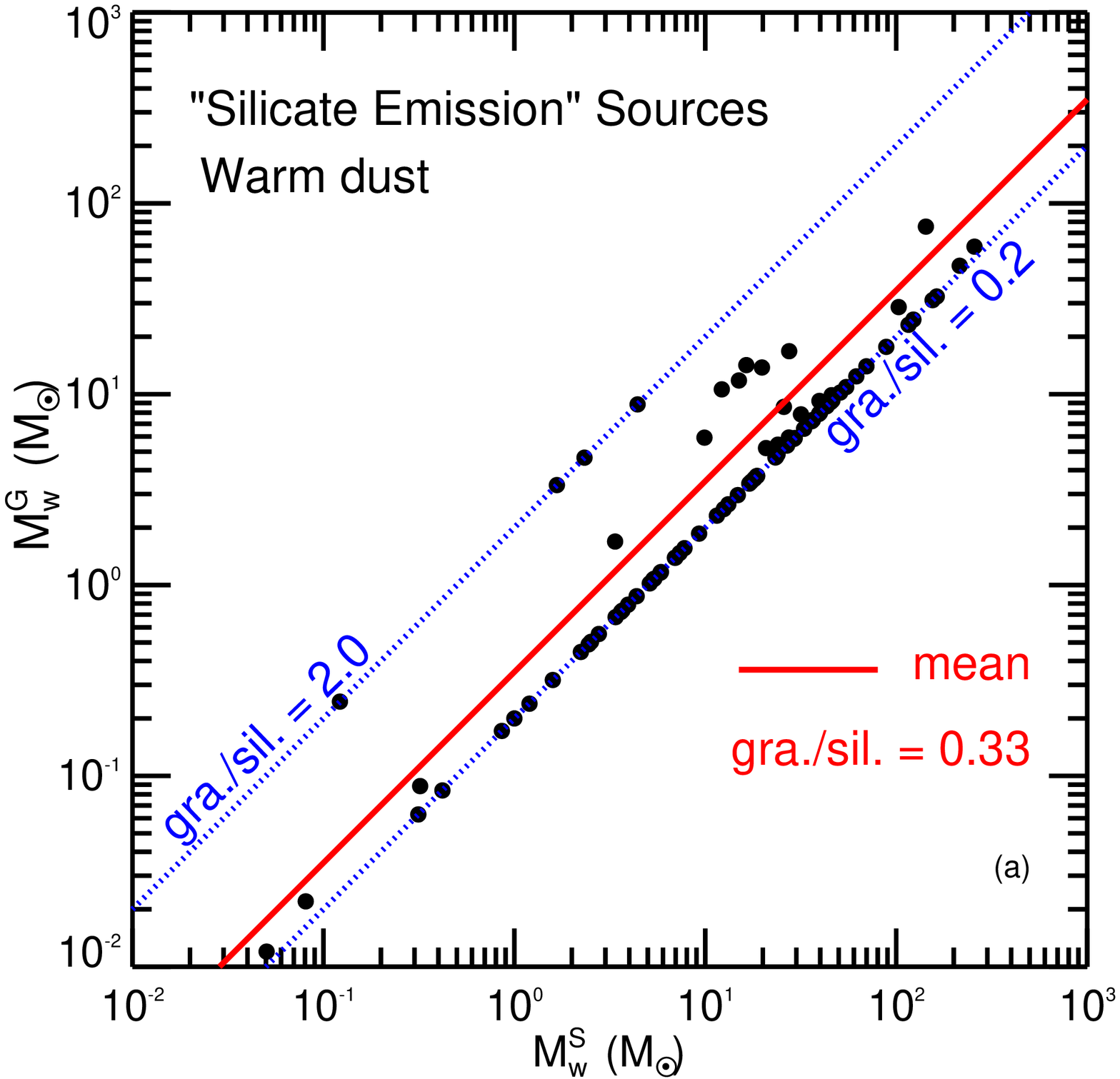}}
\resizebox{0.6\hsize}{!}{\includegraphics{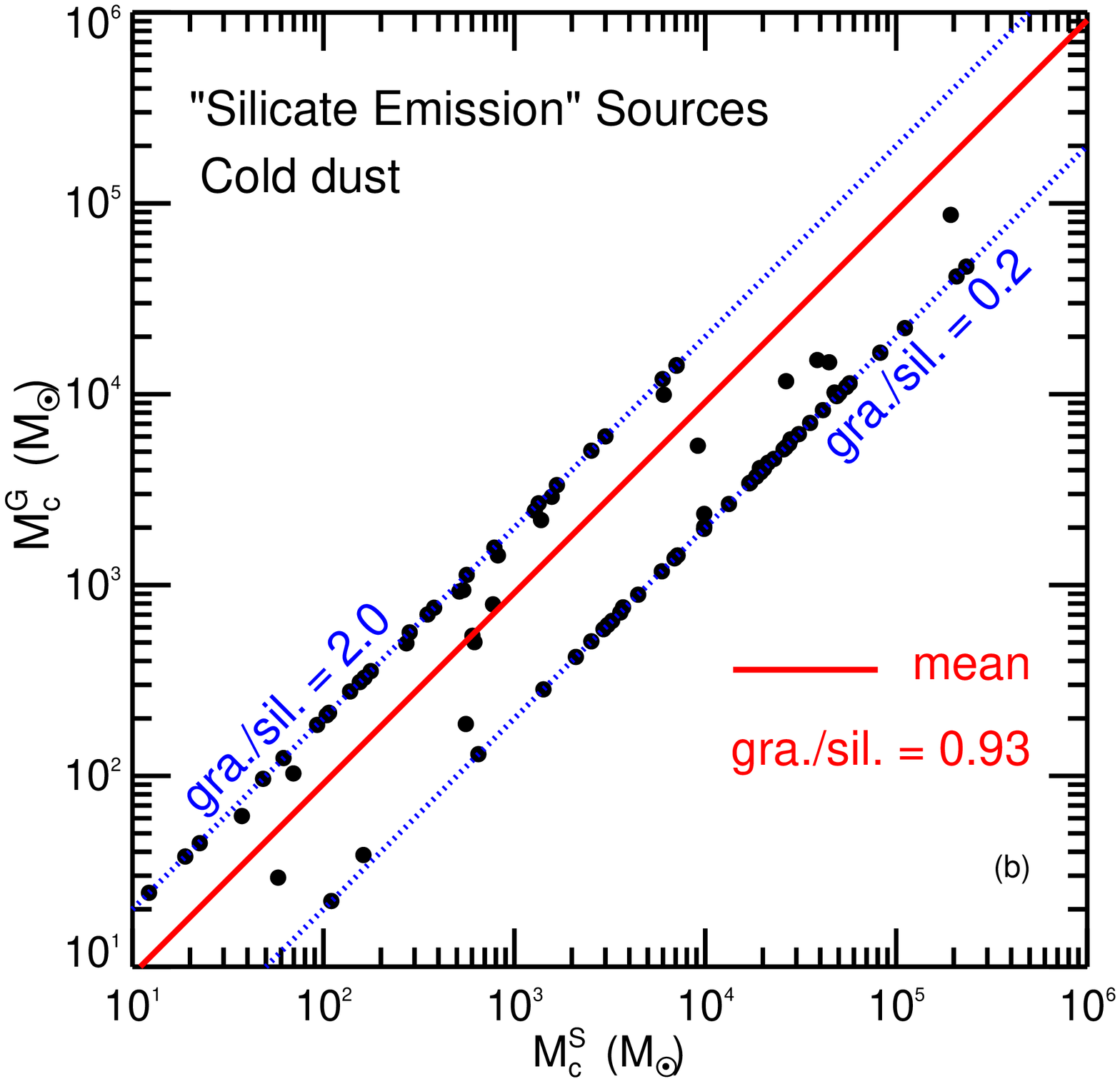}}
\caption{\footnotesize
         \label{fig:mass_se}
         Dust mass ratios of warm graphite to warm silicate 
         ($\Mwcarb/\Mwsil$; upper panel)
         and of cold graphite to cold silicate 
         ($\Mccarb/\Mcsil$; bottom panel)
         derived for those 93 
         ``silicate emission'' sources.
         The blue dotted lines show 
         the upper ($\Mccarb/\Mcsil=2.0$) and 
         lower ($\Mccarb/\Mcsil=0.2$) boundaries.
         The red solid lines show the mean mass ratios 
         derived from our model. 
         }
\end{center}
\end{figure*}

\clearpage
\begin{figure*}
\begin{center}
\resizebox{1.0\hsize}{!}{\includegraphics{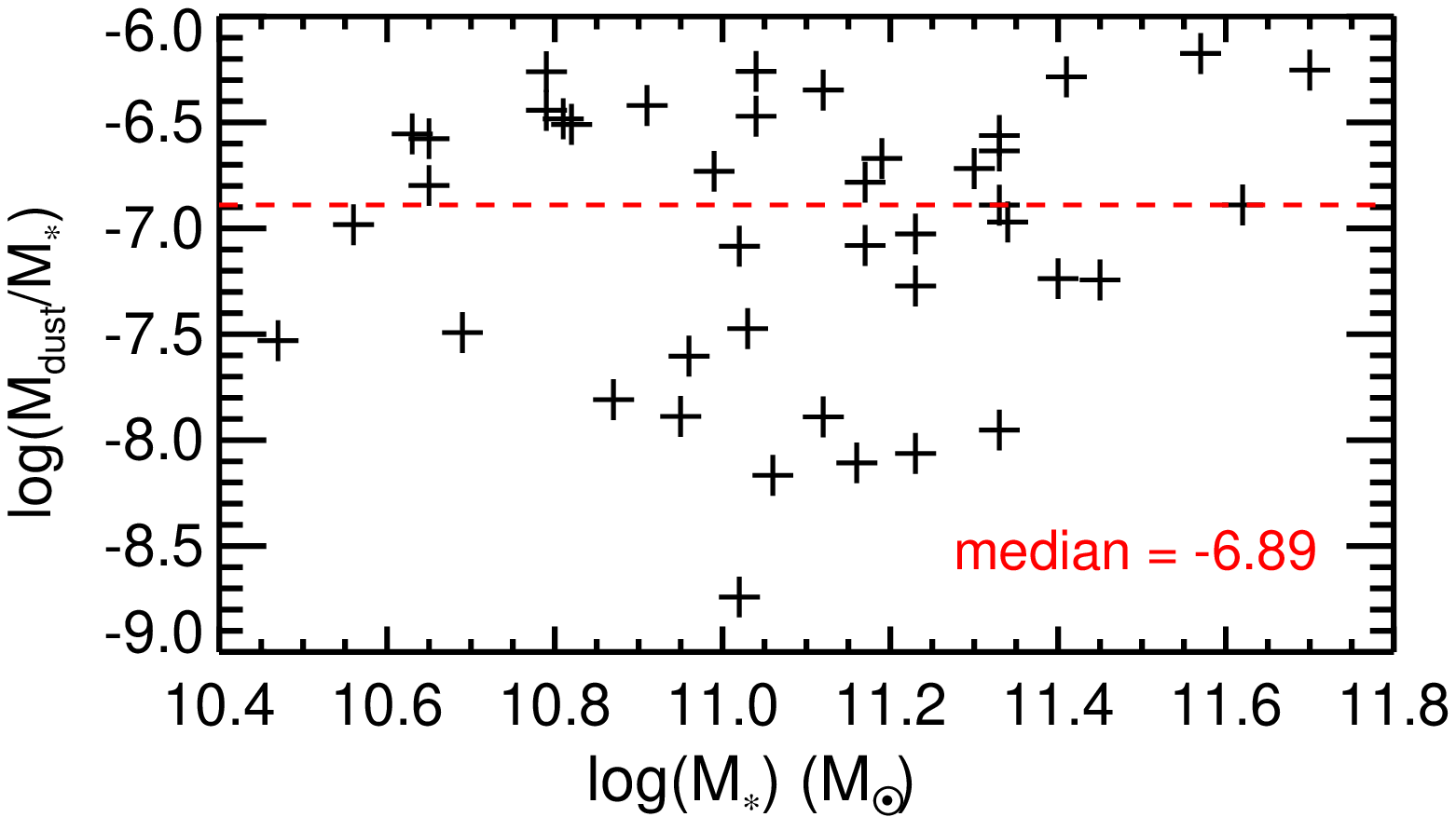}}
\caption{\footnotesize
         \label{fig:mass_dust_star}
         Comparison of the stellar mass ($M_{\star}$) 
         of the host galaxies of 39 PG quasars 
         and two {\it 2MASS} quasars
         with the total dust mass ($M_{\rm dust}$)
         obtained by summing over all four dust components
         (i.e., $M_{\rm dust} =  \Mwsil + \Mcsil + \Mwcarb +\Mccarb$).
         The horizontal dashed line plots 
         the median ratio of 
         $\langle M_{\rm dust}/M_{\star}\rangle\approx1.3\times10^{-7}$.
         }
\end{center}
\end{figure*}

\begin{figure*}
\begin{center}
\resizebox{\hsize}{!}
{\includegraphics{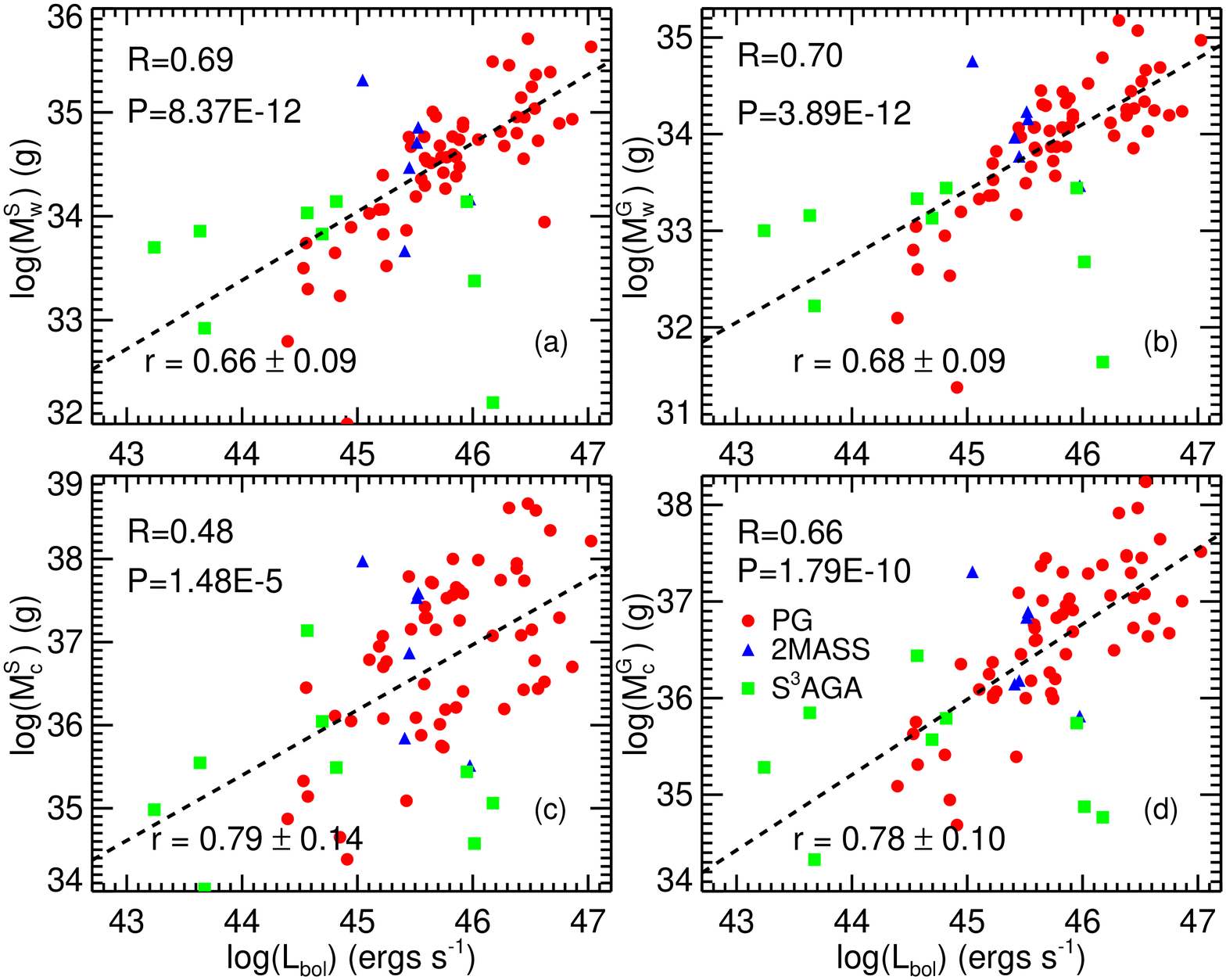}}
\caption{\footnotesize
              \label{fig:M_Lbol}
             Correlation of the bolometric luminosity 
             $L_{\rm bol}$ [approximated by 
              $7\times\lambda L_\lambda(5100\Angstrom$)]
             with the mass of 
             the warm silicate component ($\Mwsil$; a), 
             the warm graphite component ($\Mwcarb$; b),
             the cold silicate component ($\Mcsil$; c), and
             the cold graphite component ($\Mccarb$; d). 
             In each panel, the PG quasars are shown as red circles,
             the {\it 2MASS} quasars are shown as blue triangles, and
             the \saga\ AGNs are shown as green squares.
             Also labeled in each panel are the Pearson correlation
             coefficient ($R$) and the probability ($P$) of no
             correlation as well as the slope of the linear relation
             (dashed lines).
             }
\end{center}
\end{figure*}

\begin{figure*}
\begin{center}
\resizebox{\hsize}{!}
{\includegraphics{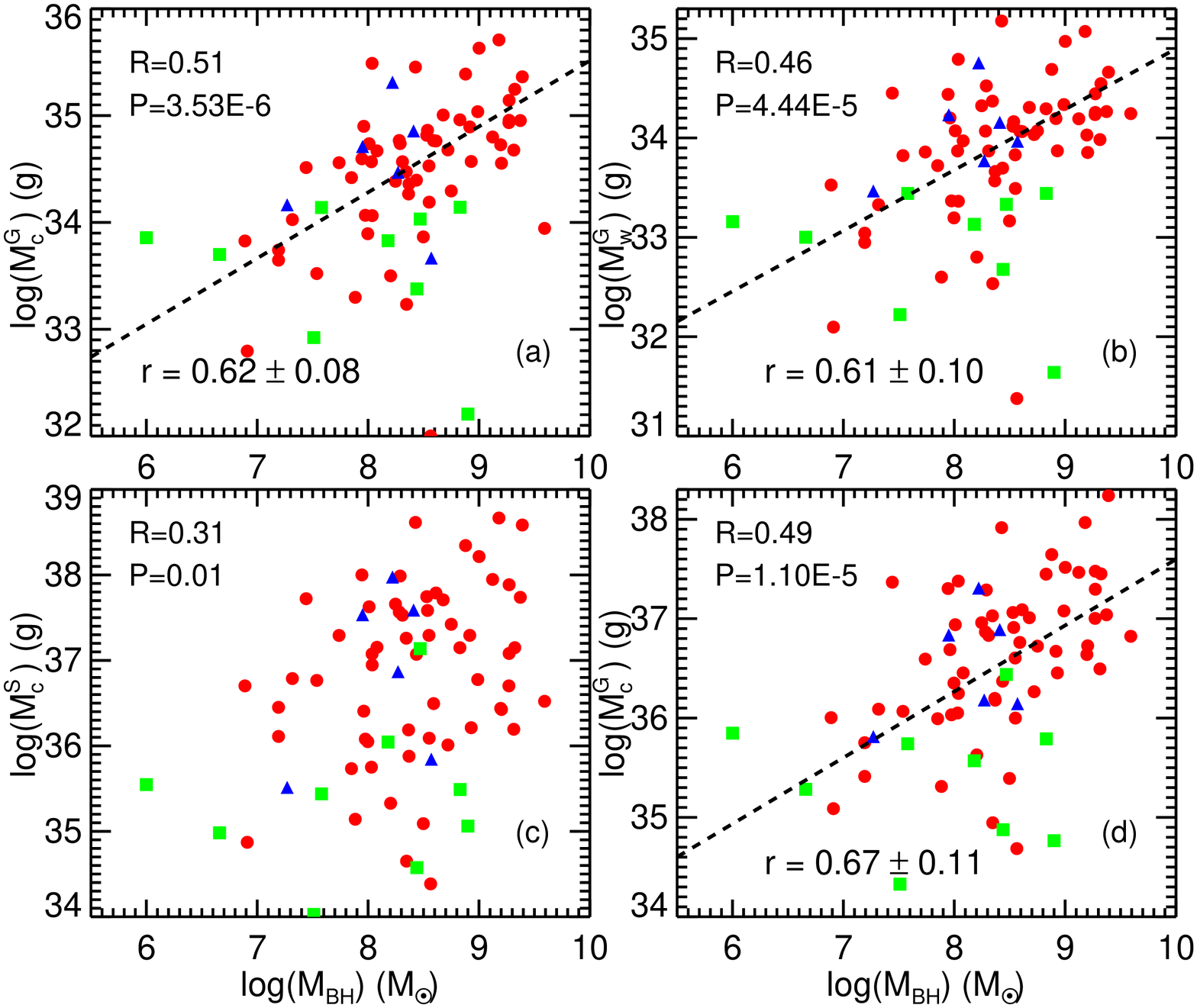}}
\caption{\footnotesize
             \label{fig:M_Mbh}
             Correlation of the black hole mass ($M_{\rm BH}$)
             with the mass of 
             warm silicate ($\Mwsil$; a), 
             warm graphite ($\Mwcarb$; b),
             cold silicate ($\Mcsil$; c), and
             cold graphite ($\Mccarb$; d). 
             In each panel, the PG quasars are shown as red circles,
             the {\it 2MASS} quasars are shown as blue triangles, and
             the \saga\ AGNs are shown as green squares.
             The Pearson correlation coefficient ($R$) 
             and the probability ($P$) of no correlation 
             as well as the slope of the linear relation
             are also labeled in each panel. 
            }
\end{center}
\end{figure*}

\begin{figure*}
\begin{center}
\resizebox{\hsize}{!}
{\includegraphics{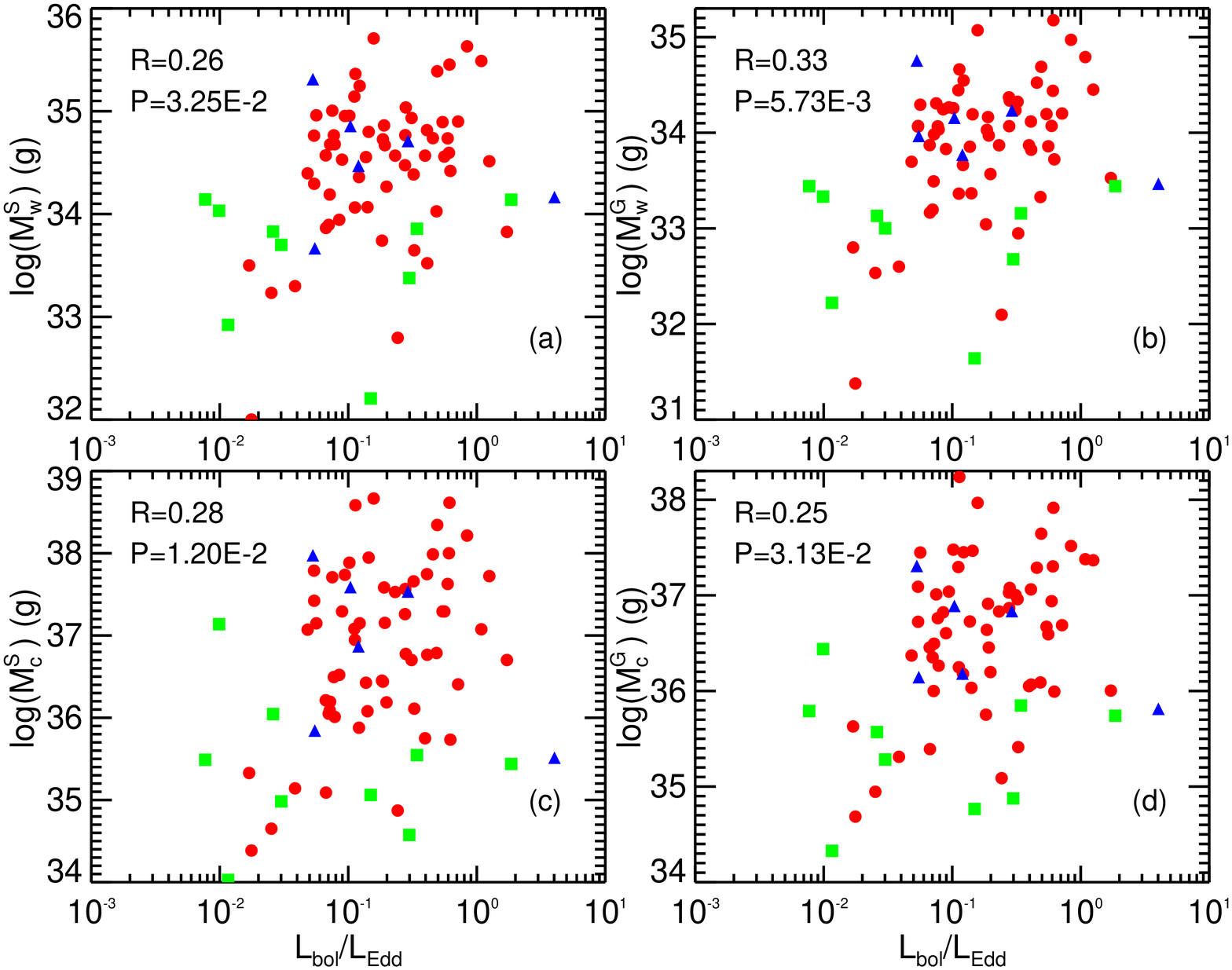}}
\caption{\footnotesize
              \label{fig:M_Lbol2Ledd}
             Correlation of the Eddington ratio 
             (represented by $L_{\rm bol}/L_{\rm Edd}$)
             with the mass of 
             warm silicate ($\Mwsil$; a), 
             warm graphite ($\Mwcarb$; b),
             cold silicate ($\Mcsil$; c), and
             cold graphite ($\Mccarb$; d). 
            In each panel, the PG quasars are shown as red circles,
            the {\it 2MASS} quasars are shown as blue triangles, and
            the \saga\ AGNs are shown as green squares.
            }
\end{center}
\end{figure*}

\begin{figure*}
\begin{center}
\resizebox{\hsize}{!}
{\includegraphics{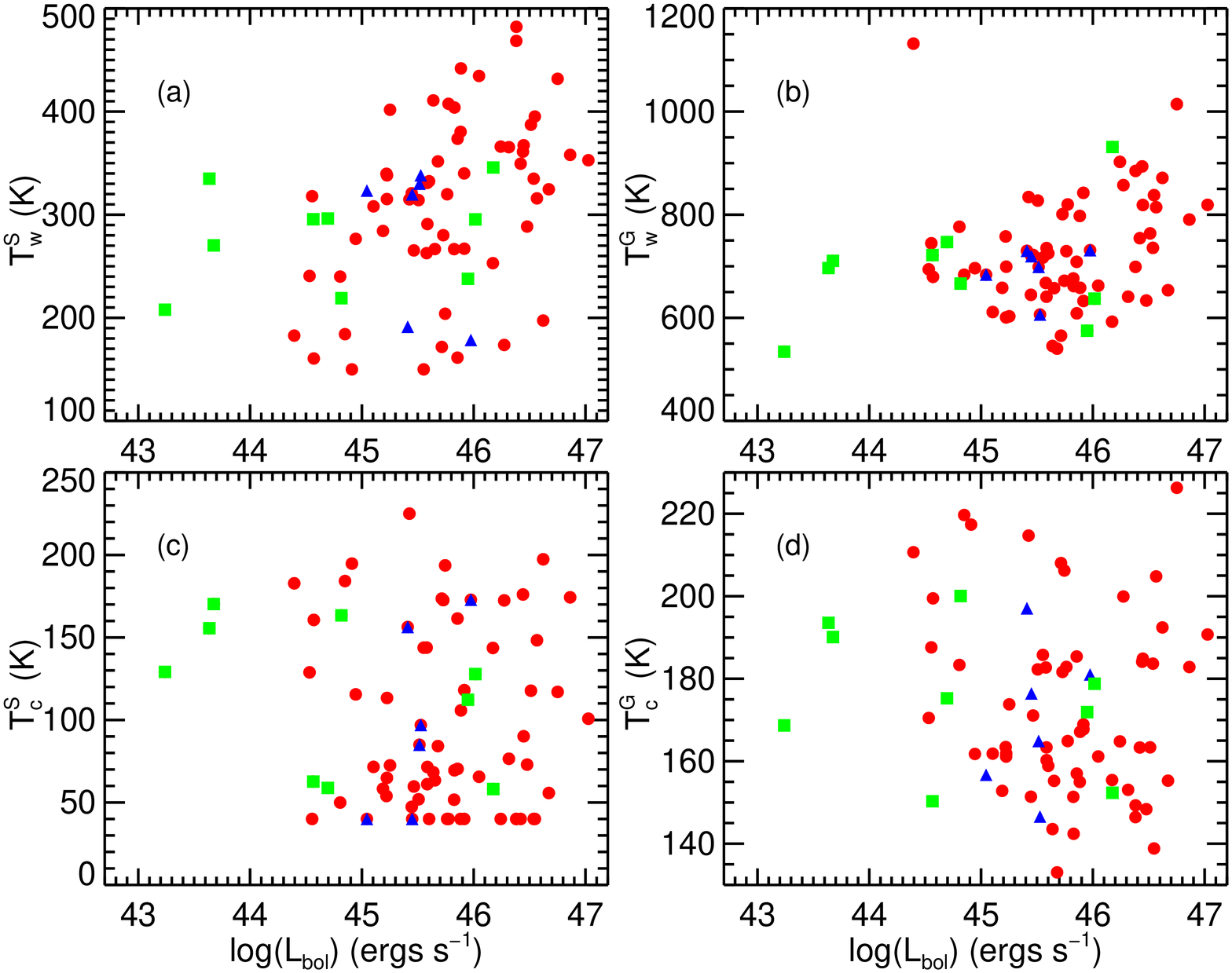}}
\caption{\footnotesize
              \label{fig:T_Lbol}
             Correlation of the bolometric luminosity 
             with the temperature of 
             the warm silicate component ($\Twsil$; a), 
             the warm graphite component ($\Twcarb$; b),
             the cold silicate component ($\Tcsil$; c), and
             the cold graphite component ($\Tccarb$; d). 
            In each panel, the PG quasars are shown as red circles,
            the {\it 2MASS} quasars are shown as blue triangles, and
            the \saga\ AGNs are shown as green squares.
}
\end{center}
\end{figure*}

\begin{figure*}
\begin{center}
\resizebox{\hsize}{!}
{\includegraphics{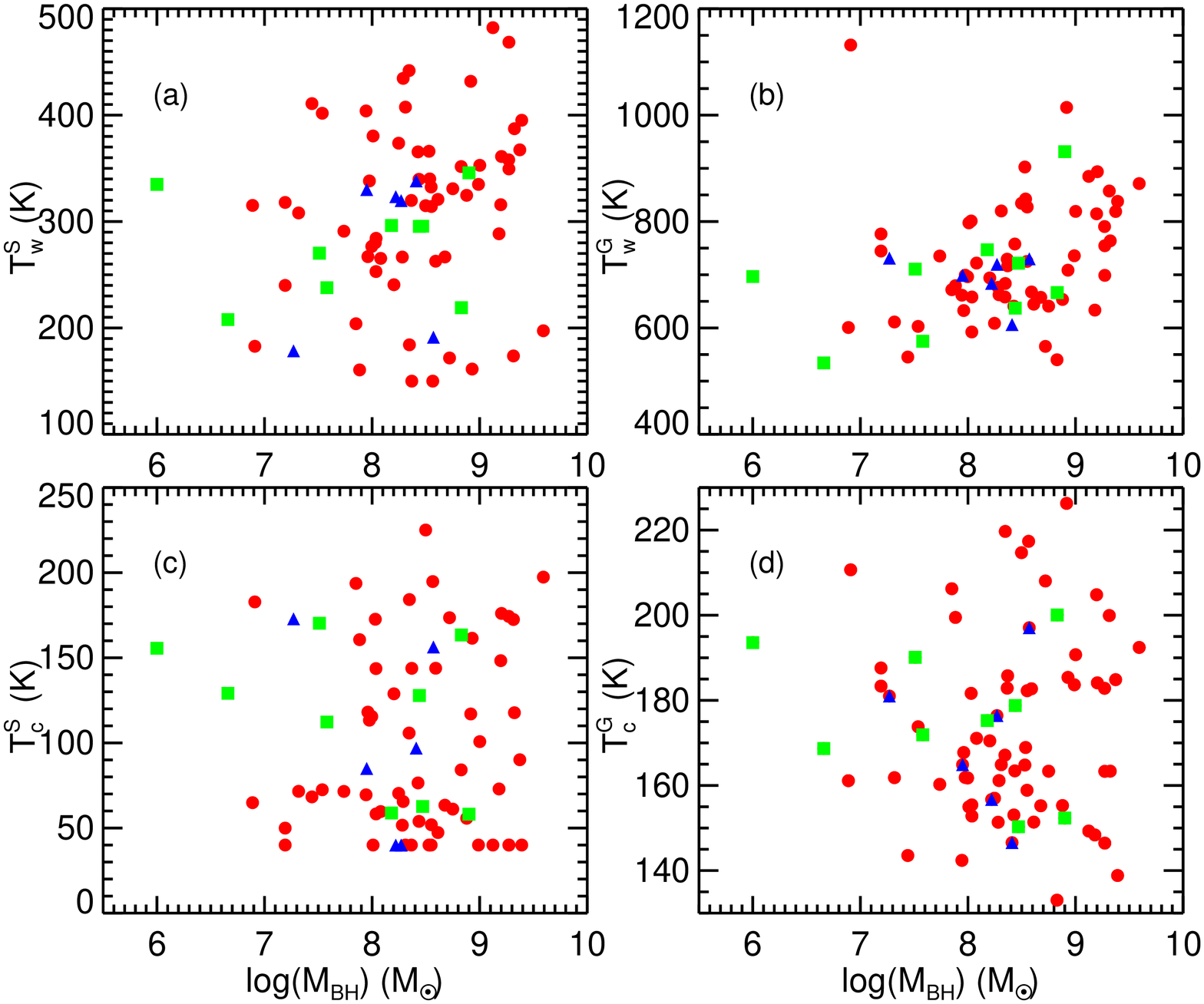}}
\caption{\footnotesize
             \label{fig:T_Mbh}
            Correlation of the black hole mass ($M_{\rm BH}$)
             with the temperature of 
             warm silicate ($\Twsil$; a), 
             warm graphite ($\Twcarb$; b),
             cold silicate ($\Tcsil$; c), and
             cold graphite ($\Tccarb$; d). 
            In each panel, the PG quasars are shown as red circles,
            the {\it 2MASS} quasars are shown as blue triangles, and
            the \saga\ AGNs are shown as green squares.
            }
\end{center}
\end{figure*}

\begin{figure*}
\begin{center}
\resizebox{\hsize}{!}
{\includegraphics{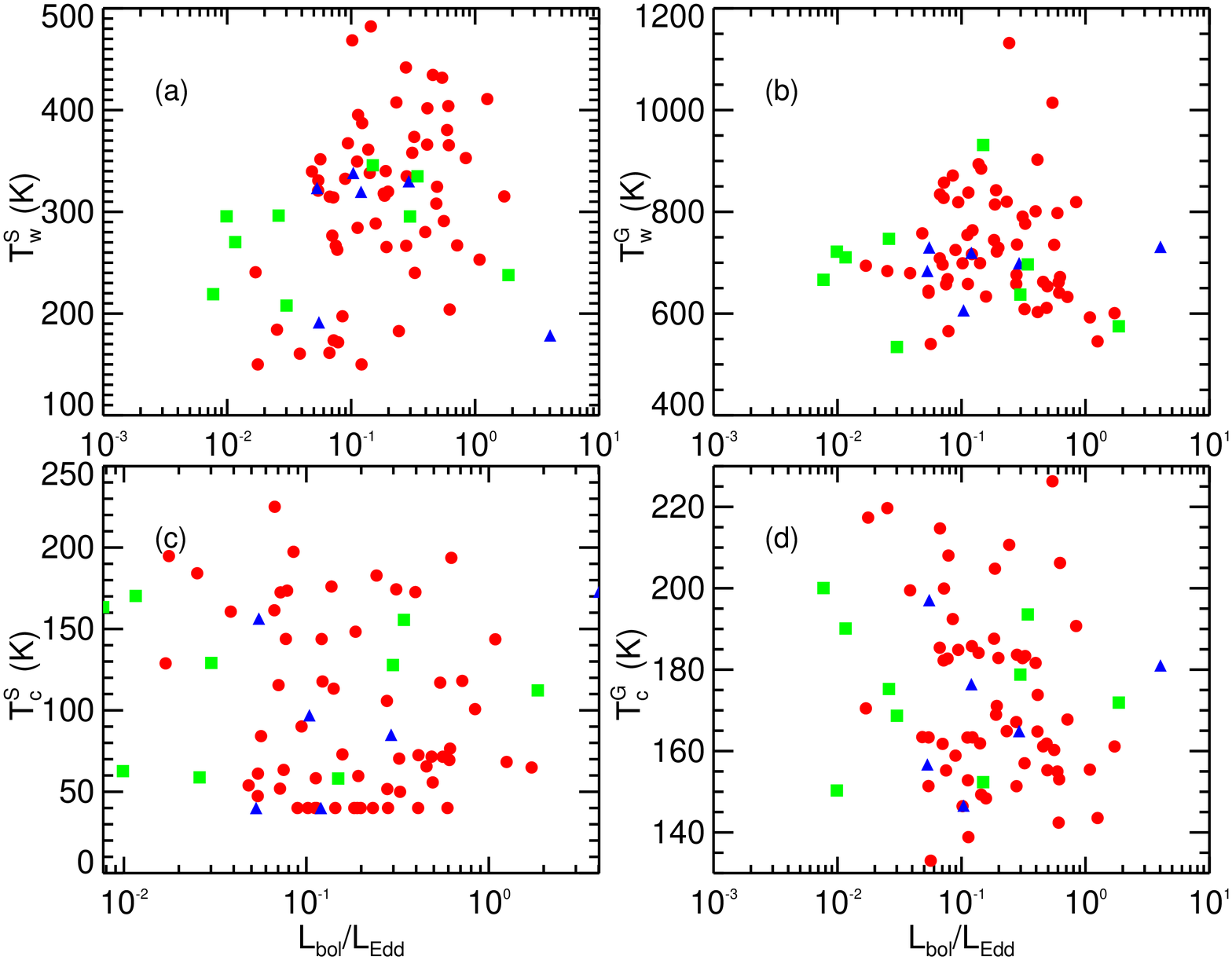}}
\caption{\footnotesize
         \label{fig:T_Lbol2Ledd}
         Correlation of the Eddington ratio 
         $L_{\rm bol}/L_{\rm Edd}$
         with the temperature of 
         warm silicate ($\Twsil$; a), 
         warm graphite ($\Twcarb$; b),
         cold silicate ($\Tcsil$; c), and
         cold graphite ($\Tccarb$; d). 
         In each panel, the PG quasars are shown as red circles,
         the {\it 2MASS} quasars are shown as blue triangles, and
         the \saga\ AGNs are shown as green squares.     
         }
\end{center}
\end{figure*}

\begin{figure*}
\begin{center}
\resizebox{0.6\hsize}{!}{\includegraphics{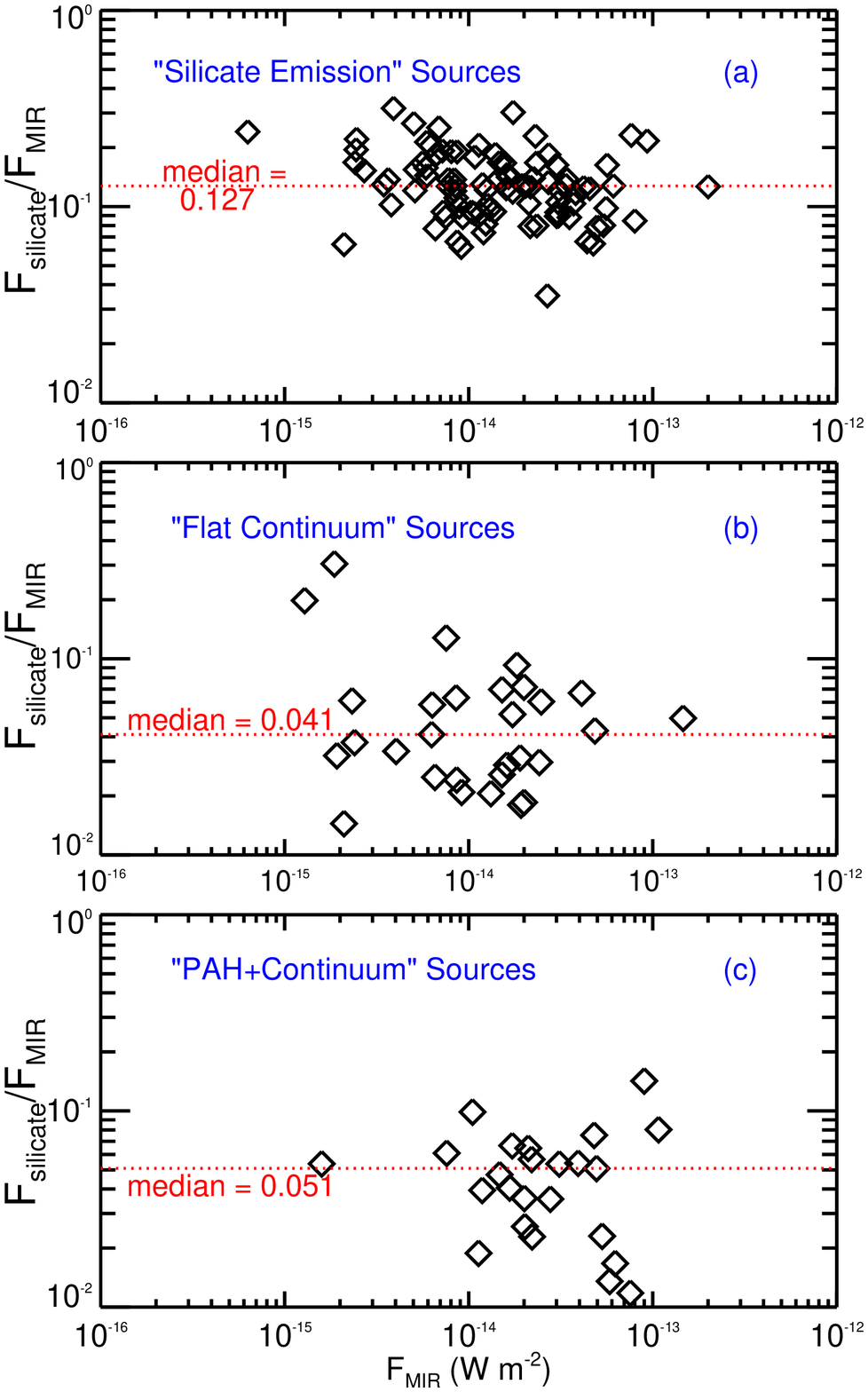}}
\caption{\footnotesize
         \label{fig:se_frac}
         Fractional fluxes emitted in 
         the 9.7 and 18$\mum$ silicate 
         features ($F_{\rm silicate}$)
         relative to the total mid-IR emission
         at $\simali$5--38$\mum$ ($F_{\rm MIR}$)
        for the ``silicate emission'' sources (a),
        the ``flat continuum'' sources (b), 
        and the ``PAH\,+\,continuum'' sources (c).
        For the latter two classes of sources,
        $F_{\rm silicate}$ is actually an upper limit. 
        The dashed horizontal line shows
        the mean fraction 
       $\langle F_{\rm silicate}/F_{\rm MIR}\rangle$. 
        }
\end{center}
\end{figure*}

\begin{figure*}[ht]
\leavevmode
\begin{center}
\resizebox{0.8\hsize}{!}{
\includegraphics[angle=90]{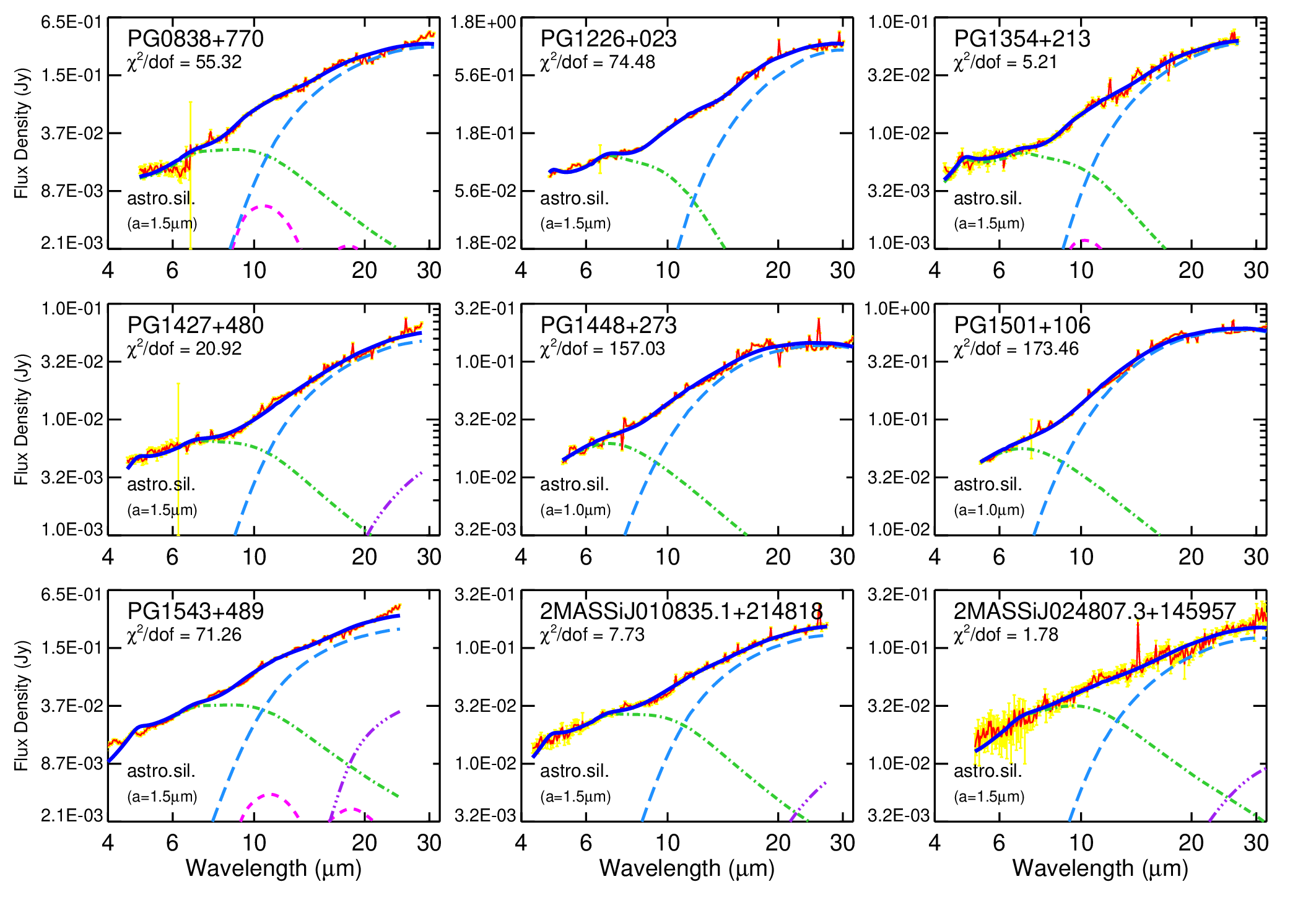}}
\caption{\footnotesize
         \label{fig:nosil_em_mod1}
          Comparison of the {\it Spitzer}/IRS spectra 
          (red solid lines) 
          of the PG quasars PG0838+770, PG1226+023, 
          PG1354+213, PG1427+480, PG1448+273, PG1501+106, 
          PG1543+489 and the {\it 2MASS} quasars 
          2MASSiJ010835.1+214818 and 2MASSiJ024807.3+145957
          which show a featureless thermal continuum
          with the model spectra (blue solid lines)
          which are the sum of warm silicate
          (magenta short dashed lines),
          cold silicate (purple dash-dot-dotted lines),
          warm graphite (green dash-dotted lines),
          and cold graphite (light blue long dashed lines).
          Also shown are the observed $1\sigma$ errors 
          (yellow vertical lines).
          }
\end{center}
\end{figure*}

\begin{figure*}[ht]
\leavevmode
\figurenum{\ref{fig:nosil_em_mod1}}
\begin{center}
\resizebox{0.8\hsize}{!}{
\includegraphics[angle=90]{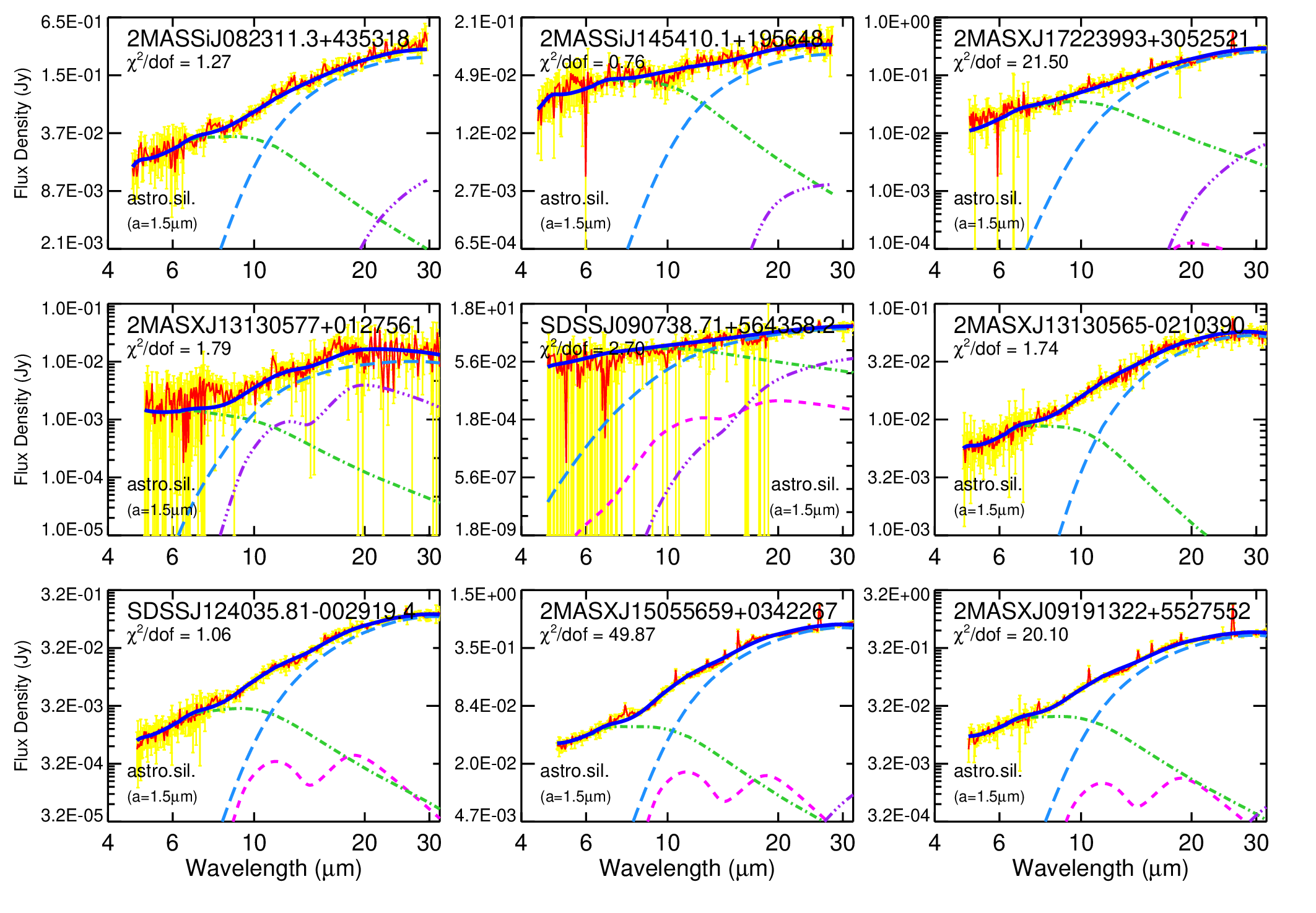}}
\caption{\footnotesize
         Continued,  
         but for the {\it 2MASS} quasars 
         2MASSiJ082311.3+435318, and 2MASSiJ145410.1+195648 
         and the \saga\ AGNs 
         2MASXJ17223993+3052521, 2MASXJ13130577+0127561, 
         SDSSJ090738.71+564358.2, 2MASXJ13130565-0210390, 
         SDSSJ124035.81-002919.4, 2MASXJ15055659+0342267, 
         and 2MASXJ09191322+5527552.  
         }
\end{center}
\end{figure*}

\begin{figure*}[ht]
\leavevmode
\figurenum{\ref{fig:nosil_em_mod1}}
\begin{center}
\resizebox{0.8\hsize}{!}{
\includegraphics[angle=90]{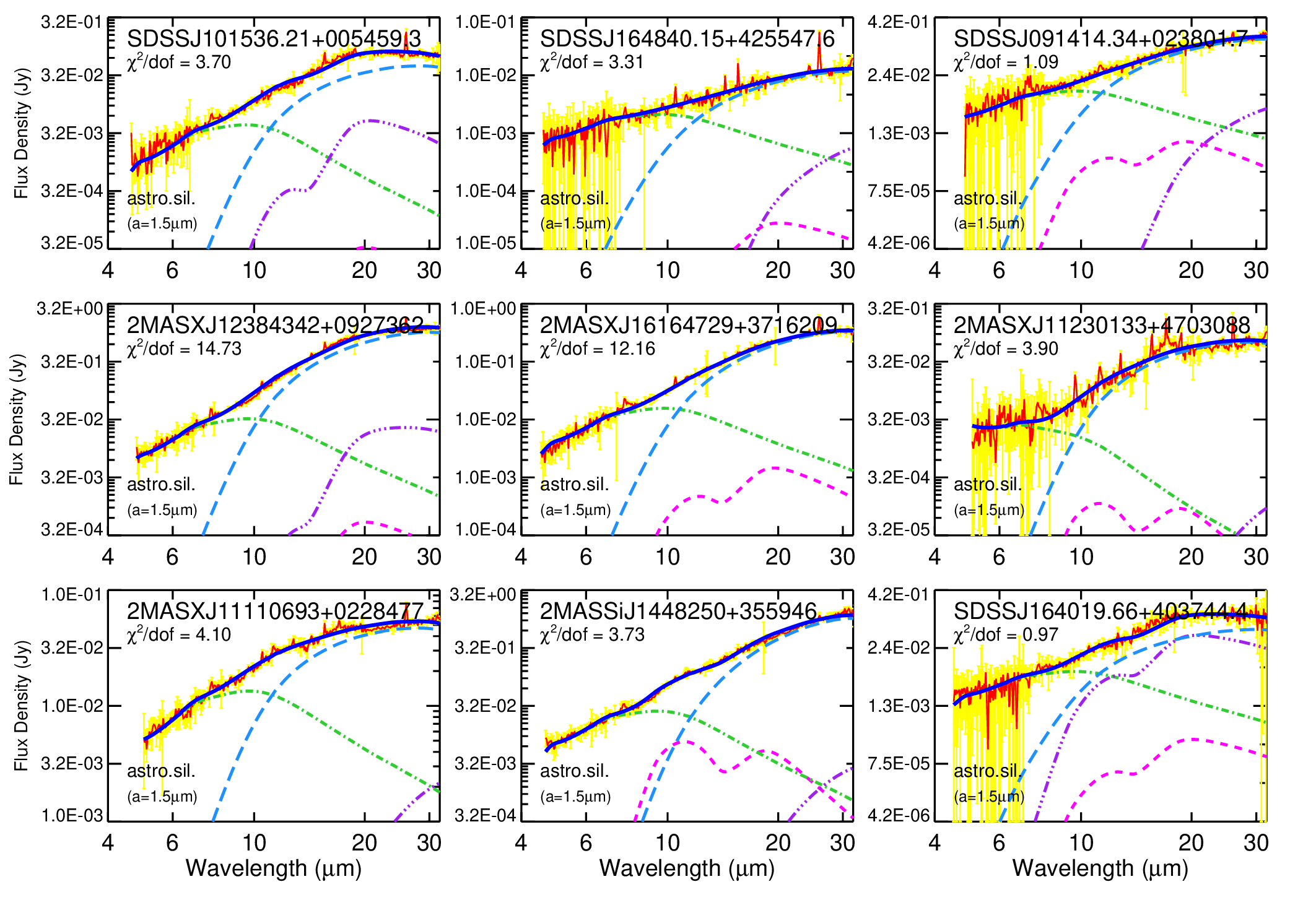}}
\caption{\footnotesize
         Continued,  
         but for the \saga\ AGNs, 
         SDSSJ101536.21+005459.3, SDSSJ164840.15+425547.6, 
         SDSSJ091414.34+023801.7, 2MASXJ12384342+0927362, 
         2MASXJ16164729+3716209, 2MASXJ11230133+4703088, 
         2MASXJ11110693+0228477, 2MASSiJ1448250+355946, 
         and SDSSJ164019.66+403744.4.
         }
\end{center}
\end{figure*}

\begin{figure*}[ht]
\leavevmode
\figurenum{\ref{fig:nosil_em_mod1}}
\begin{center}
\resizebox{0.8\hsize}{!}{
\includegraphics[angle=90]{f19d.eps}}
\caption{\footnotesize
         Continued,  
         but for the \saga\ AGNs 
         SDSSJ104058.79+581703.3, UGC05984 and UGC06527.         
         }
\end{center}
\end{figure*}

\begin{figure*}
\begin{center}
\resizebox{0.6\hsize}{!}{\includegraphics{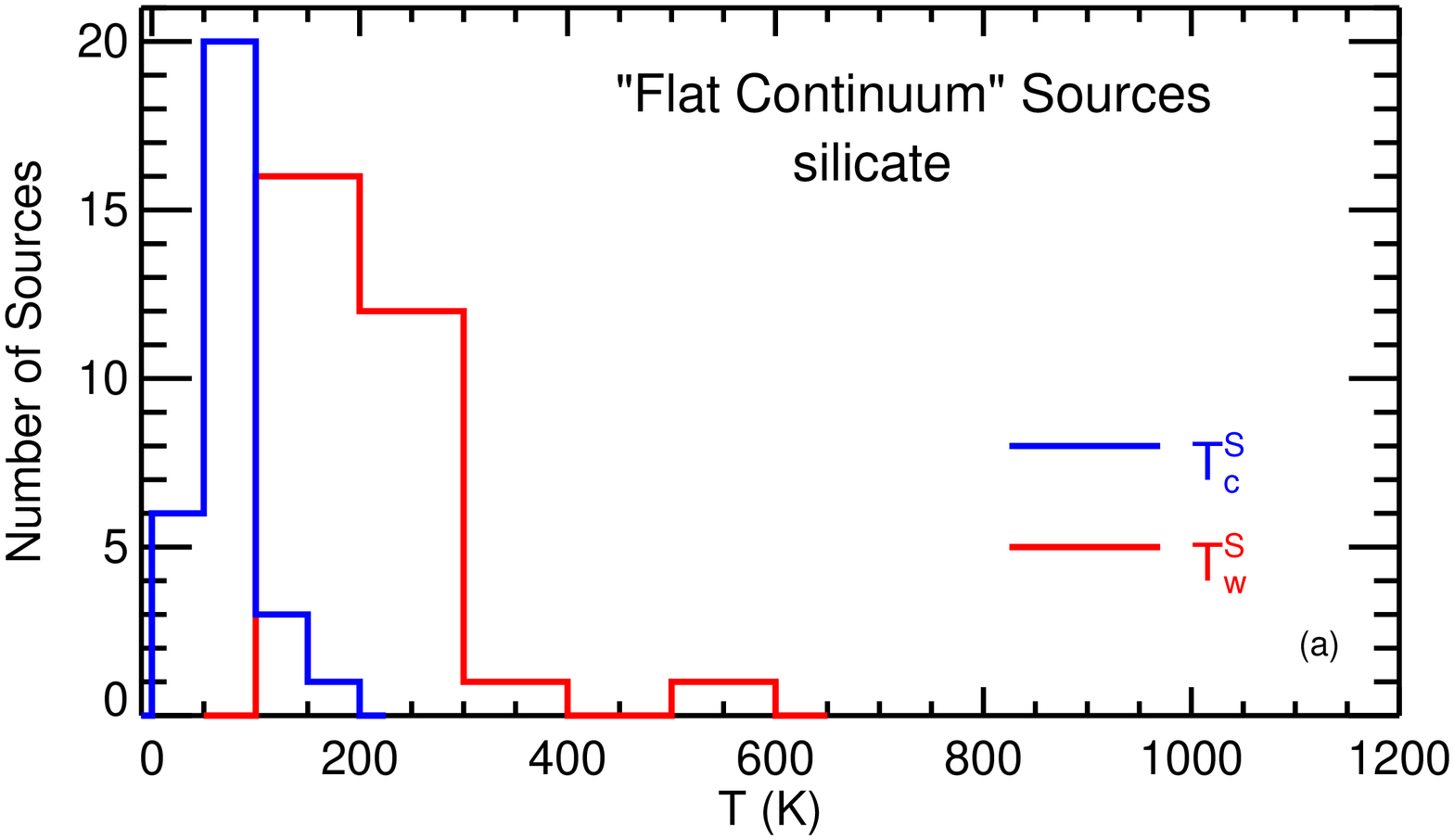}}
\resizebox{0.6\hsize}{!}{\includegraphics{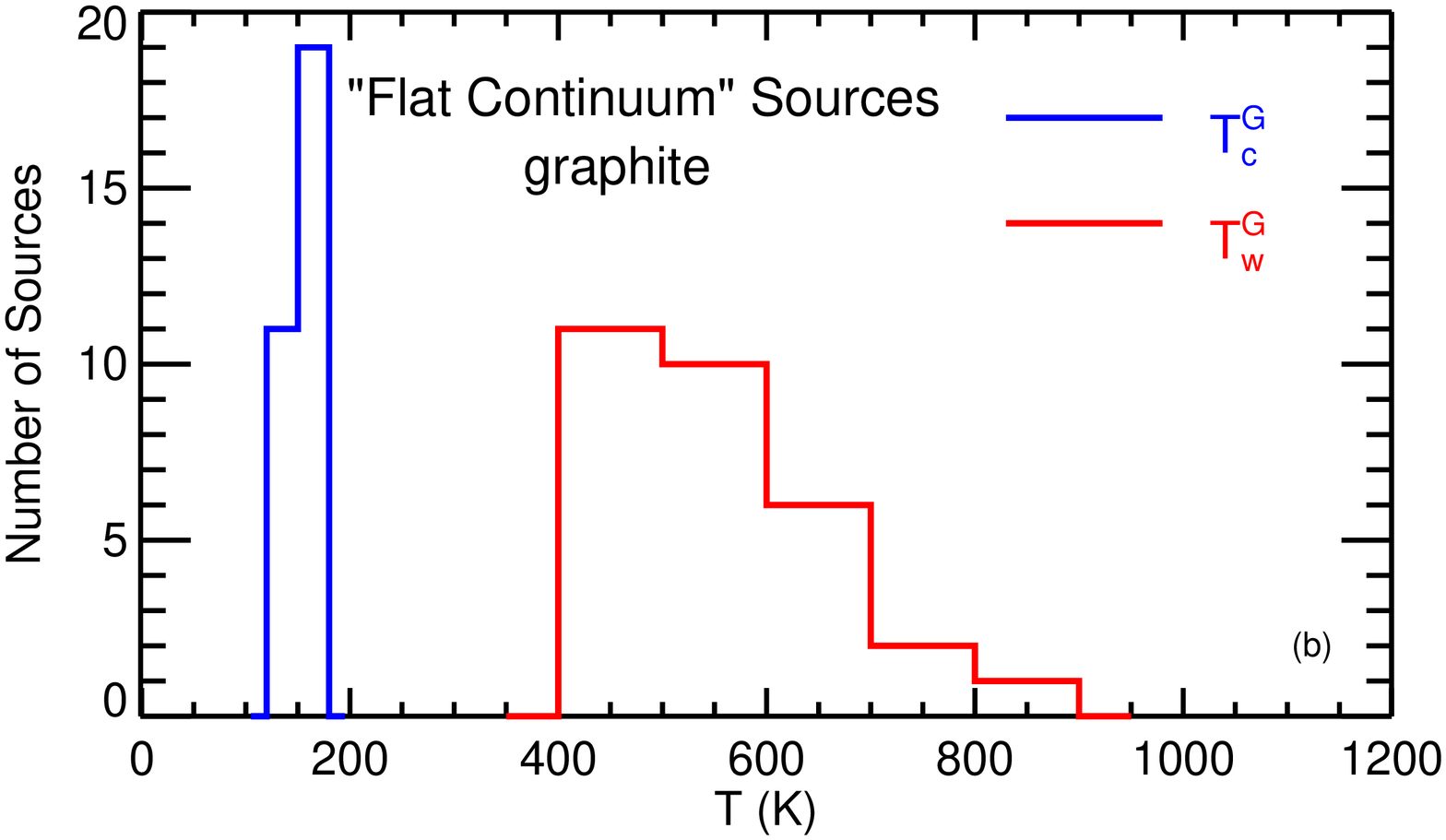}}
\caption{\footnotesize
         \label{fig:temperature_fl}
         Same as Figure~\ref{fig:temperature_se}
         but for those 30 ``flat continuum'' sources 
         which show a featureless thermal continuum.
         }
\end{center}
\end{figure*}

\begin{figure*}
\begin{center}
\resizebox{0.6\hsize}{!}{\includegraphics{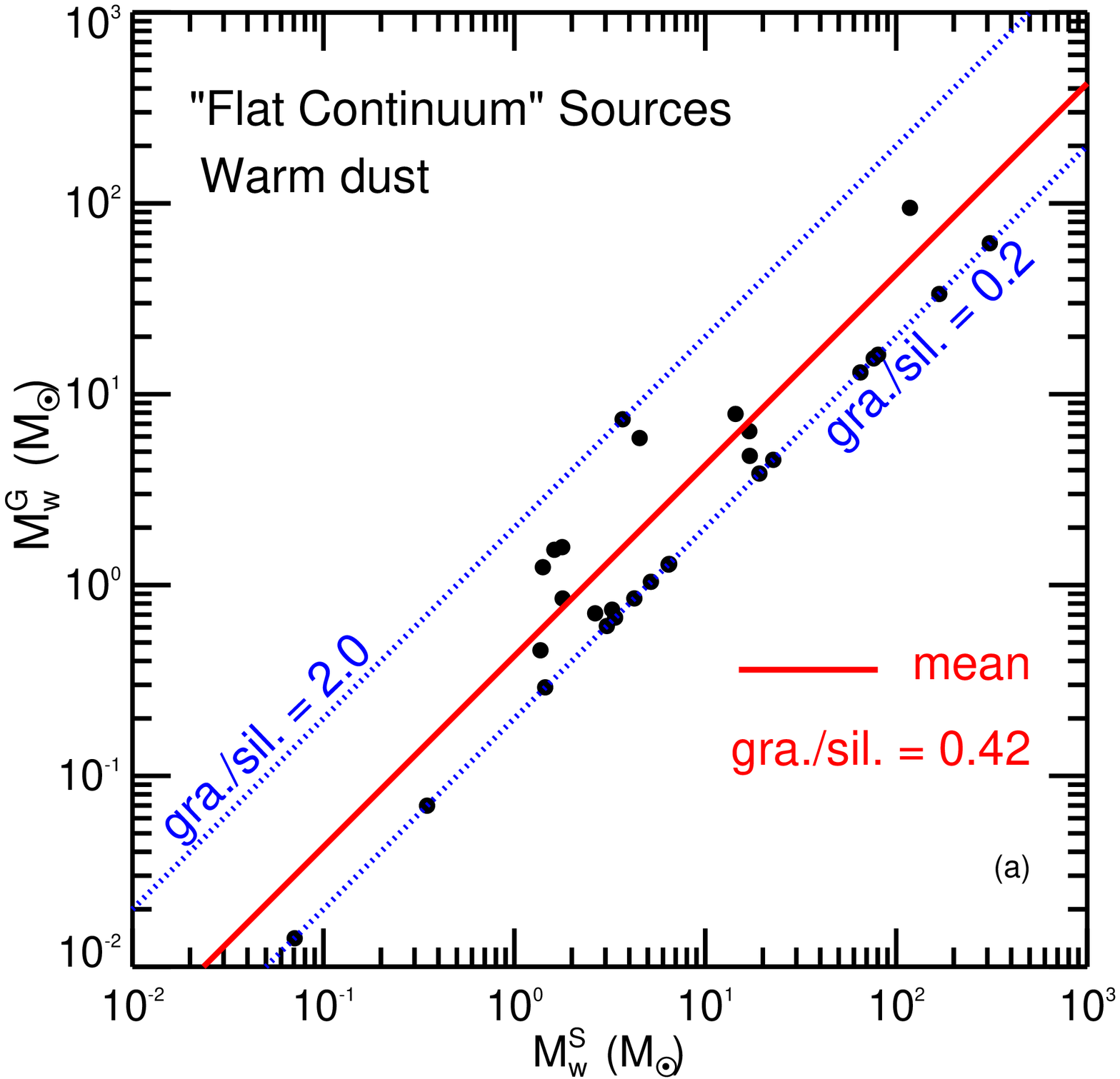}}
\resizebox{0.6\hsize}{!}{\includegraphics{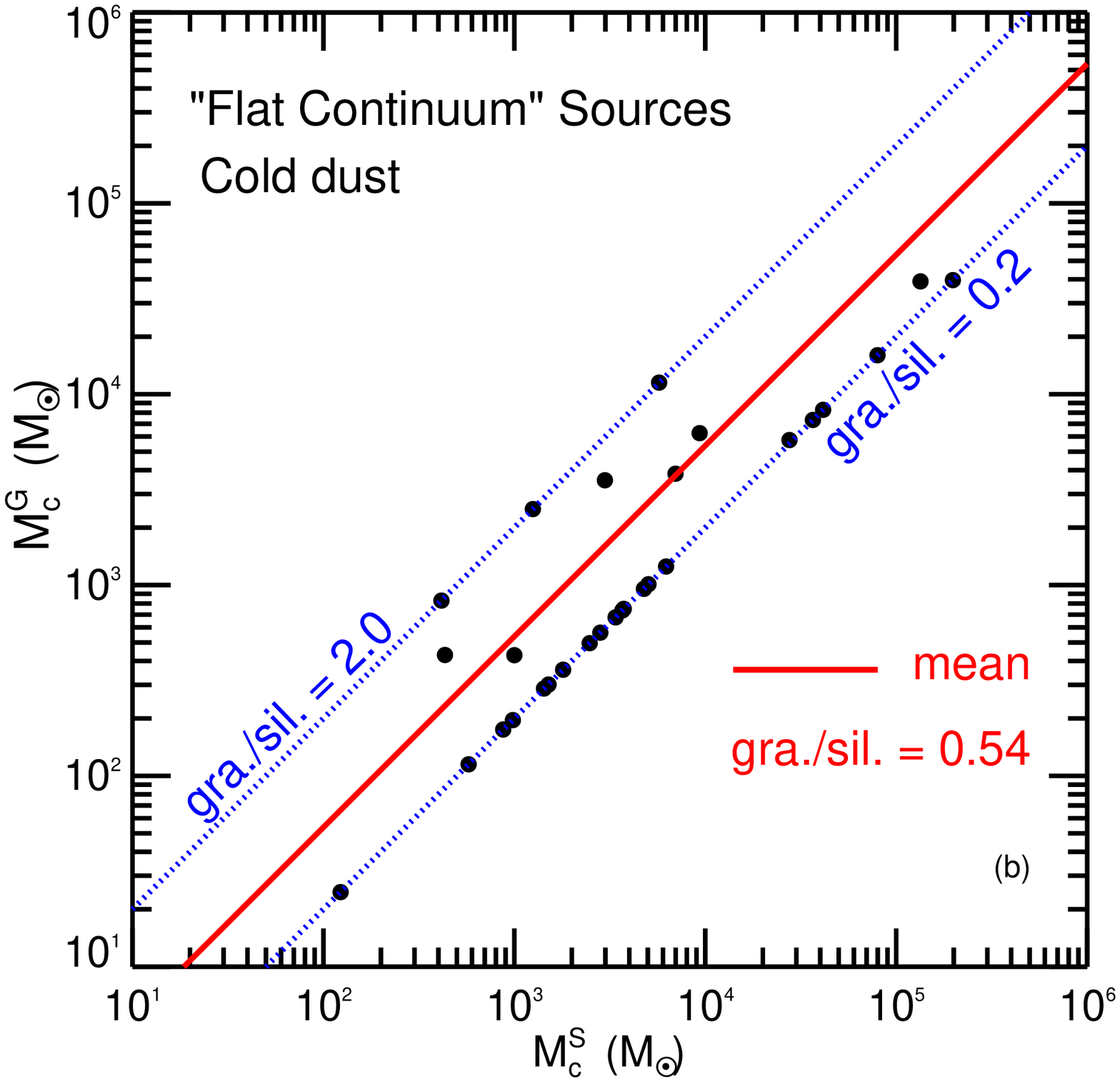}}
\caption{\footnotesize
         \label{fig:mass_fl} 
          Same as Figure~\ref{fig:mass_se}
          but for the 30 ``flat continuum'' sources
          which show a featureless thermal continuum.
         }
\end{center}
\end{figure*}

\begin{figure*}[ht]
\leavevmode
\begin{center}
\resizebox{0.8\hsize}{!}{
\includegraphics[angle=90]{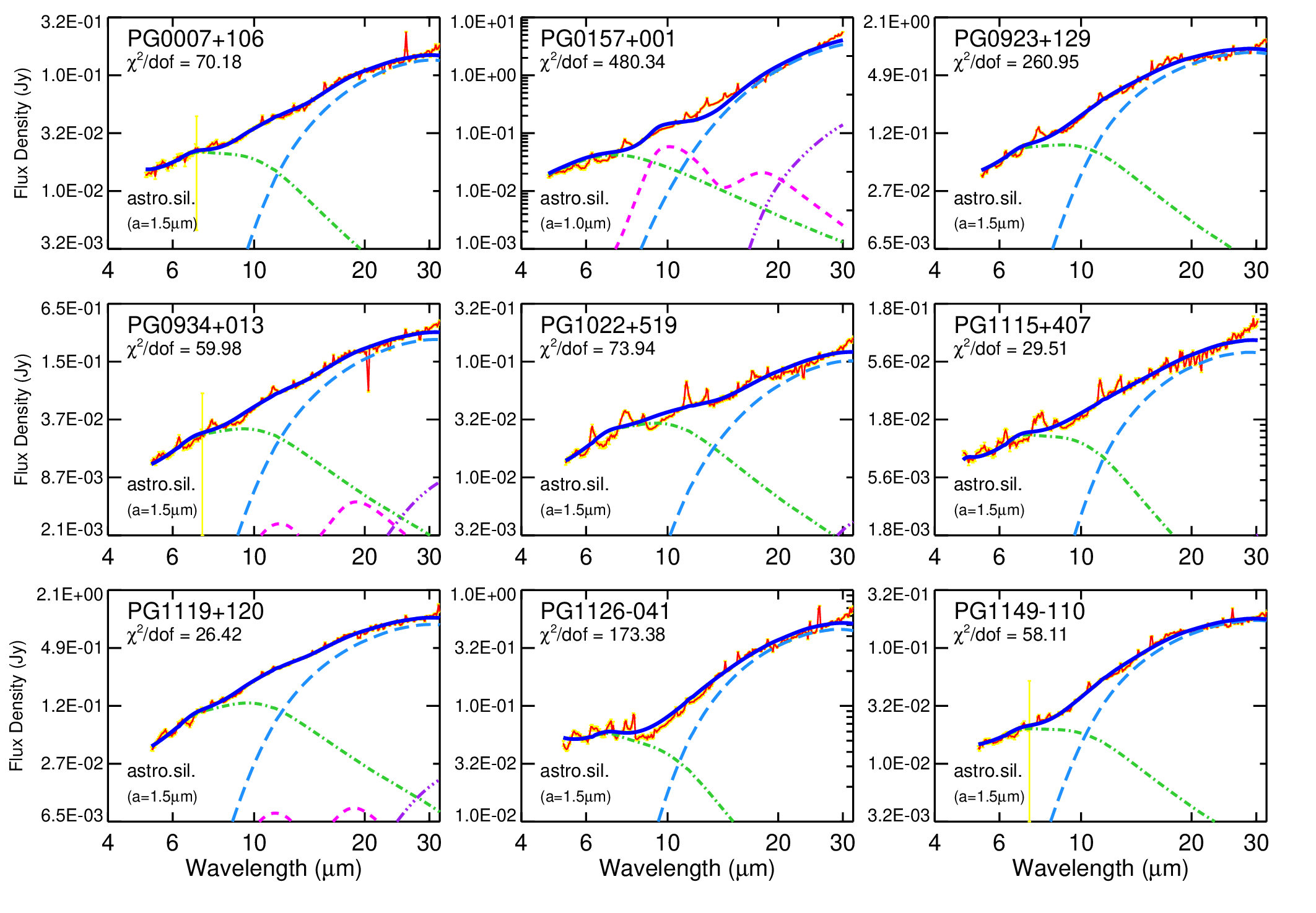}}
\caption{\footnotesize
        \label{fig:PAH_em_mod1}
          Comparison of the {\it Spitzer}/IRS spectra 
          (red solid lines) 
          of the PG quasars PG0007+106, PG0157+001, 
          PG0923+129, PG0934+013, PG1022+519, 
          PG1115+407, PG1119+120, PG1126-041, 
          and PG1149-110 which show a thermal continuum
          superimposed by PAH features
          with the model spectra (blue solid lines)
          which are the sum of warm silicate
          (magenta short dashed lines),
          cold silicate (purple dash-dot-dotted lines),
          warm graphite (green dash-dotted lines),
          and cold graphite (light blue long dashed lines).
          Also shown are the observed $1\sigma$ errors 
          (yellow vertical lines).
          }
\end{center}
\end{figure*}

\begin{figure*}[ht]
\leavevmode
\figurenum{\ref{fig:PAH_em_mod1}}
\begin{center}
\resizebox{0.8\hsize}{!}{
\includegraphics[angle=90]{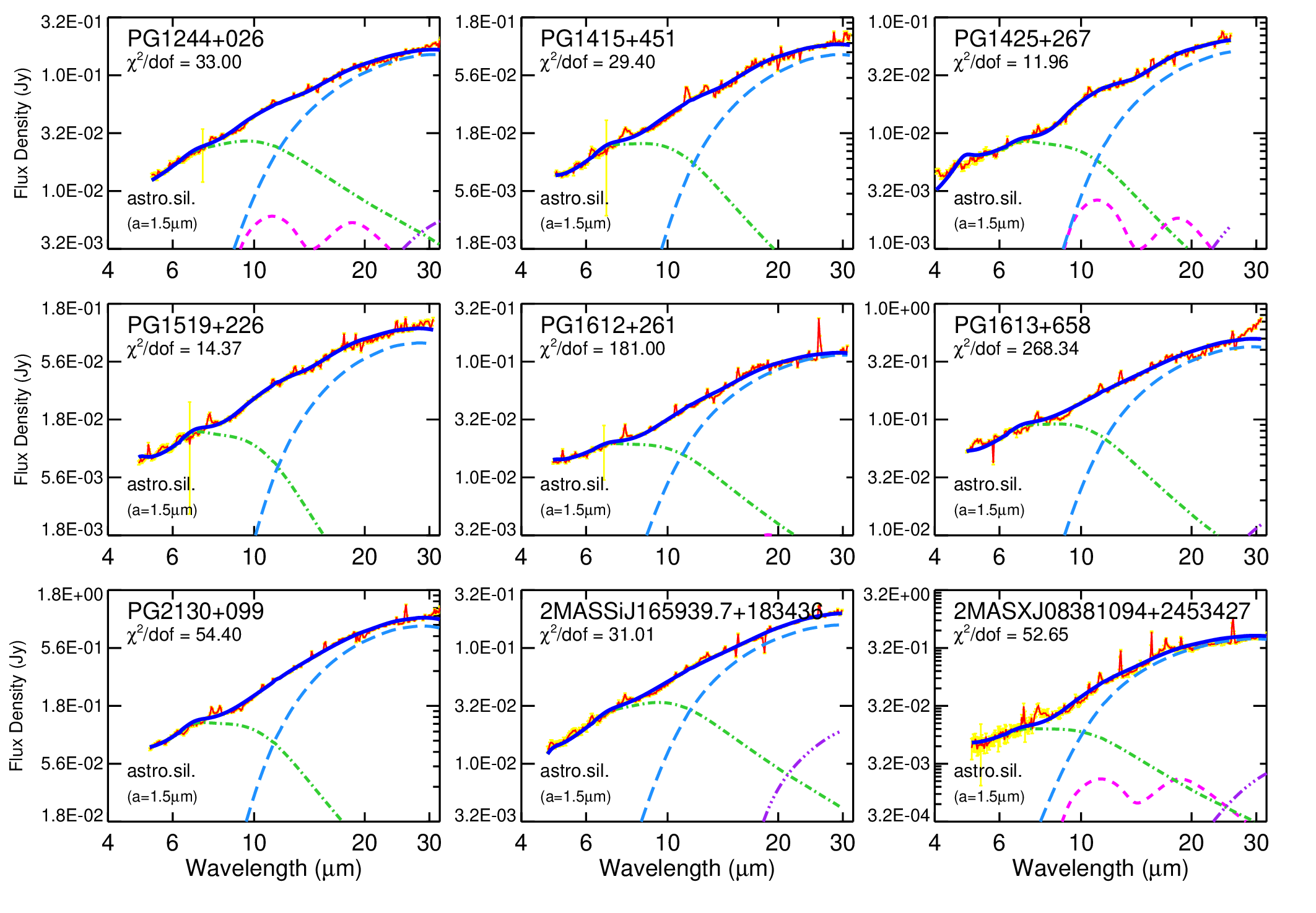}}
\caption{\footnotesize
             Continued,
             but for the PG quasars 
             PG1244+026, PG1415+451, PG1425+267, PG1519+226, 
             PG1612+261, PG1613+658, PG2130+099, 
             the {\it 2MASS} quasar 
             2MASSiJ165939.7+183436, and 
             the \saga\ AGN 2MASXJ08381094+2453427. 
             }
\end{center}
\end{figure*}

\begin{figure*}[ht]
\figurenum{\ref{fig:PAH_em_mod1}}  
\begin{center}
\resizebox{0.8\hsize}{!}{
\includegraphics[angle=90]{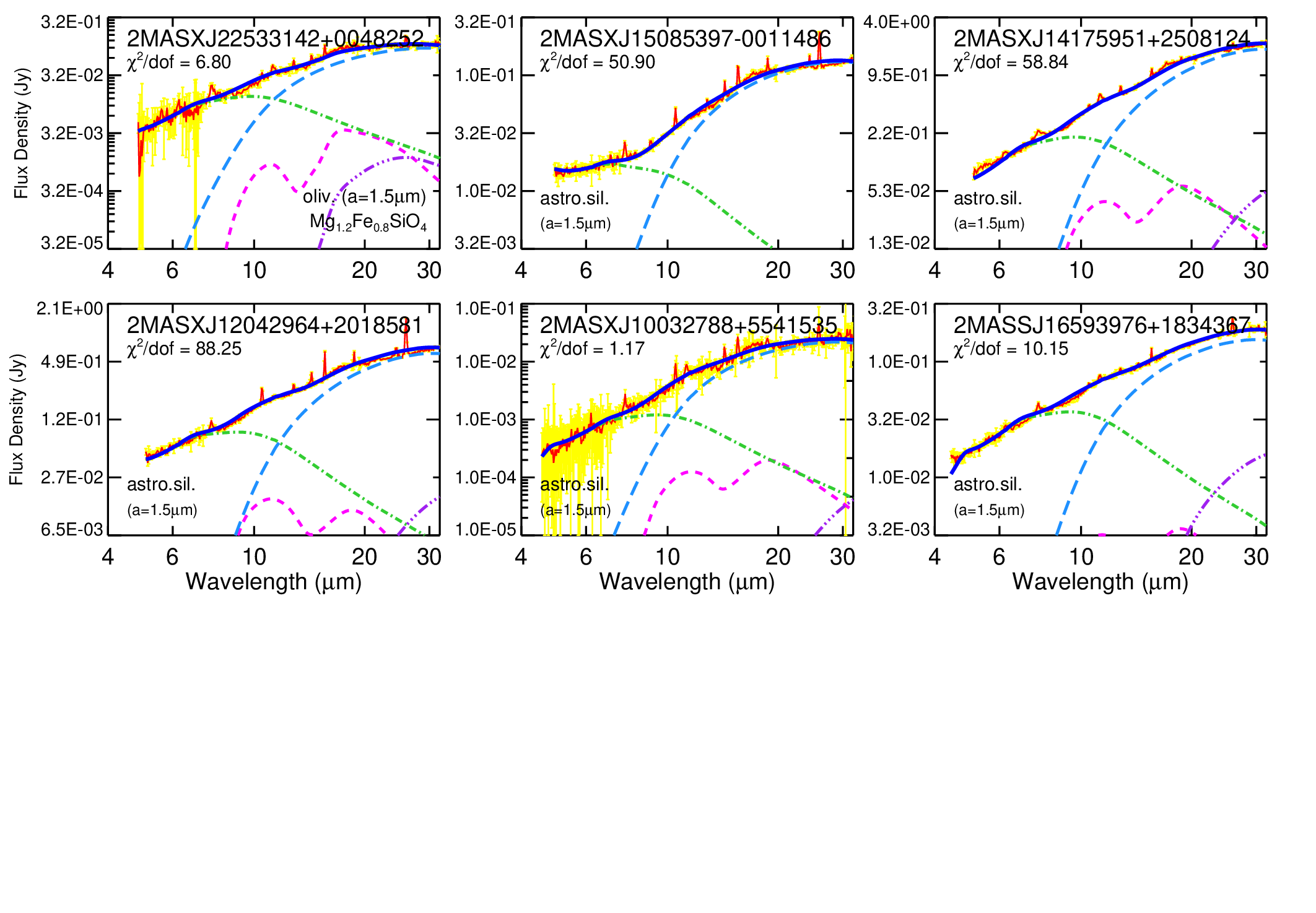}}
\caption{\footnotesize
             Continued, 
             but for the \saga\ AGNs 
             2MASXJ22533142+0048252, 2MASXJ15085397-0011486, 
             2MASXJ14175951+2508124, 2MASXJ12042964+2018581, 
             2MASXJ10032788+5541535, and 2MASSJ16593976+1834367.  
             }
\end{center}
\end{figure*}

\clearpage

\begin{figure*}[ht]
\begin{center}
\resizebox{0.6\hsize}{!}{\includegraphics{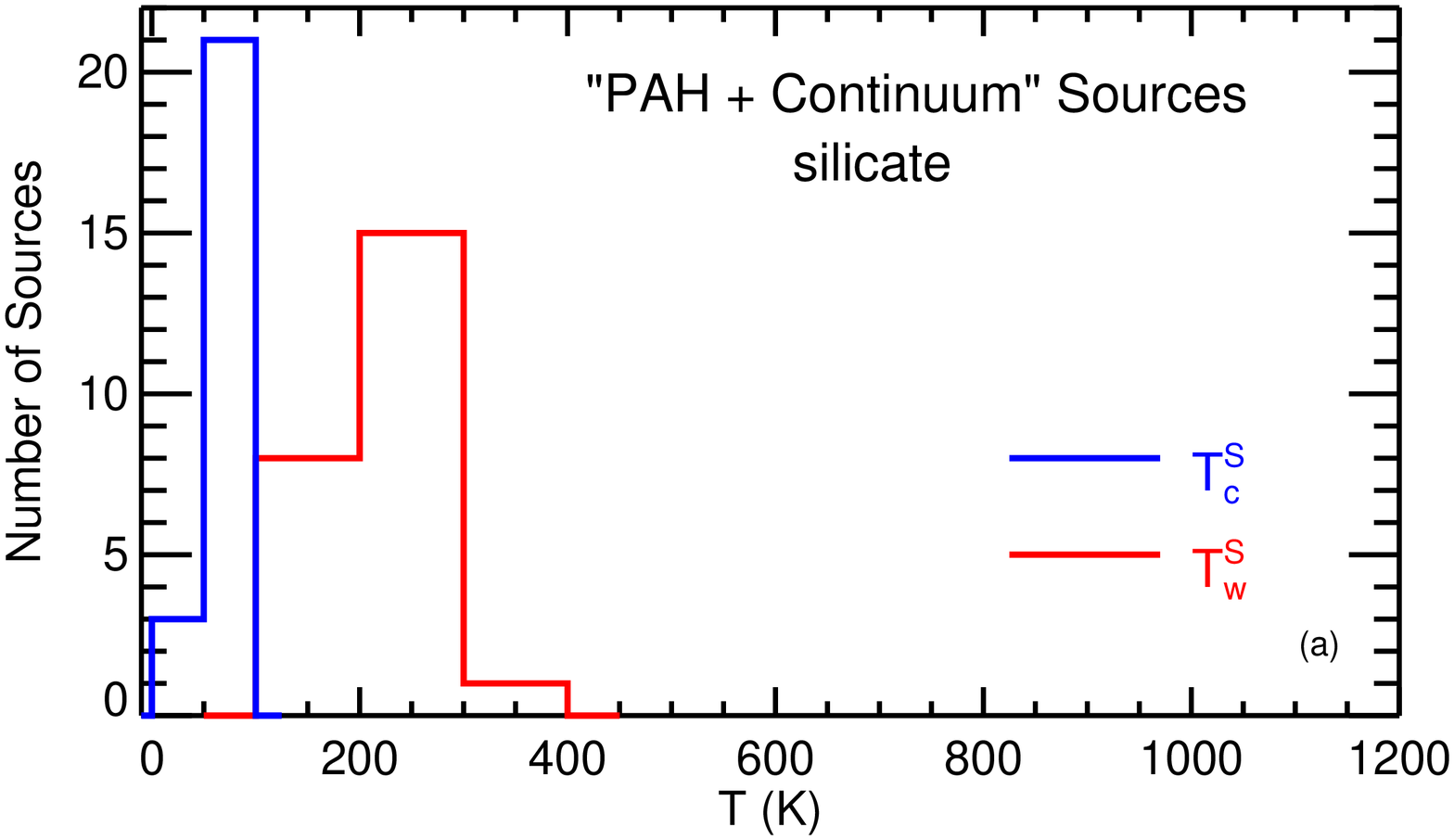}}
\resizebox{0.6\hsize}{!}{\includegraphics{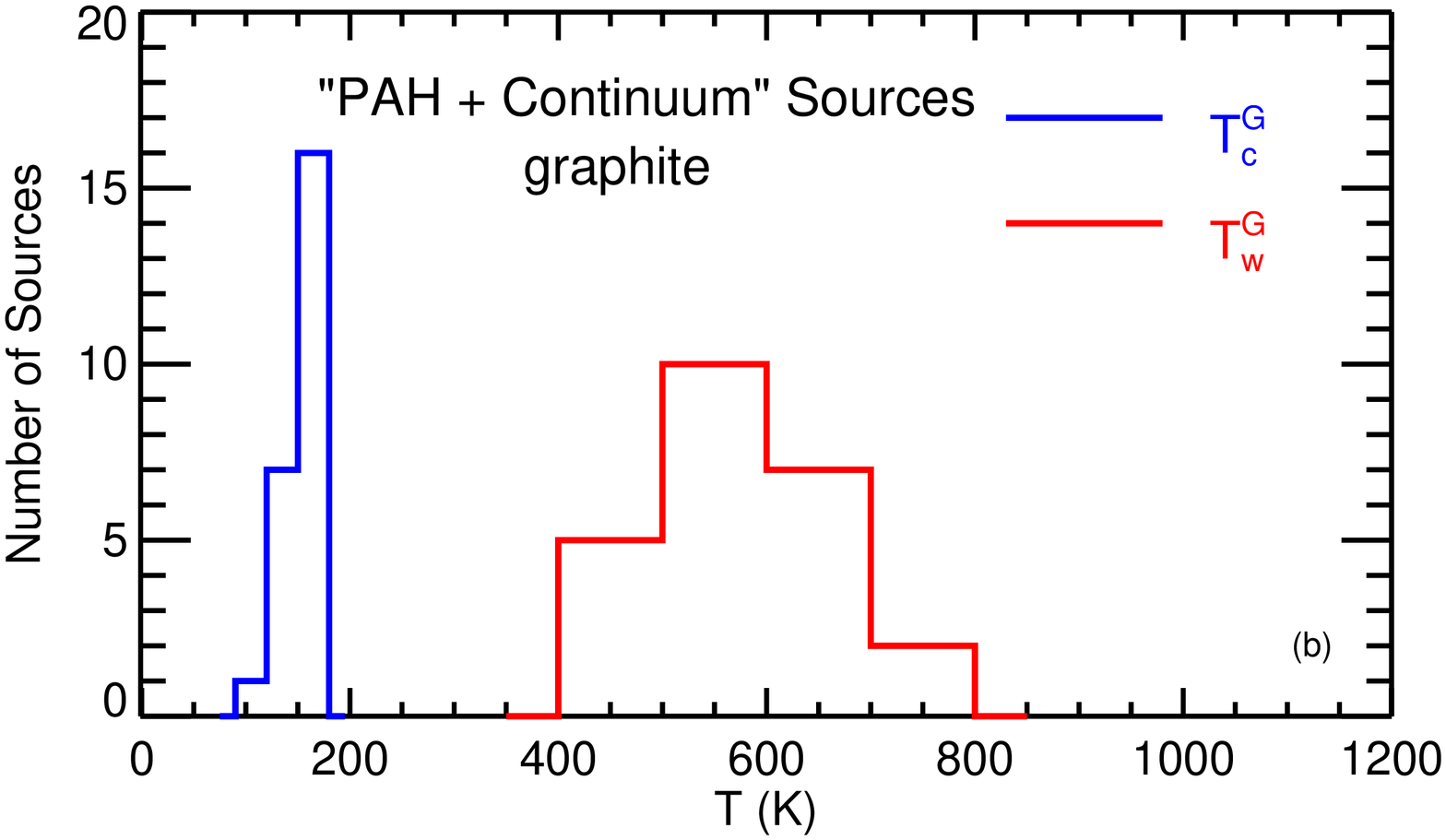}}
\caption{\footnotesize
         \label{fig:temperature_pah}
         Same as Figure~\ref{fig:temperature_se}
         but for those 24 sources 
         which show a thermal continuum
         superimposed with PAH features.
         }
\end{center}
\end{figure*}

\begin{figure*}[ht]
\begin{center}
\resizebox{0.6\hsize}{!}{\includegraphics{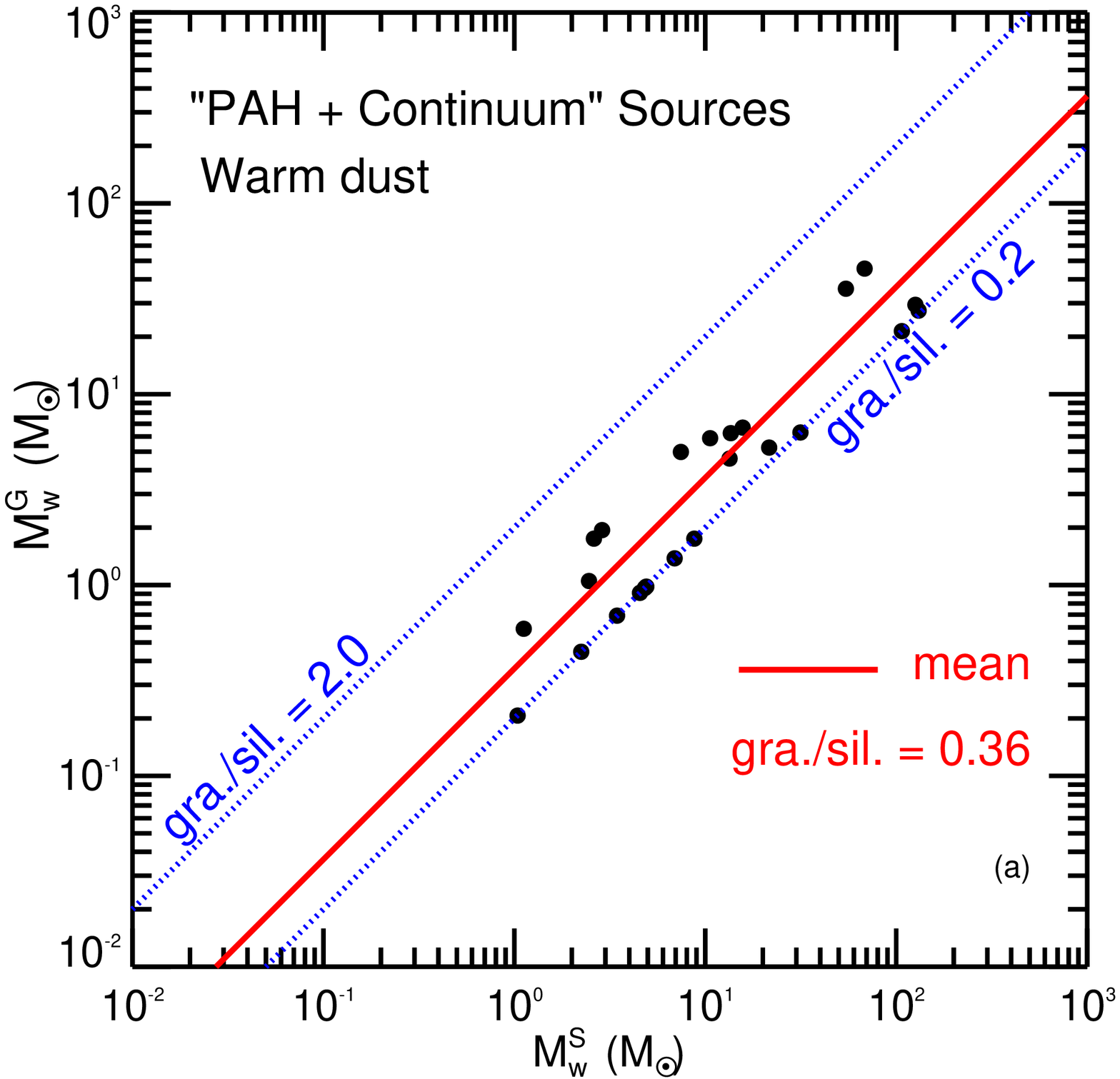}}
\resizebox{0.6\hsize}{!}{\includegraphics{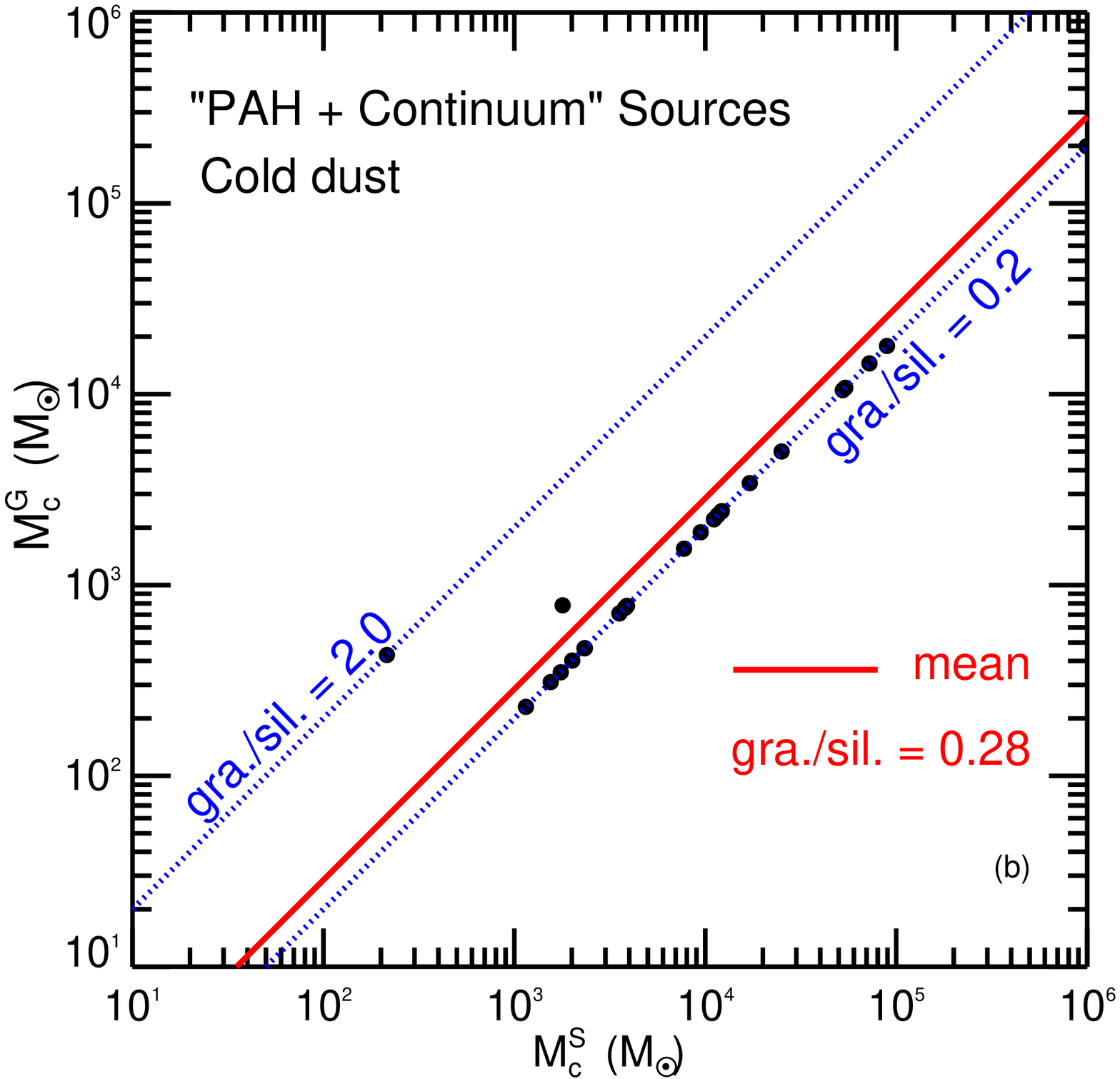}}
\caption{\footnotesize
         \label{fig:mass_pah}
          Same as Figure~\ref{fig:mass_se}
          but for the 24 sources 
          which show a featureless thermal continuum
          superimposed with PAH features.
         }
\end{center}
\end{figure*}

\clearpage

\begin{landscape}
\begin{deluxetable}{lccccccc}
\tabletypesize\tiny
\tablecolumns{8}
\tablecaption{\label{tab:silem93para}
              Basic Parameters of All 93 Sources 
              from Our PG Quasar Sample, 
              {\it 2MASS} Quasar Sample, 
              and \saga\ AGN Sample.
              These Sources All Show Silicate Emission 
              around 9.7 and 18$\mum$. 
              }
\tablewidth{0pt}
\tablehead{
\colhead{Source} & \colhead{R.A.} & \colhead{Dec.} & \colhead{Redshift} & \colhead{Type} & 
\colhead{$\rm \lambda L_{\lambda}(5100 \AA)$} & 
\colhead{$\lg(M_{\rm BH})$} & 
\colhead{Reference} \\
\colhead{} & \colhead{} & \colhead{} & \colhead{} & \colhead{} & 
\colhead{($\rm ergs\,s^{-1}$)} & \colhead{($\rm M_{\odot}$)} &  \colhead{} \\
\colhead{(1)} & \colhead{(2)} & \colhead{(3)} & \colhead{(4)} & \colhead{(5)} & 
\colhead{(6)} & \colhead{(7)} & \colhead{(8)}    
}
\startdata 
PG0003+158               & 00h05m59.20s & +16d09m49.0s   &  0.450   &  1.0  &  $  46.018^{ +0.033  }_{  -0.036  }$  & $  9.270^{+ 0.088  }_{ -0.110 }$ &  Ves2006 \\
PG0003+199               & 00h06m19.52s & +20d12m10.5s   &  0.025   &  1.0  &  $  43.710^{ +0.011  }_{  -0.011  }$  & $  7.192^{+ 0.081  }_{ -0.099 }$ &  Mar2003 \\
PG0026+129               & 00h29m13.60s &  +13d16m03.0s  &  0.142   &  1.0  &  $  44.900^{ +0.070  }_{  -0.070  }$  & $  7.850^{+ 0.120  }_{ -0.120 }$ &  N1987, Kaspi2000 \\ 
PG0043+039               & 00h45m47.27s &  +04d10m24.4s  &  0.384   &  1.0  &  $  45.537^{ +0.030  }_{  -0.032  }$  & $  9.123^{+ 0.085  }_{ -0.105 }$ &  Ves2006 \\
PG0049+171               & 00h51m54.80s &  +17d25m58.4s  &  0.064   &  1.0  &  $  44.004^{ +0.011  }_{  -0.011  }$  & $  8.347^{+ 0.079  }_{ -0.097 }$ &  Ves2006 \\
PG0050+124               & 00h53m34.94s &  +12d41m36.2s  &  0.061   &  1.0  &  $  44.794^{ +0.097  }_{  -0.126  }$  & $  7.441^{+ 0.093  }_{ -0.119 }$ &  Ves2006 \\
PG0052+251               & 00h54m52.10s &  +25d25m38.0s  &  0.155   &  1.0  &  $  44.870^{ +0.070  }_{  -0.070  }$  & $  8.720^{+ 0.100  }_{ -0.100 }$ &  N1987, Kaspi2000 \\
PG0804+761               & 08h10m58.60s &  +76d02m42.0s  &  0.100   &  1.0  &  $  44.930^{ +0.070  }_{  -0.070  }$  & $  8.310^{+ 0.010  }_{ -0.010 }$ &  N1987, Kaspi2000 \\
PG0844+349               & 08h47m42.40s &  +34d45m04.0s  &  0.064   &  1.0  &  $  44.380^{ +0.010  }_{  -0.010  }$  & $  7.975^{+ 0.082  }_{ -0.101 }$ &  Mar2003 \\
PG0921+525               & 09h25m12.87s &  +52d17m10.5s  &  0.035   &  1.0  &  $  43.550^{ +0.120  }_{  -0.120  }$  & $  6.910^{+ 0.120  }_{ -0.120 }$ &  SG1983, WPM1999\\
PG0923+201               & 09h25m54.72s &  +19d54m05.1s  &  0.190   &  1.0  &  $  45.038^{ +0.018  }_{  -0.019  }$  & $  8.009^{+ 0.082  }_{ -0.101 }$ &  Ves2006 \\
PG0947+396               & 09h50m48.39s &  +39d26m50.5s  &  0.206   &  1.0  &  $  44.808^{ +0.020  }_{  -0.021  }$  & $  8.677^{+ 0.081  }_{ -0.099 }$ &  Ves2006 \\
PG0953+414               & 09h56m52.39s &  +41d15m22.3s  &  0.239   &  1.0  &  $  45.300^{ +0.060  }_{  -0.060  }$  & $  8.270^{+ 0.060  }_{ -0.090 }$ &  Ves2002 \\
PG1001+054               & 10h04m20.14s &  +05d13m00.5s  &  0.161   &  1.0  &  $  44.711^{ +0.017  }_{  -0.017  }$  & $  7.738^{+ 0.081  }_{ -0.099 }$ &  Ves2006 \\
PG1004+130               & 10h07m26.10s &  +12d48m56.2s  &  0.240   &  1.0  &  $  45.536^{ +0.022  }_{  -0.023  }$  & $  9.272^{+ 0.084  }_{ -0.104 }$ &  Ves2006 \\
PG1011-040               & 10h14m20.69s &  -04d18m40.5s  &  0.058   &  1.0  &  $  44.259^{ +0.012  }_{  -0.012  }$  & $  7.317^{+ 0.079  }_{ -0.097 }$ &  Ves2006 \\
PG1012+008               & 10h14m54.90s &  +00d33m37.4s  &  0.185   &  1.0  &  $  45.011^{ +0.021  }_{  -0.022  }$  & $  8.247^{+ 0.082  }_{ -0.101 }$ &  Ves2006 \\
PG1048-090               & 10h51m29.90s &  -09d18m10.0s  &  0.344   &  1.0  &  $  45.596^{ +0.027  }_{  -0.029  }$  & $  9.203^{+ 0.085  }_{ -0.105 }$ &  Ves2006 \\
PG1049-005               & 10h51m51.44s &  -00d51m17.7s  &  0.357   &  1.0  &  $  45.633^{ +0.028  }_{  -0.030  }$  & $  9.180^{+ 0.085  }_{ -0.106 }$ &  Ves2006 \\
PG1048+342               & 10h51m43.90s &  +33d59m26.7s  &  0.167   &  1.0  &  $  44.708^{ +0.018  }_{  -0.019  }$  & $  8.369^{+ 0.081  }_{ -0.099 }$ &  Ves2006 \\
PG1100+772               & 11h04m13.69s &  +76d58m58.0s  &  0.313   &  1.0  &  $  45.575^{ +0.026  }_{  -0.027  }$  & $  9.272^{+ 0.085  }_{ -0.105 }$ &  Ves2006 \\
PG1103-006               & 11h06m31.77s &  -00d52m52.5s  &  0.425   &  1.0  &  $  45.667^{ +0.033  }_{  -0.036  }$  & $  9.323^{+ 0.086  }_{ -0.107 }$ &  Ves2006 \\
PG1114+445               & 11h17m06.40s &  +44d13m33.3s  &  0.144   &  1.0  &  $  44.734^{ +0.017  }_{  -0.017  }$  & $  8.591^{+ 0.081  }_{ -0.099 }$ &  Ves2006 \\
PG1116+215               & 11h19m08.68s &  +21d19m18.0s  &  0.177   &  1.0  &  $  45.397^{ +0.018  }_{  -0.019  }$  & $  8.529^{+ 0.083  }_{ -0.103 }$ &  Ves2006 \\
PG1121+422               & 11h24m39.18s &  +42d01m45.0s  &  0.234   &  1.0  &  $  44.883^{ +0.022  }_{  -0.023  }$  & $  8.030^{+ 0.081  }_{ -0.100 }$ &  Ves2006 \\
PG1151+117               & 11h53m49.27s &  +11d28m30.4s  &  0.176   &  1.0  &  $  44.756^{ +0.020  }_{  -0.021  }$  & $  8.549^{+ 0.081  }_{ -0.099 }$ &  Ves2006 \\
PG1202+281               & 12h04m42.11s &  +27d54m11.8s  &  0.165   &  1.0  &  $  44.601^{ +0.027  }_{  -0.029  }$  & $  8.612^{+ 0.081  }_{ -0.099 }$ &  Ves2006 \\
PG1211+143               & 12h14m17.70s &  +14d03m12.6s  &  0.085   &  1.0  &  $  45.071^{ +0.014  }_{  -0.014  }$  & $  7.961^{+ 0.082  }_{ -0.101 }$ &  Ves2006 \\
PG1216+069               & 12h19m20.93s &  +06d38m38.5s  &  0.334   &  1.0  &  $  45.721^{ +0.027  }_{  -0.028  }$  & $  9.196^{+ 0.085  }_{ -0.106 }$ &  Ves2006 \\
PG1229+204               & 12h32m03.60s &  +20d09m29.2s  &  0.063   &  1.0  &  $  44.100^{ +0.012  }_{  -0.012  }$  & $  7.997^{+ 0.080  }_{ -0.098 }$ &  Mar2003 \\
PG1259+593               & 13h01m12.93s &  +59d02m06.7s  &  0.472   &  1.0  &  $  45.906^{ +0.034  }_{  -0.037  }$  & $  8.917^{+ 0.087  }_{ -0.109 }$ &  Ves2006 \\
PG1302-102               & 13h05m33.01s &  -10d33m19.4s  &  0.286   &  1.0  &  $  45.827^{ +0.024  }_{  -0.026  }$  & $  8.879^{+ 0.086  }_{ -0.107 }$ &  Ves2006 \\
PG1307+085               & 13h09m47.00s &  +08d19m48.2s  &  0.155   &  1.0  &  $  45.010^{ +0.028  }_{  -0.028  }$  & $  8.930^{+ 0.096  }_{ -0.123 }$ &  Mar2003 \\
PG1309+355               & 13h12m17.80s &  +35d15m21.0s  &  0.183   &  1.0  &  $  45.041^{ +0.019  }_{  -0.020  }$  & $  8.344^{+ 0.082  }_{ -0.100 }$ &  Ves2006\\
PG1310-108               & 13h13m05.78s &  -11d07m42.4s  &  0.035   &  1.0  &  $  43.725^{ +0.010  }_{  -0.011  }$  & $  7.884^{+ 0.079  }_{ -0.097 }$ &  Ves2006 \\
PG1322+659               & 13h23m49.52s &  +65d41m48.2s  &  0.168   &  1.0  &  $  44.980^{ +0.098  }_{  -0.126  }$  & $  8.281^{+ 0.094  }_{ -0.120 }$ &  Ves2006 \\
PG1341+258               & 13h43m56.75s &  +25d38m47.7s  &  0.087   &  1.0  &  $  44.344^{ +0.097  }_{  -0.126  }$  & $  8.037^{+ 0.092  }_{ -0.117 }$ &  Ves2006 \\
PG1351+640               & 13h53m15.83s &  +63d45m45.7s  &  0.087   &  1.0  &  $  44.835^{ +0.014  }_{  -0.015  }$  & $  8.828^{+ 0.081  }_{ -0.099 }$ &  Ves2006 \\
PG1352+183               & 13h54m35.69s &  +18d05m17.5s  &  0.158   &  1.0  &  $  44.816^{ +0.017  }_{  -0.017  }$  & $  8.423^{+ 0.081  }_{  -0.099}$ &  Ves2006  \\
PG1402+261               & 14h05m16.21s &  +25d55m34.1s  &  0.164   &  1.0  &  $  44.983^{ +0.017  }_{  -0.018  }$  & $  7.944^{+ 0.081  }_{ -0.100 }$ &  Ves2006 \\
PG1404+226               & 14h06m21.89s &  +22d23m46.6s  &  0.098   &  1.0  &  $  44.379^{ +0.017  }_{  -0.018  }$  & $  6.889^{+ 0.080  }_{ -0.098 }$ &  Ves2006 \\
PG1411+442               & 14h13m48.33s &  +44d00m14.0s  &  0.089   &  1.0  &  $  44.620^{ +0.014  }_{  -0.014  }$  & $  8.080^{+ 0.086  }_{ -0.107 }$ &  Mar2003 \\
PG1416-129               & 14h19m03.80s &  -13d10m44.0s  &  0.129   &  1.0  &  $  45.135^{ +0.037  }_{  -0.041  }$  & $  9.045^{+ 0.083  }_{ -0.103 }$ &  Ves2006\\
PG1426+015               & 14h29m06.59s &  +01d17m06.5s  &  0.086   &  1.0  &  $  44.740^{ +0.070  }_{  -0.070  }$  & $  8.750^{+ 0.100  }_{ -0.100 }$ &  N1987, Kaspi2000\\
PG1435-067               & 14h38m16.16s &  -06d58m21.3s  &  0.129   &  1.0  &  $  44.918^{ +0.036  }_{  -0.040  }$  & $  8.365^{+ 0.082  }_{ -0.102 }$ &  Ves2006 \\
PG1444+407               & 14h46m45.94s &  +40d35m05.8s  &  0.267   &  1.0  &  $  45.203^{ +0.023  }_{  -0.024  }$  & $  8.289^{+ 0.083  }_{ -0.102 }$ &  Ves2006 \\
PG1512+370               & 15h14m43.04s &  +36d50m50.4s  &  0.371   &  1.0  &  $  45.602^{ +0.030  }_{  -0.032  }$  & $  9.373^{+ 0.085  }_{ -0.106 }$ &  Ves2006 \\
PG1534+580               & 15h35m52.36s &  +57d54m09.2s  &  0.030   &  1.0  &  $  43.687^{ +0.010  }_{   -0.011 }$  & $  8.203^{+ 0.080  }_{ -0.097 }$ &  Ves2006\\
PG1535+547               & 15h36m38.36s &  +54d33m33.2s  &  0.038   &  1.0  &  $  43.961^{ +0.010  }_{  -0.011  }$  & $  7.192^{+ 0.079  }_{ -0.097 }$ &  Ves2006\\
PG1545+210               & 15h47m43.54s &  +20d52m16.6s  &  0.266   &  1.0  &  $  45.428^{ +0.023  }_{  -0.024  }$  & $  9.314^{+ 0.084  }_{ -0.104 }$ &  Ves2006 \\
PG1552+085               & 15h54m44.58s &  +08d22m21.5s  &  0.119   &  1.0  &  $  44.407^{ +0.015  }_{  -0.015  }$  & $  7.537^{+ 0.080  }_{ -0.099 }$ &  Ves2006 \\
PG1617+175               & 16h20m11.29s &  +17d24m27.7s  &  0.114   &  1.0  &  $  44.375^{ +0.069  }_{  -0.082  }$  & $  8.436^{+ 0.115  }_{ -0.191 }$ &  Kaspi2000 \\
PG1626+554               & 16h27m56.12s &  +55d22m31.5s  &  0.133   &  1.0  &  $  44.580^{ +0.026  }_{   0.028  }$  & $  8.498^{+ 0.081  }_{ -0.099 }$ &  Ves2006 \\
PG1700+518               & 17h01m24.80s &  +51d49m20.0s  &  0.292   &  1.0  &  $  45.470^{ +0.010  }_{  -0.010  }$  & $  8.427^{+ 0.109  }_{ -0.146 }$ &  Mar2003 \\
PG1704+608               & 17h04m41.38s &  +60d44m30.5s  &  0.371   &  1.0  &  $  45.702^{ +0.030  }_{  -0.032  }$  & $  9.391^{+ 0.086  }_{ -0.107 }$ &  Ves2006 \\
PG2112+059               & 21h14m52.57s &  +06d07m42.5s  &  0.466   &  1.0  &  $  46.181^{ +0.034  }_{  -0.037  }$  & $  9.001^{+ 0.089  }_{ -0.112 }$ &  Ves2006 \\
PG2209+184               & 22h11m53.89s &  +18d41m49.9s  &  0.070   &  1.0  &  $  44.469^{ +0.012  }_{  -0.013  }$  & $  8.766^{+ 0.080  }_{ -0.098 }$ &  Ves2006\\
PG2214+139               & 22h17m12.26s &  +14d14m20.9s  &  0.067   &  1.0  &  $  44.662^{ +0.097  }_{  -0.126  }$  & $  8.551^{+ 0.093  }_{ -0.118 }$ &  Ves2006\\
PG2233+134               & 22h36m07.68s &  +13d43m55.3s  &  0.325   &  1.0  &  $  45.327^{ +0.027  }_{  -0.028  }$  & $  8.036^{+ 0.083  }_{ -0.103 }$ &  Ves2006 \\
PG2251+113               & 22h54m10.40s &  +11d36m38.3s  &  0.323   &  1.0  &  $  45.692^{ +0.026  }_{  -0.028  }$  & $  8.989^{+ 0.085  }_{ -0.106 }$ &  Ves2006 \\
PG2304+042               & 23h07m02.91s &  +04d32m57.2s  &  0.042   &  1.0  &  $  44.066^{ +0.097  }_{  -0.126  }$  & $  8.564^{+ 0.092  }_{ -0.117 }$ &  Ves2006\\
PG2308+098               & 23h11m17.76s &  +10d08m15.5s  &  0.432   &  1.0  &  $  45.777^{ +0.101  }_{  -0.131  }$  & $  9.592^{+ 0.098  }_{ -0.126 }$ &  Ves2006 \\
2MASSiJ081652.2+425829   & 08h16m52.24s &  +42d58m29.4s  &  0.235   &  1.0  &  $  44.605^{ +0.002  }_{  -0.002  }$  & $  8.27^{+ 0.01   }_{ -0.01  }$ &  Shen2011 \\
2MASSiJ095504.5+170556   & 09h55m04.55s &  +17d05m56.4s  &  0.139   &  1.0  &  $  45.130^{ +0.000  }_{  -0.002  }$  & $  7.27^{+ 0.38   }_{ -0.38  }$ &  Shen2011 \\
2MASSiJ130005.3+163214   & 13h00m05.35s &  +16d32m14.8s  &  0.080   &  1.0  &     --                                  &    --                              &     --   \\
2MASSiJ132917.5+121340   & 13h29m17.52s &  +12d13m40.2s  &  0.203   &  1.0  &     --                                  &    --                              &     --   \\
2MASSiJ1402511+263117    & 14h02m51.20s &  +26d31m17.6s  &  0.187   &  1.0  &  $  44.565^{ +0.039  }_{  -0.039  }$  & $  8.57^{+ 0.03   }_{ -0.03  }$ &  Shen2011\\
2MASSiJ145608.6+275008   & 14h56m08.65s &  +27d50m08.8s  &  0.250   &  1.0  &  $  44.670^{ +0.015  }_{  -0.015  }$  & $  7.95^{+ 0.01   }_{ -0.01  }$ &  Shen2011\\
2MASSiJ151653.2+190048   & 15h16m53.23s &  +19d00m48.3s  &  0.190   &  1.0  &  $  44.200^{ +0.000  }_{  -0.000  }$  & $  8.22^{+ 0.00   }_{ -0.00  }$ &  Shen2011\\
2MASSiJ151901.5+183804   & 15h19m01.48s &  +18d38m04.9s  &  0.187   &  1.0  &     --                                  &    --                              &   --  \\
2MASSiJ154307.7+193751   & 15h43m07.78s &  +19d37m51.8s  &  0.228   &  1.5  &     --                                  &    --                              &  -- \\
2MASSiJ222221.1+195947   & 22h22m21.14s &  +19d59m47.1s  &  0.211   &  1.0  &     --                                  &    --                              &  -- \\
2MASSiJ223742.6+145614   & 22h37m42.60s &  +14d56m14.0s  &  0.277   &  1.0  &     --                                  &    --                              &  -- \\
2MASSiJ234259.3+134750   & 23h42m59.36s &  +13d47m50.4s  &  0.299   &  1.5  &  $  44.682^{ +0.005  }_{  -0.005  }$  & $  8.41^{+ 0.02   }_{ -0.02  }$ &  Shen2011 \\
2MASSiJ234449.5+122143   & 23h44m49.56s &  +12d21m43.1s  &  0.199   &  1.0  &     --                                  &    --                              &  -- \\
2MASXJ09210862+4538575   & 09h21m08.06s &  +45d38m57.0s  &  0.175   &  1.0  &  $  43.972^{ +0.004  }_{ -0.004   }$  & $ 8.83^{+  0.04  }_{ -0.04  }$ &   Shen2011\\
2MASXJ00370409-0109081   & 00h37m04.01s &  -01d09m08.0s  &  0.074   &  1.0  &  $  45.170^{ +0.000  }_{ -0.000   }$  & $ 8.44^{+  0.50  }_{ -0.50  }$ &   Shen2011\\
2MASXJ02335161+0108136   & 02h33m51.60s &  +01d08m14.0s  &  0.022   &  1.0  &     --                                  &    --                              &   --\\
2MASXJ07582810+3747121   & 07h58m28.01s &  +37d47m12.0s  &  0.041   &  1.0  &  $  45.330^{ +0.000  }_{ -0.000   }$  & $ 8.90^{+  0.53  }_{ -0.53  }$ &   Tri2013\\
2MASXiJ0208238-002000    & 02h08m23.08s &  -00d20m01.0s  &  0.074   &  1.0  &     --                                  &   --                               &   -- \\
2MASXJ02061600-0017292   & 02h06m16.00s &  -00d17m29.0s  &  0.043   &  1.0  &  $  43.850^{ +0.000  }_{ -0.000   }$  & $ 8.18^{+  0.00  }_{ -0.00  }$ &   Du2014 \\
2MASXJ10493088+2257523   & 10h49m30.90s &  +22d57m52.0s  &  0.033   &  1.0  &     --                                  &     --                             &   -- \\
2MASXJ12485992-0109353   & 12h48m59.90s &  -01d09m35.0s  &  0.089   &  1.0  &     --                                  &   --                               &   --  \\
2MASXJ14070036+2827141   & 14h07m00.40s &  +28d27m15.0s  &  0.077   &  1.0  &     --                                  &   --                               &   --  \\
2MASXJ02143357-0046002   & 02h14m33.50s &  -00d46m00.0s  &  0.026   &  1.0  &     --                                  &   --                               &   --   \\
2MASXJ09234300+2254324   & 09h23m43.00s &  +22d54m33.0s  &  0.033   &  1.0  &  $  45.104^{ +0.000  }_{ -0.000   }$  & $ 7.58^{+  0.04  }_{ -0.04  }$ &   Dasyra.2011\\
2MASXJ12170991+0711299   & 12h17m09.90s &  +07d11m30.0s  &  0.008   &  1.0  &  $  42.080^{ +0.000  }_{ -0.000   }$  & $  7.71^{+  0.00  }_{ -0.00  }$ &   Woo2012\\
2MASXJ12232410+0240449   & 12h23m24.10s &  +02d40m45.0s  &  0.024   &  1.0  &  $  42.830^{ +0.006  }_{ -0.006   }$  & $ 7.51^{+  0.07  }_{ -0.07  }$ &   HK2014\\
2MASXJ13381586+0432330   & 13h38m15.90s &  +04d32m33.0s  &  0.023   &  1.0  &  $  45.146^{ +0.000  }_{ -0.000   }$  & $  7.56^{+  0.00  }_{ -0.00  }$ &   WZ2007 \\
2MASXJ13495283+0204456   & 13h49m52.80s &  +02d04m45.0s  &  0.033   &  1.0  &  $  42.394^{ +0.000  }_{ -0.000   }$  & $ 6.66^{+  0.00  }_{ -0.00  }$ &   Zhu2010\\
2MASXJ23044349-0841084   & 23h04m43.50s &  -08d41m09.0s  &  0.047   &  1.0  &  $  43.720^{ +0.000  }_{ -0.000   }$  & $ 8.47^{+  0.00  }_{ -0.00  }$ &   WL2004\\
SDSSJ115138.24+004946.4  & 11h51m38.20s &  +00d49m47.0s  &  0.195   &  1.0  &  $  42.790^{ +0.040  }_{ -0.040   }$  & $ 6.00^{+  0.00  }_{ -0.00  }$ &   GH2008\\
SDSSJ170246.09+602818.8  & 17h02m46.10s &  +60d28m19.0s  &  0.069   &  1.0  &     --                                  &    --                              &  --\\
\enddata
\tablecomments{
Column (1): AGN name; 
Column (2): Right ascension of the AGN; 
Column (3): Declination of the AGN; 
Column (4): Redshift $z$; 
Column (5): Optical classification derived 
            from the emission line ratio of the AGN: 
            ``1.0'' for type 1 AGN with broad emission lines,
            ``2.0'' for type 2 AGN with only narrow emission lines, 
            values between 1.0 and 2.0 for intermediate types;
Column (6): Power emitted at $\lambda = 5100\Angstrom$; 
Column (7): Black hole mass (in unit of solar mass $M_{\odot}$); 
Column (8): References from which we collect 
            $\lambda L_{\lambda}(5100\Angstrom)$ and $M_{\rm BH}$ --- 
                    N1987 = Neugebauer \etal (1987);
                    Mar2003  = Marziani \etal (2003);
                    SG1983  = Schmidt \& Green (1983);
                    WPM1999 = Wandel, Peterson \& Malkan (1999); 
                    Kaspi2000 = Kaspi \etal (2000); 
                    WL2004 = Wu \& Liu (2004); 
                    WZ2007 = Wang \& Zhang (2007); 
                    GH2008 = Greene \& Ho (2008); 
                    Ves2006 = Vestergaard \etal (2006);
                    Dong2010 = Dong \etal (2010); 
                    Sani2010 = Sani \etal (2010); 
                    Zhu2010 = Zhu \etal (2010); 
                    Dasyra2011 = Dasyra \etal (2011); 
                    Shen2011 = Shen \etal (2011); 
                    Woo2012 = Woo \etal (2012); 
                    Tri2013 = Trichas \etal (2013); 
                    Rose2013 = Rose \etal (2013); 
                    Du2014 = Du \etal 2014; 
                    HK2014 = Ho \& Kim (2014). 
}
\end{deluxetable}
\end{landscape}

\begin{landscape}
\begin{deluxetable}{lccccccc}
\tabletypesize\tiny
\tablecolumns{8}
\tablecaption{\label{tab:conti30para}
              Basic Parameters of 30 Sources 
              from Our PG Quasar Sample, 
              {\it 2MASS} Quasar Sample, 
              and \saga\ AGN Sample 
              Which Show No Silicate Emission
              but a Featureless Thermal Continuum
              }
\tablewidth{0pt}
\tablehead{
\colhead{Source} & \colhead{R.A.} & \colhead{Dec.} & \colhead{Redshift} & \colhead{Type} & 
\colhead{$\rm \lambda L_{\lambda}(5100 \AA)$} & 
\colhead{$\lg(M_{\rm BH})$} & 
\colhead{Reference} \\
\colhead{} & \colhead{} & \colhead{} & \colhead{} & \colhead{} & 
\colhead{($\rm ergs\,s^{-1}$)} & \colhead{($\rm M_{\odot}$)} &  \colhead{} \\
\colhead{(1)} & \colhead{(2)} & \colhead{(3)} & \colhead{(4)} & \colhead{(5)} & 
\colhead{(6)} & \colhead{(7)} & \colhead{(8)}    
}
\startdata 
PG0838+770               & 08h44m45.26s &  +76d53m09.5s  &  0.131   &  1.0  &  $  44.727^{ +0.015  }_{  -0.015  }$  & $  8.154^{+ 0.080  }_{ -0.099 }$ &  Ves2006 \\
PG1226+023               & 12h29m06.70s &  +02d03m08.6s  &  0.158   &  1.0  &  $  46.060^{ +0.014  }_{  -0.014  }$  & $  9.262^{+ 0.130  }_{ -0.186 }$ &  Mar2003 \\
PG1354+213               & 13h56m32.80s &  +21d03m52.4s  &  0.300   &  1.0  &  $  44.977^{ +0.072  }_{  -0.086  }$  & $  8.627^{+ 0.088  }_{ -0.110 }$ &  Ves2006 \\
PG1427+480               & 14h29m43.07s &  +47d47m26.2s  &  0.221   &  1.0  &  $  44.759^{ +0.021  }_{  -0.022  }$  & $  8.088^{+ 0.081  }_{ -0.099 }$ &  Ves2006 \\
PG1448+273               & 14h51m08.76s &  +27d09m26.9s  &  0.065   &  1.0  &  $  44.482^{ +0.011  }_{  -0.011  }$  & $  6.970^{+ 0.080  }_{ -0.098 }$ &  Ves2006 \\
PG1501+106               & 15h04m01.20s &  +10d26m16.2s  &  0.036   &  1.0  &  $  44.285^{ +0.010  }_{  -0.011  }$  & $  8.523^{+ 0.079  }_{ -0.097 }$ &  Ves2006 \\
PG1543+489               & 15h45m30.24s &  +48d46m09.1s  &  0.400   &  1.0  &  $  45.445^{ +0.037  }_{  -0.041  }$  & $  7.998^{+ 0.085  }_{ -0.105 }$ &  Ves2006 \\
2MASSiJ010835.1+214818   & 01h08m35.10s &  +21d48m18.0s  &  0.285   &  1.9  &     --                                  &    --                              &   --      \\
2MASSiJ024807.3+145957   & 02h48m07.36s &  +14d59m57.7s  &  0.072   &  1.0  &  $  43.910^{ +0.170  }_{  -0.170  }$  & $ 7.46^{+  0.00  }_{ -0.00  }$ &   Rose2013\\
2MASSiJ082311.3+435318   & 08h23m11.27s &  +43d53m18.5s  &  0.182   &  1.5  &     --                                  &    --                              &   --\\
2MASSiJ145410.1+195648   & 14h54m10.17s &  +19d56m48.7s  &  0.243   &  1.9  &     --                                  &    --                              &   --\\
2MASXJ17223993+3052521   & 17h22m39.90s &  +30d52m53.0s  &  0.043   &  1.0  &     --                                  &    --                              &   --\\
2MASXJ13130577+0127561   & 13h13m05.80s &  +01d27m56.0s  &  0.029   &  1.0  &     --                                  &    --                              &   --\\ 
2MASXJ13130565-0210390   & 13h13m05.70s &  -02d10m39.0s  &  0.084   &  1.0  &     --                                  &    --                              &   --\\
2MASXJ15055659+0342267   & 15h05m56.50s &  +03d42m26.0s  &  0.036   &  1.0  &  $   43.450^{ +0.000  }_{  -0.000  }$  & $  7.55^{+ 0.00   }_{ -0.00  }$ &  Dong2010 \\ %
2MASXJ09191322+5527552   & 09h19m13.20s &  +55d27m55.0s  &  0.049   &  1.0  &     --                                  &    --                              &  --\\
2MASXJ12384342+0927362   & 12h38m43.40s &  +09d27m37.0s  &  0.083   &  1.0  &     --                                  &    --                              &  -- \\
2MASXJ16164729+3716209   & 16h16m47.30s &  +37d16m21.0s  &  0.152   &  1.0  &     --                                  &    --                              &  --\\
2MASXJ11230133+4703088   & 11h23m01.30s &  +47d03m09.0s  &  0.025   &  1.0  &     --                                  &    --                              &  --  \\
2MASXJ11110693+0228477   & 11h11m06.90s &  +02d28m48.0s  &  0.035   &  1.0  &     --                                  &    --                              &  --\\
2MASSiJ1448250+355946    & 14h48m25.10s &  +35d59m47.0s  &  0.113   &  1.0  &     --                                  &    --                              &  -- \\
SDSSJ090738.71+564358.2  & 09h07m38.70s &  +56d43m58.0s  &  0.099   &  1.0  &     --                                  &    --                              &   --\\
SDSSJ124035.81-002919.4  & 12h40m35.80s &  -00d29m19.0s  &  0.081   &  1.0  &                                         &    --                              &   --\\
SDSSJ101536.21+005459.3  & 10h15m36.20s &  +00d54m59.0s  &  0.120   &  2.0  &     --                                  &    --                              &  --\\
SDSSJ164840.15+425547.6  & 16h48m40.10s &  +42d55m48.0s  &  0.129   &  1.0  &     --                                  &    --                              &  --\\
SDSSJ091414.34+023801.7  & 09h14m14.30s &  +02d38m02.0s  &  0.073   &  2.0  &     --                                  &    --                              &  --\\
SDSSJ164019.66+403744.4  & 16h40m19.70s &  +40d37m45.0s  &  0.151   &  1.0  &     --                                  &    --                              &  --\\
SDSSJ104058.79+581703.3  & 10h40m58.70s &  +58d17m04.0s  &  0.071   &  1.0  &     --                                  &    --                              &  --\\
UGC05984                 & 10h52m16.70s &  +30d03m55.0s  &  0.035   &  2.0  &      --                                 &      --                            &  -- \\
UGC06527                 & 11h32m37.60s &  +52d56m53.0s  &  0.026   &  1.0  &      --                                 &      --                            &  -- \\
\enddata
\tablecomments{
Column (1): AGN name; 
Column (2): Right ascension of the AGN; 
Column (3): Declination of the AGN; 
Column (4): Redshift $z$; 
Column (5): Optical classification derived 
            from the emission line ratio of the AGN: 
            ``1.0'' for type 1 AGN with broad emission lines,
            ``2.0'' for type 2 AGN with only narrow emission lines, 
            values between 1.0 and 2.0 for intermediate types;
Column (6): Power at $\lambda = 5100\Angstrom$; 
Column (7): Black hole mass (in unit of solar mass $M_{\odot}$); 
Column (8): References from which we collect 
            $\lambda L_{\lambda}(5100\Angstrom)$ 
            and $M_{\rm BH}$ --- 
            Mar2003  = Marziani \etal (2003);
            Ves2006 = Vestergaard \etal (2006);
            Dong2010 = Dong \etal (2010); 
            Rose2013 = Rose \etal (2013). 
            }
\end{deluxetable}
\end{landscape}

\begin{landscape}
\begin{deluxetable}{lccccccc}
\tabletypesize\tiny
\tablecolumns{8}
\tablecaption{\label{tab:pah24para}
              Basic Parameters of 24 Sources 
              from Our PG Quasar Sample, 
              {\it 2MASS} Quasar Sample, 
              and \saga\ AGN Sample 
              Which Show a Thermal Continuum
              Superimposed with PAH Features
              }
\tablewidth{0pt}
\tablehead{
\colhead{Source} & \colhead{R.A.} & \colhead{Dec.} & \colhead{Redshift} & \colhead{Type} & 
\colhead{$\rm \lambda L_{\lambda}(5100 \AA)$} & 
\colhead{$\lg(M_{\rm BH})$} & 
\colhead{Reference} \\
\colhead{} & \colhead{} & \colhead{} & \colhead{} & \colhead{} & 
\colhead{($\rm ergs\,s^{-1}$)} & \colhead{($\rm M_{\odot}$)} &  \colhead{} \\
\colhead{(1)} & \colhead{(2)} & \colhead{(3)} & \colhead{(4)} & \colhead{(5)} & 
\colhead{(6)} & \colhead{(7)} & \colhead{(8)}    
}
\startdata 
PG0007+106               & 00h10m31.01s &  +10d58m29.5s  &  0.089   &  1.0  &  $  44.816^{ +0.014  }_{  -0.015  }$  & $  8.728^{+ 0.081  }_{ -0.099 }$ &  Ves2006\\
PG0157+001               & 01h59m50.21s &  +00d23m40.6s  &  0.164   &  1.0  &  $  44.975^{ +0.017  }_{  -0.018  }$  & $  8.166^{+ 0.081  }_{ -0.100 }$ &  Ves2006\\
PG0923+129               & 09h26m03.29s &  +12d44m03.6s  &  0.029   &  1.0  &  $  43.860^{ +0.097  }_{  -0.125  }$  & $  8.598^{+ 0.092  }_{ -0.117 }$ &  Ves2006\\
PG0934+013               & 09h37m01.03s &  +01d05m43.5s  &  0.050   &  1.0  &  $  43.875^{ +0.097  }_{  -0.126  }$  & $  7.041^{+ 0.092  }_{ -0.117 }$ &  Ves2006\\
PG1022+519               & 10h25m31.28s &  +51d40m34.9s  &  0.045   &  1.0  &  $  43.696^{ +0.097  }_{  -0.126  }$  & $  7.145^{+ 0.092  }_{ -0.117 }$ &  Ves2006\\
PG1115+407               & 11h18m30.29s &  +40d25m54.0s  &  0.154   &  1.0  &  $  44.619^{ +0.017  }_{  -0.018  }$  & $  7.667^{+ 0.080  }_{ -0.099 }$ &  Ves2006\\
PG1119+120               & 11h21m47.10s &  +11d44m18.3s  &  0.049   &  1.0  &  $  44.132^{ +0.012  }_{  -0.012  }$  & $  7.470^{+ 0.079  }_{ -0.097 }$ &  Ves2006\\
PG1126-041               & 11h29m16.66s &  -04d24m07.6s  &  0.060   &  1.0  &  $  44.385^{ +0.012  }_{  -0.012  }$  & $  7.749^{+ 0.080  }_{ -0.098 }$ &  Ves2006\\
PG1149-110               & 11h52m03.54s &  -11d22m24.3s  &  0.049   &  1.0  &  $  44.107^{ +0.097  }_{  -0.126  }$  & $  7.924^{+ 0.092  }_{ -0.117 }$ &  Ves2006\\
PG1244+026               & 12h46m35.25s &  +02d22m08.8s  &  0.048   &  1.0  &  $  43.801^{ +0.030  }_{  -0.032  }$  & $  6.523^{+ 0.080  }_{ -0.099 }$ &  Ves2006 \\
PG1415+451               & 14h17m00.70s &  +44d56m06.0s  &  0.114   &  1.0  &  $  44.561^{ +0.017  }_{  -0.018  }$  & $  8.014^{+ 0.080  }_{ -0.098 }$ &  Ves2006\\
PG1425+267               & 14h27m35.61s &  +26d32m14.5s  &  0.366   &  1.0  &  $  45.761^{ +0.100  }_{  -0.130  }$  & $  9.734^{+ 0.097  }_{ -0.126 }$ &  Ves2006\\
PG1519+226               & 15h21m14.26s &  +22d27m43.9s  &  0.137   &  1.0  &  $  44.710^{ +0.019  }_{  -0.020  }$  & $  7.942^{+ 0.081  }_{ -0.099 }$ &  Ves2006\\
PG1612+261               & 16h14m13.20s &  +26d04m16.2s  &  0.131   &  1.0  &  $  44.717^{ +0.026  }_{  -0.028  }$  & $  8.058^{+ 0.081  }_{ -0.100 }$ &  Ves2006\\
PG1613+658               & 16h13m57.18s &  +65d43m09.6s  &  0.129   &  1.0  &  $  44.650^{ +0.070  }_{  -0.070  }$  & $  8.990^{+ 0.10   }_{ -0.10  }$ &  N1987, Kaspi2000\\
PG2130+099               & 21h32m27.81s &  +10d08m19.5s  &  0.063   &  1.0  &  $  44.280^{ +0.011  }_{  -0.011  }$  & $  7.947^{+ 0.081  }_{ -0.099 }$ &  Mar2003\\
2MASSiJ165939.7+183436   & 16h59m39.77s &  +18d34m36.8s  &  0.170   &  1.0  &  $  44.241^{ +0.006  }_{  -0.006  }$  & $  8.950^{+  0.03  }_{  -0.03 }$ &  Shen2011 \\
2MASXJ08381094+2453427   & 08h38m10.90s &  +24d53m43.0s  &  0.029   &  1.0  &     --                                  &    --                              &   -- \\
2MASXJ22533142+0048252   & 22h53m31.40s &  +00d48m26.0s  &  0.072   &  1.0  &     --                                  &    --                              &   -- \\
2MASXJ15085397-0011486   & 15h08m53.90s &  -00d11m49.0s  &  0.054   &  1.0  &     --                                  &    --                              &   -- \\
2MASXJ14175951+2508124   & 14h17m59.50s &  +25d08m12.0s  &  0.016   &  1.0  &     --                                  &    --                              &   -- \\
2MASXJ12042964+2018581   & 12h04m29.70s &  +20d18m58.0s  &  0.023   &  1.0  &     --                                  &    --                              &   -- \\
2MASXJ10032788+5541535   & 10h03m27.90s &  +55d41m54.0s  &  0.146   &  2.0  &     --                                  &    --                              &   -- \\
2MASSJ16593976+1834367   & 16h59m39.80s &  +18d34m37.0s  &  0.171   &  1.0  &     --                                  &    --                              &   -- \\
\enddata
\tablecomments{
Column (1): AGN name; 
Column (2): Right ascension of the AGN; 
Column (3): Declination of the AGN; 
Column (4): Redshift $z$; 
Column (5): Optical classification derived 
            from the emission line ratio of the AGN: 
            ``1.0'' for type 1 AGN with broad emission lines,
            ``2.0'' for type 2 AGN with only narrow emission lines, 
            values between 1.0 and 2.0 for intermediate types;
Column (6): Power at $\lambda = 5100\Angstrom$; 
Column (7): Black hole mass (in unit of solar mass $M_{\odot}$); 
Column (8): References from which we collect 
            $\lambda L_{\lambda}(5100\Angstrom)$ 
            and $M_{\rm BH}$ --- 
            N1987 = Neugebauer \etal (1987);
            Mar2003  = Marziani \etal (2003)
            Kaspi2000 = Kaspi \etal (2000); 
            Ves2006 = Vestergaard \etal (2006);
            Shen2011 = Shen \etal (2011). 
            }
\end{deluxetable}
\end{landscape}


\end{document}